\documentclass[runningheads,orivec]{llncs}

\usepackage[dvipsnames]{xcolor}
\usepackage[T1]{fontenc}
\usepackage{anyfontsize}
\usepackage{textcomp}
\usepackage[utf8x]{inputenc}

\usepackage{amsmath}
\usepackage{amssymb}
\usepackage{url}
\usepackage{csquotes}
\usepackage{longtable}

\usepackage{orcidlink}

\usepackage{algorithm}
\usepackage{algpseudocode}

\usepackage{mdframed}

\usepackage[bibliography=common]{apxproof}

\usepackage{paralist}

\usepackage{graphicx}

\usepackage{adjustbox}
\usepackage{varwidth}

\usepackage{microtype}

\usepackage[capitalise]{cleveref}
\crefname{problem}{Problem}{Problems}
\crefname{definition}{Def.}{Defs.}
\usepackage{stmaryrd}

\usepackage{nicefrac}
\usepackage{underscore}
\usepackage{booktabs}
\usepackage{wrapfig}

\usepackage{cite}

\makeatletter
\newsavebox{\@brx}
\newcommand{\llangle}[1][]{\savebox{\@brx}{\(\m@th{#1\langle}\)}%
  \mathopen{\copy\@brx\kern-0.5\wd\@brx\usebox{\@brx}}}
\newcommand{\rrangle}[1][]{\savebox{\@brx}{\(\m@th{#1\rangle}\)}%
  \mathclose{\copy\@brx\kern-0.5\wd\@brx\usebox{\@brx}}}
\makeatother

\newtheoremrep{theorem}{Theorem}
\newtheoremrep{lemma}{Lemma}
\newtheoremrep{proposition}{Proposition}

\renewcommand{\paragraph}[1]{\smallskip\noindent\emph{#1}~}
\renewcommand{\subsubsection}[1]{\medskip\noindent\textbf{#1}~}

\usepackage{tikz, pgf, tikz-cd}
\usetikzlibrary{automata, positioning, shapes.geometric, shapes.symbols, patterns, fit}
\tikzset{elliptic state/.style={draw,ellipse}}
\tikzstyle{block} = [rectangle, draw,
    text width=2.5cm, text centered, rounded corners, minimum height=4em]
\tikzstyle{decision} = [diamond, draw,
    text width=1.5cm, text badly centered, inner sep=2pt]

\tikzset{every state/.style={inner sep=0pt, minimum size=15pt}}
\tikzset{pMC/.style={node distance=1.5cm, on grid, auto, initial text=,every label/.style={font=\footnotesize}}}
\usepackage{pgfplots}
\pgfplotsset{compat=newest}
\usepgfplotslibrary{groupplots}
\usepgfplotslibrary{polar}
\usepgfplotslibrary{statistics}
\usepgfplotslibrary{fillbetween}

\usepackage{xspace}
\newcounter{goodsmileys}
\newcounter{badsmileys}
\newcommand{\Smiley}[2]{%
	\begin{tikzpicture}[-,scale=#2]
	\newcommand*{\SmileyRadius}{1.0}%
	\pgfmathsetmacro{\eyeX}{0.5*\SmileyRadius*cos(30)}
	\pgfmathsetmacro{\eyeY}{0.3*\SmileyRadius*sin(30)}
	\draw [line width=0.28mm] (\eyeX-0.25,\eyeY) -- (\eyeX-0.25,\eyeY+0.4);
	\draw [line width=0.28mm] (-\eyeX+0.25,\eyeY) -- (-\eyeX+0.25,\eyeY+0.4);

	\pgfmathsetmacro{\xScale}{2*\eyeX/180}
	\pgfmathsetmacro{\yScale}{1.0*\eyeY}
	\draw[line width=0.28mm, domain=-\eyeX:\eyeX]
	plot ({\x},{
		-0.1+#1*0.15 %
		-#1*1.75*\yScale*(sin((\x+\eyeX)/\xScale))-\eyeY});
	\end{tikzpicture}%
}%
\newcommand{\good}{%
  \stepcounter{goodsmileys}%
  \Smiley{0.6}{0.22}\xspace%
}
\newcommand{\bad}{%
  \stepcounter{badsmileys}%
  \Smiley{-0.6}{0.22}\xspace%
}
\newcommand{\opt}{%
    \mathop{\mathrm{opt}}
}
\newcommand{\isub}{%
    \mathrm{isub}
}

\newcommand\M{\mathcal{M}}
\newcommand\D{\mathcal{D}}
\newcommand\I{\mathcal{I}}
\newcommand\MC{\mathrm{MC}}

\newcommand{\elimstate}[1]{\langle#1\rangle}
\newcommand{\intervals}{\mathrm{Int}(\mathbb{Q})}

\newcommand{\ttra}{\mathtt{t}}

\usepackage{pifont}
\newcommand{\cmark}{\ding{51}}%
\newcommand{\xmark}{\ding{55}}%
\newcommand{\yes}{\textcolor{ForestGreen}{\cmark}}
\newcommand{\no}{\textcolor{red}{\xmark}}

\usepackage[strut=off]{subcaption}

\begin{document}
\title{Generalized Parameter Lifting: Finer Abstractions for Parametric Markov Chains\thanks{This work has been partially funded by the NWO Veni Grant ProMiSe (222.147), DFG RTG
2236/2 (UnRAVeL), the KI-Starter Project ‘Verifying
AI Systems under Partial Observability’ of the Ministry
of Culture and Science of the German State of North
Rhine-Westphalia and the European Union’s Horizon 2020
research and innovation programme under the Marie
Skłodowska-Curie grant agreement No. 101008233 (MIS-
SION). }}
\author{%
  Linus~Heck\inst{1}\orcidlink{0000-0002-4774-7609} \and Tim~Quatmann\orcidlink{0000-0002-2843-5511}\inst{2} \and Jip~Spel\orcidlink{0000-0002-9113-2791}\inst{2} \and \\ Joost-Pieter~Katoen\inst{2}\orcidlink{0000-0002-6143-1926} \and  Sebastian~Junges\orcidlink{0000-0003-0978-8466}\inst{1}
}
\institute{Radboud University, Nijmegen, the Netherlands\\ \email{\{linus.heck,sebastian.junges\}@ru.nl} \and RWTH Aachen University, Aachen, Germany\\ \email{\{tim.quatmann,jip.spel,katoen\}@cs.rwth-aachen.de}}

\maketitle

\begin{abstract}
Parametric Markov chains (pMCs) are Markov chains (MCs) with symbolic probabilities.
A pMC encodes a family of MCs, where each member is obtained by replacing parameters with constants.
The parameters allow encoding dependencies between transitions, which sets pMCs apart from interval MCs.
The verification problem for pMCs asks whether each MC in the corresponding family satisfies a given temporal specification.
The state-of-the-art approach for this problem is parameter lifting (PL)---an abstraction-refinement loop that abstracts the pMC to a non-parametric model analyzed with standard probabilistic model checking techniques.
This paper presents two key improvements to tackle the main limitations of PL.
First, we introduce generalized parameter lifting (GPL) to lift various restrictive assumptions made by PL.
Second, we present a big-step transformation algorithm that reduces parameter dependencies in pMCs and, therefore, results in tighter approximations.
Experiments show that GPL is widely applicable and that the big-step transformation accelerates pMC verification by up to orders of magnitude.
\end{abstract}

\begin{toappendix}
\section{Proofs}
\label{app:proofs}
\end{toappendix}

\section{Introduction}%
\label{sec:intro}
\emph{Markov chains} (MCs) describe system behavior under probabilistic uncertainty: They are used to model hardware circuits with faults, network communication over unreliable channels, and randomized protocols for distributed systems. Given an MC, probabilistic model checking tools like Storm~\cite{DBLP:journals/sttt/HenselJKQV22} or Prism~\cite{DBLP:conf/cav/KwiatkowskaNP11} can determine, e.g., the probability of a system failure or the expected time until a successful packet transmission.
However, verification results are only valid for fixed transition probabilities---which may not be known exactly---and it is unclear how sensitive results are to perturbations of these probabilities.

\paragraph{Parametric MCs.}
A variety of \emph{uncertain MCs} allow representing uncertainty about the probabilities as a first-class citizen~\cite{DBLP:journals/sttt/BadingsSSJ23}.
Prominent examples are \emph{interval MCs} (iMCs)~\cite{DBLP:journals/ai/GivanDG03, DBLP:conf/lics/JonssonL91}, where transition probabilities are given by intervals, and \emph{parametric MCs} (pMCs)~\cite{DBLP:journals/fac/LanotteMT07, DBLP:conf/ictac/Daws04}.  This paper improves the ability to verify the latter. 
In pMCs, we consider a finite set of symbols, called \emph{parameters}. 
Contrary to (parameter-free) MCs, transition probabilities are polynomial functions over these parameters.
By replacing the parameters with fixed values, we obtain MCs.
A pMC is a generator for a set of MCs, given by all possible parameter instantiations.
The main advantage of pMCs over iMCs is their ability to model \emph{dependencies} between different states: by using the same parameter, we can encode that, e.g., the probability of successful network transmission is dependent on the value of a counter on the receiver. Dependencies are crucial for encoding finite memory policies in partially observable MDPs (POMDPs) as pMCs~\cite{DBLP:conf/uai/Junges0WQWK018}.

\begin{figure}[t]
    \begin{subfigure}{0.33\linewidth}\centering
            \begin{tikzpicture}[pMC]
                \centering
                \node[state, initial below] (s0) {$s_0$};
                \node[state] (s1) [right of=s0] {$s_1$};
                \node[state] (s2) [right of=s1] {$s_2$};
                \node[state] (s3) [below of=s2, yshift=0.5cm] {$\good$};
                \node[state, draw=lightgray, text=lightgray] (bad) [below of=s1, yshift=0.5cm] {\(\bad\)};

                \path[->, draw=lightgray] (s0) edge (bad);
                \path[->, draw=lightgray] (s1) edge (bad);
                \path[->, draw=lightgray] (s2) edge (bad);

                \path[->] (s0) edge node[above] {$p$} (s1);
                \path[->] (s1) edge node[above] {$1-p$} (s2);
                \path[->] (s2) edge node[left] {$q$} (s3);
                \path[->] (s3) edge[loop left] node[left] {$1$} (s3);
                \path[->, draw=lightgray] (bad) edge[loop left] node[left, text=lightgray] {$1$} (bad);
            \end{tikzpicture}
        \caption{parametric Markov chain}
        \label{fig:expmc}
    \end{subfigure}%
    \begin{subfigure}{0.33\linewidth}\centering
            \begin{tikzpicture}[pMC]
                \centering
                \node[state, initial below] (s0) {$s_0$};
                \node[state] (s1) [right of=s0] {$s_1$};
                \node[state] (s2) [right of=s1] {$s_2$};
                \node[state] (s3) [below of=s2, yshift=0.5cm] {$\good$};
                \node[state, draw=lightgray, text=lightgray] (bad) [below of=s1, yshift=0.5cm] {\(\bad\)};

                \path[->, draw=lightgray] (s0) edge (bad);
                \path[->, draw=lightgray] (s1) edge (bad);
                \path[->, draw=lightgray] (s2) edge (bad);

                \path[->] (s0) edge node[above] {$0.4$} (s1);
                \path[->] (s1) edge node[above] {$0.6$} (s2);
                \path[->] (s2) edge node[left] {$0.7$} (s3);
                \path[->] (s3) edge[loop left] node[left] {$1$} (s3);
                \path[->, draw=lightgray] (bad) edge[loop left] node[left, text=lightgray] {$1$} (bad);
            \end{tikzpicture}
        \caption{Markov chain}
        \label{fig:exmc}
    \end{subfigure}%
    \begin{subfigure}{0.33\linewidth}\centering
            \begin{tikzpicture}[pMC]
                \centering
                \node[state, initial below] (s0) {$s_0$};
                \node[state] (s1) [right of=s0] {$s_1$};
                \node[state] (s2) [right of=s1] {$s_2$};
                \node[state] (s3) [below of=s2, yshift=0.5cm] {$\good$};
                \node[state, draw=lightgray, text=lightgray] (bad) [below of=s1, yshift=0.5cm] {\(\bad\)};

                \path[->, draw=lightgray] (s0) edge (bad);
                \path[->, draw=lightgray] (s1) edge (bad);
                \path[->, draw=lightgray] (s2) edge (bad);

                \path[->] (s0) edge node[above] {\scriptsize$[0.3, 0.6]$} (s1);
                \path[->] (s1) edge node[above] {\scriptsize$[0.4, 0.7]$} (s2);
                \path[->] (s2) edge node[left] {\scriptsize$[0.6, 0.7]$} (s3);
                \path[->] (s3) edge[loop left] node[left] {$1$} (s3);
                \path[->, draw=lightgray] (bad) edge[loop left] node[left, text=lightgray] {$1$} (bad);
            \end{tikzpicture}
        \caption{interval Markov chain}
        \label{fig:eximc}
    \end{subfigure}
    \caption{Different types of (uncertain) MCs}
    \vspace{-3mm}
\end{figure}
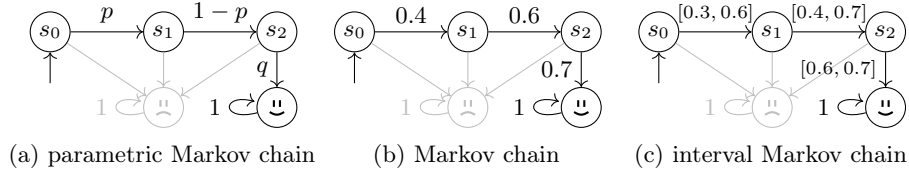

\begin{example}
    Consider the pMC $\D$ in \cref{fig:expmc} over parameters $p$ and $q$. 
Replacing them in $\D$ using a parameter instantiation $u \colon p \mapsto 0.4, q \mapsto 0.7$ yields the MC in \cref{fig:exmc}.
We can also replace the parameters with intervals given by a parameter \emph{region} \(R = [0.3, 0.6] \times [0.6, 0.7]\), which yields the iMC in \cref{fig:eximc}.
\label{ex:pmcexample}
\end{example}

\paragraph{Decision problems for pMCs.} 
Parameter instantiations are mappings from parameters to their domain.
A pMC $\D$ and an instantiation $u$ together define an instantiated MC $\D[u]$.
Regions describe sets of parameter instantiations with a geometric interpretation as rectangular sets of points in Euclidean space.
Given a pMC $\D$, a region $R$, and a temporal specification $\varphi$, two classical problems on pMCs are \emph{feasibility}: \emph{Is there a parameter instantiation $u \in R$ such that $\D[u]$ satisfies $\varphi$?} and its dual problem, \emph{verification}: \emph{Does $\D[u]$ satisfy $\varphi$ for every instantiation $u \in R$?} The verification problem is particularly relevant to demonstrate that a system is robust against perturbations of the parameter assignments and it is a subroutine to parameter space partitioning~\cite[Section~9]{DBLP:journals/fmsd/JungesAHJKQV24}.  The feasibility problem is $\exists\mathbb{R}$-complete~\cite{DBLP:journals/jcss/JungesK0W21}, i.e., it is as hard as answering whether a multivariate polynomial has a real-valued root~\cite{DBLP:journals/corr/abs-2407-18006}, while the verification problem is co-$\exists\mathbb{R}$-complete. In contrast, verifying iMCs is possible in polynomial time~\cite{DBLP:conf/cav/PuggelliLSS13}. 

\begin{example}
\label{ex:verificationexample}
    Consider \(\D\) and \(R\) from \cref{ex:pmcexample}. Two verification problem instances are:
\emph{Is the probability to reach \(\good\) in \(\D\) below 20\% for
        all instantiations in \(R\)?}
 and 
\emph{Is the probability also below 15\%}?
The former holds, as the global maximum probability in \(R\) is 17.5\% at \(u \colon p \mapsto 0.5, q \mapsto 0.7\).
For the latter problem, \(u\) is a counterexample.
On the other hand, the iMC in \cref{fig:eximc} violates both specifications as its maximum probability is 29.4\%.
In the pMC, the instantiated transition probabilities at states $s_0$ and $s_1$ are dependent as both refer to the same parameter $p$.
Such global dependencies are no longer present in the iMC.
 \end{example}

\paragraph{Practically solving pMCs.}
Practically solving feasibility positively only requires making a good guess, for which various incomplete approaches handling thousands of parameter exist~\cite{DBLP:journals/tac/CubuktepeJJKT22,DBLP:conf/vmcai/HeckSJMK22}. 
For the verification problem, the literature considers two approaches:
either an encoding as a nonlinear equation system solved by a constraint solver or \emph{parameter lifting (PL)}---an abstraction-refinement algorithm. 
For anything but toy examples, the latter is the only viable approach~\cite{DBLP:journals/fmsd/JungesAHJKQV24}.
Given a pMC $\D$ and a region $R$, PL replaces possible parameter instantiations with nondeterministic choices by constructing a (non-parametric) Markov decision process (MDP).
The resulting MDP $\mathcal{M}_\mathrm{abstr}$ yields an abstraction of the instantiated MCs of $\D$ in $R$: If $\mathcal{M}_\mathrm{abstr}$ satisfies the specification $\varphi$, then $\varphi$ also holds for every instantiation $\D[u]$, $u \in R$.
If $\varphi$ does not hold in $\mathcal{M}_\mathrm{abstr}$, the abstraction is refined by partitioning $R$ into smaller subregions $R = R_1 \cup \dots \cup R_m$ that are verified individually.
The key enabler of PL in practice is that the abstraction can be efficiently analyzed. However, the applicability of PL is often limited:
\begin{compactenum}[(i)]
\item PL is only applicable to pMC $\D$ and its region $R$, if transitions of $\D$ are monotonic functions, and $R$ is well defined and graph preserving, i.e., the instantiated model $\D[u]$ for any $u \in R$ is guaranteed to be a valid MC and the topology of the underlying graphs is invariant under all instantiations.
\item The MDP abstraction in PL discards any parameter dependencies between states, often leading to an immense number of required refinement steps.
\end{compactenum}

\noindent
\emph{We improve the original PL approach and present solutions to both shortcomings.}
Both improvements build on the same conceptual basis: using iMCs instead of MDPs as the abstraction layer in the abstraction-refinement loop.

\paragraph{Fewer restrictions with generalized parameter lifting.}
As a first step, we reformulate the PL abstraction in terms of iMCs (\cref{section:imcs}).
We call this conservative generalization of the original  (standard) PL approach \emph{generalized parameter lifting}  (GPL).
GPL is the first approach that can verify every induced Markov chain of a given pMC:
By using iMCs, we support arbitrary (potentially non-monotonic) parametric transition functions in the input pMC.
Furthermore, the iMC formalism supports verifying regions for which some instantiations do not yield an MC (\cref{sec:illdefined}).
Finally, a novel and tailored variation of end component elimination for iMCs (\cref{section:notgraphperserving}) yields support for regions that are not graph preserving.
GPL thus relaxes these restrictions for PL.
This has significant practical implications as outlined in \cref{sec:overview}.
In particular, the support for regions that are not graph preserving and/or not well defined enables mixing families of MCs---such as software product lines~\cite{DBLP:conf/tacas/CeskaJJK19, DBLP:journals/fac/ChrszonDKB18}---with continuous parameters (\Cref{sec:families}).

\paragraph{The big-step transformation yields finer abstractions.}
The abstraction of pMCs into either MDPs (for standard PL) or iMCs (for GPL) discards dependencies between transition probabilities at different states, often leading to coarse abstractions (see \cref{ex:verificationexample}).
We remedy this by a novel \emph{big-step transformation} step, which is a pMC-to-pMC transformation that merges transitions over some fixed parameter (\cref{section:transformation}).
This transformation, inspired by flip-hoisting techniques on probabilistic programs \cite{DBLP:journals/corr/abs-2110-10284}, reduces the number of dependencies while preserving specification satisfaction. The subsequently executed GPL algorithm provides much tighter approximations and thus requires far fewer refinement steps.
GPL with this transformation enabled can provide speedups up to multiple orders of magnitude.  
As the big-step transformation results in pMCs with arbitrary transition functions, it is enabled by GPL's capability to verify such pMCs.

\paragraph{Contributions.}
This paper introduces \emph{generalized} parameter lifting (GPL) and the \emph{big-step} transformation for pMCs:
In contrast to standard PL, GPL can prove specifications for \emph{every} induced Markov chain of \emph{any} given pMC, including pMCs with nonlinear transitions, e.g., from \cite{DBLP:journals/dc/VolkBKA22}.
Additionally, GPL is the first scalable approach for the verification of families of pMCs.
The experiments show that GPL retains the scalability of standard PL.
The big-step transformation is introduced as part of a novel framework of \emph{pMC-to-pMC transformations}, which also captures state elimination \cite{DBLP:conf/ictac/Daws04}. It yields finer abstractions in GPL and thereby significantly reduces the number of refinement steps.
This allows verifying pMCs with many parameters, including a benchmark from \cite{DBLP:conf/atva/QuatmannD0JK16} that was out of reach and is solved in seconds with the big-step transformation.

\section{Problem Statement}\label{sec:problem}
We fix an ordered finite set \(V = \{ p_1, \dots, p_n \} \) of \emph{parameters} with subset of discrete parameters $V_D \subseteq V$ and \emph{domain} $\mathbb{D}_p = \mathbb{Z}$ if $p \in V_D$ and $\mathbb{D}_p = \mathbb{R}$ otherwise.
A \emph{parameter instantiation} is a mapping $u \colon V \to \mathbb{R}$---or equivalently a vector $u \in \mathbb{R}^n$---with $u(p) \in \mathbb{D}_p$.
 $\mathbb{D}^V = \mathbb{D}_{p_1} \times \dots \times \mathbb{D}_{p_n} \subseteq \mathbb{R}^n$ is the set of all parameter instantiations.
\( \mathbb{Q}[V] \) is the set of (multivariate)  polynomials over \( V \) with rational coefficients.
 \(f[u] \in \mathbb{R} \) denotes the evaluation of polynomial \( f \in \mathbb{Q}[V] \) at $u \in \mathbb{D}^{V}$.
The set of closed intervals with rational boundaries is given by $\intervals = \{ [a,b] \mid a,b \in \mathbb{Q}, a\le b\}$.
An $n$-dimensional \emph{region} \(R = (I_1 \times \dots \times I_n) \cap \mathbb{D}^V \) is a product of intervals $I_1, \dots, I_n \in \intervals$ restricted to parameter instantiations.
\begin{definition}[Parametric Markov chain]
    \label{def:pmc}
    A \emph{parametric Markov chain (pMC)} is a tuple \( \D = (S, V, s_I, \mathcal{P})\) with finite set \(S\) of states and parameters \( V \), initial state \( s_I \in S \), and transition function \(\mathcal{P}\colon S \times S \rightarrow \mathbb{Q}[V] \cup [0,1]\).
\( \D \) is a \emph{Markov Chain (MC)} if $\mathcal{P}(s,s') \in [0,1]$ and $\sum_{s'' \in S} \mathcal{P}(s,s'') = 1$ for all $s,s' \in S$.
\end{definition}
We may drop the variable set \( V \) for MCs and write them as \(\M = (S,s_I,\mathcal{P}) \).
An instantiation \(u \in \mathbb{D}^V\) is \emph{well defined} for a pMC \(\D = (S, V, s_I, \mathcal{P})\), if the \emph{instantiated pMC} \( \D[u] = (S, V, s_I, \mathcal{P}_u) \) with \(\mathcal{P}_u(s,s') = \mathcal{P}(s,s')[u] \) is an MC.
A region \(R\) induces a potentially infinite family of instantiated pMCs. 
We write  \(wd_\D(R) = \{u \in R \mid \D[u] \text{ is an MC} \} \) for the well-defined instantiations in \( R \) and drop the subscript \(\D\) if it is clear.
Region \( R \) is \emph{well defined} if \( wd_\D(R) = R \) and \emph{graph preserving} if for all $u,u' \in R$, $s,s' \in S$: \( \mathcal{P}(s,s')[u] = 0 \) iff \( \mathcal{P}(s,s')[u'] = 0\).

The transition function of an MC \(\M = (S, s_I, \mathcal{P})\) defines a probability distribution \( \mathcal{P}(s,\cdot) \) for the direct successor of each state $s \in S$.
We lift the distributions to a probability measure \( {\Pr}^{\M} \) (or simply \( \Pr \) if \( \M \) is clear) on measurable sets of infinite paths in the usual way, see~\cite{DBLP:books/daglib/0020348}.
\(\Pr(s \leadsto \good)\) is the probability to eventually reach a given set of target states  \(\good \subseteq S\) starting from $s \in S$.
We denote by \(\bad \subseteq S\) the set of all states \(s\) where \(\Pr(s \leadsto \good) = 0\).
A \emph{BSCC} is a strongly connected set of states where no outside state is reachable.
A \emph{(reachability probability) specification} is given by \(\varphi = \mathbb{P}_{\sim \lambda}(\Diamond \good)\), where \({\sim} \in \{<, \leq, \geq, >\}\).
An MC \(\M\) satisfies the specification \(\varphi\), written \(\M \vDash \varphi\), if \({\Pr}^{\M}(s_I \leadsto \good) \sim \lambda\).
Our goal is to verify \(\varphi\) for \emph{all} induced MCs in a region.
\vspace{-0.2em}
\begin{mdframed}%
     Given a pMC \(\mathcal{D}\), a region \(R\), and a specification \(\varphi  = \mathbb{P}_{\sim \lambda}(\Diamond \good)\),
     does \( \mathcal{D}[u] \vDash \varphi\) hold for all Markov chains \(\mathcal{D}[u]\) with \(u \in wd_\D(R)\)?
 \end{mdframed} %
\vspace{-0.2em}
\noindent Note that \(R\) does not have to be well defined. Our results generalize to expected rewards in a straightforward way.
All proofs can be found in \cref{app:proofs}.

\section{Our Approach in a Nutshell}
\label{sec:overview}%
\begin{figure}[t]
    \centering
\begin{minipage}{0.53\textwidth}
    \centering
    \begin{tikzpicture}[
        node distance=1.2cm and 1.5cm,
        box/.style={draw, rectangle, minimum width=1.8cm, minimum height=0.5cm, align=center},
        arrow/.style={-{Latex}, thick},
        fitbox/.style={draw, inner sep=0.5cm},
        gplbox/.style={draw=gray, inner sep=1cm, dashed},
        every node/.style={scale=0.7}
      ]
    \node[box, style={minimum height=1.5cm}] (refinement) {Region\\Refinement};
    \node[right=2cm of refinement, yshift=-0.4cm] (abstractionmid) {};
    \node[above=0.9cm of abstractionmid] (abstraction) {};
    \node[above=0.85cm of abstractionmid] (abstractionlabel) {Abstraction};
    \node[box, above=0.25 of abstractionmid.center] (constructimc) {Construct\\iMC \(\isub_{R'}(\D)\)};
    \node[box, below=0.25cm of abstractionmid.center] (checkimc) {\(\isub_{R'}(\D) \vDash \varphi\)?};
    \node[right=.7cm of abstractionmid] (spacer) {};
    \node[right=.3cm of abstractionmid, text=gray, yshift=2cm] (gpl) {GPL};

    \node[above=0.3cm of refinement] (regionr) {Region $R$};
    \node[box, right=1.2cm of constructimc.center, align=center] (bigstep) {Big-Step\\Transfomation};
    \node[above=0.3cm of bigstep] (pmcd) {pMC $\D$};
    \node[right=1.2cm of checkimc.center] (spec) {Specification $\varphi$};
    \node[below=0.3cm of refinement] (specholds) {\yes / \no};
    \node[fit=(constructimc)(checkimc)(abstraction), fitbox] (abstractionbox) {};
    
    \node[fit=(spacer)(refinement)(constructimc)(checkimc)(abstraction), gplbox] (abstractionbox) {};

    \draw[arrow] (regionr) -- (refinement);
    \draw[arrow] (pmcd) -- (bigstep);
    \draw[arrow] (bigstep) -- (constructimc);
    \draw[arrow] (spec) -- (checkimc);
    \draw[arrow] (constructimc) -- (checkimc);
    \draw[arrow] (refinement) -- (specholds);
    \draw[arrow, bend left=10] ([yshift=0.25cm] refinement.east) to node[above] {$R'\subseteq R$} (constructimc.west);
    \draw[arrow, bend left=20] (checkimc.west) to node[below] {\yes / \no} ([yshift=-0.25cm] refinement.east);

    \end{tikzpicture}
    \vspace{-6pt}
    \captionof{figure}{
        Generalized Parameter Lifting Abstraction-Refinement Loop%
        \label{fig:arloop}%
    }
\end{minipage}%
    \hfill%
    \begin{minipage}{0.45\textwidth}
        \centering
        \tiny
        \setlength{\tabcolsep}{4pt} 
        \begin{tabular}{l c c}
            \toprule
            & \textbf{Standard PL} & \textbf{GPL} \\
            \midrule
            \textbf{Abstraction} & MDPs & iMCs \\
            \addlinespace[0.5em]
            \textbf{pMCs} & monotonic & arbitrary \\
            \addlinespace[0.5em]
            \textbf{Parameters} &  
                \begin{tabular}{@{}c@{}}must be \\continuous\end{tabular}
            &
                \begin{tabular}{@{}c@{}}discrete or\\continuous\end{tabular}
            \\
            \addlinespace[0.5em]
            \textbf{Regions} & 
                \begin{tabular}{@{}c@{}}must be\\well defined and\\graph preserving\end{tabular}
            &
                \begin{tabular}{@{}c@{}}arbitrary\\hyperintervals\end{tabular}
            \\
            \bottomrule
        \end{tabular}
        \vspace{0pt}
        \captionof{table}{%
            Comparison of Standard PL and Generalized PL%
            \label{tab:comparison}%
        }
    \end{minipage}
\end{figure}
We present generalized parameter lifting (GPL) and the big-step transformation.
We first outline the steps of GPL and then compare to the approach in~\cite{DBLP:conf/atva/QuatmannD0JK16}.

\subsection{Overview}
\cref{fig:arloop} illustrates the approach, which is an instantiation of an abstraction-refinement loop that reduces solving the co-$\exists\mathbb{R}$-hard verification problem for pMCs by iteratively solving a set of iMCs.
We also give a pseudocode description in \cref{app:pseudocode}.
As a running example, we use the pMC $\D$ from \Cref{fig:expmc}, region $R=[0.3,0.6]\times[0.6,0.7]$ and specification $\varphi =  \mathbb{P}_{< 0.2}(\Diamond \good)$ as in \Cref{ex:pmcexample,ex:verificationexample}.
Our goal is to verify that \( \mathcal{D}[u] \vDash \varphi\) holds for all MCs \(\mathcal{D}[u]\) with \(u \in wd_\D(R)\)---simply written as \( \D,R \vDash \varphi\).

\paragraph{pMC abstraction via iMCs.}
To show that \( \D,R \vDash \mathbb{P}_{< 0.2}(\Diamond \good)\),
\emph{GPL computes an upper bound on the reachability probability} by evaluating the iMC, written $\isub_R(\D)$, that substitutes the functions in \(\D\)'s transitions with their intervals in the region~\(R\).
\Cref{fig:eximc} shows $\isub_R(\D)$ for our running example.
This iMC is a proper abstraction: For any well-defined instantiation $u \in wd_\D(R)$, the instantiated MC $\D[u]$ can also be generated by the iMC $\isub_R(\D)$.
However, the iMC also captures MCs that do not correspond to any instantiated MC $\D[u]$ of pMC $\D$.
Recall from \Cref{ex:verificationexample} that $\D,R \vDash \mathbb{P}_{< 0.2}(\Diamond \good)$.
The specification does not hold for the iMC abstraction since the maximal probability to reach $\good$ in the iMC is 0.294---achieved by picking the upper interval boundary for all transitions along the single path to $\good$.
This a is a counterexample to the specification, but it is \emph{spurious} since it is impossible to instantiate the pMC in the same way.

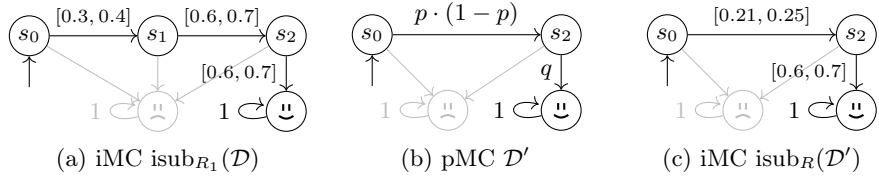
\begin{figure}[t]
    \begin{subfigure}{0.35\linewidth}\centering
            \begin{tikzpicture}[pMC]
                \node[state, initial below] (s0) {$s_0$};
                \node[state] (s1) [right of=s0,xshift=2mm] {$s_1$};
                \node[state] (s2) [right of=s1,xshift=2mm] {$s_2$};
                \node[state] (s3) [below of=s2, yshift=0.5cm] {$\good$};
                \node[state, draw=lightgray, text=lightgray] (bad) [below of=s1, yshift=0.5cm] {\(\bad\)};

                \path[->, draw=lightgray] (s0) edge (bad);
                \path[->, draw=lightgray] (s1) edge (bad);
                \path[->, draw=lightgray] (s2) edge (bad);

                \path[->] (s0) edge node[above] {\scriptsize$[0.3, 0.4]$} (s1);
                \path[->] (s1) edge node[above] {\scriptsize$[0.6, 0.7]$} (s2);
                \path[->] (s2) edge node[left] {\scriptsize$[0.6, 0.7]$} (s3);
                \path[->] (s3) edge[loop left] node[left] {$1$} (s3);
                \path[->, draw=lightgray] (bad) edge[loop left] node[left, text=lightgray] {$1$} (bad);
            \end{tikzpicture}
        \caption{iMC \(\isub_{R_1}(\D)\)}
        \label{fig:eximc2}
    \end{subfigure}%
    \begin{subfigure}{0.32\linewidth}                \centering
            \begin{tikzpicture}[pMC]
                \node[state, initial below] (s0) {$s_0$};
                \node[state] (s2) [right of=s0, xshift=1cm] {$s_2$};
                \node[state] (s3) [below of=s2, yshift=0.5cm] {$\good$};
                \node[state, draw=lightgray, text=lightgray] (bad) [below of=s0, yshift=0.5cm, xshift=1cm] {\(\bad\)};

                \path[->, draw=lightgray] (s0) edge (bad);
                \path[->, draw=lightgray] (s2) edge (bad);

                \path[->] (s0) edge node[above] {$p \cdot (1-p)$} (s2);
                \path[->] (s2) edge node[left] {$q$} (s3);
                \path[->] (s3) edge[loop left] node[left] {$1$} (s3);
                \path[->, draw=lightgray] (bad) edge[loop left] node[left, text=lightgray] {$1$} (bad);
            \end{tikzpicture}
        \caption{pMC \(\D'\)}
        \label{fig:pmcbigstep}
    \end{subfigure}%
    \begin{subfigure}{0.32\linewidth}\centering
            \begin{tikzpicture}[pMC]
                \node[state, initial below] (s0) {$s_0$};
                \node[state] (s2) [right of=s0, xshift=1cm] {$s_2$};
                \node[state] (s3) [below of=s2, yshift=0.5cm] {$\good$};
                \node[state, draw=lightgray, text=lightgray] (bad) [below of=s0, yshift=0.5cm, xshift=1cm] {\(\bad\)};

                \path[->, draw=lightgray] (s0) edge (bad);
                \path[->, draw=lightgray] (s2) edge (bad);

                \path[->] (s0) edge node[above] {\scriptsize$[0.21, 0.25]$} (s2);
                \path[->] (s2) edge node[left] {\scriptsize$[0.6, 0.7]$} (s3);
                \path[->] (s3) edge[loop left] node[left] {$1$} (s3);
                \path[->, draw=lightgray] (bad) edge[loop left] node[left, text=lightgray] {$1$} (bad);
            \end{tikzpicture}
        \caption{iMC \(\isub_R(\D')\)}
        \label{fig:imcbigstep}
    \end{subfigure}
    \caption{More Markov models}
    \label{fig:pmcexample}
\end{figure}

\paragraph{Region refinement.}
GPL employs a divide-and-conquer refinement.
Whenever a region $R$ can not be verified through abstraction, it is \emph{split} into smaller subregions $R = R_1 \cup \dots \cup R_m$.
We have $\D,R \models \varphi$ iff $\D,R_i \models \varphi$ for all $1\le i \le m$.
The initial verification problem thus reduces to verify \(\varphi\) for a series of subregions.
Smaller regions $R_i \subsetneq R$ intuitively yield a refined abstraction, because the interval transitions for iMC $\isub_{R_i}(\D)$ are tighter.
Such a split can be done repeatedly until
the subregions are sufficiently small
to conclude that \(\D, R \models \varphi\)
or
we find some \(u \in wd_\D(R_i)\), e.g., by sampling,
s.t. \(\D[u] \not\models \varphi\).
See~\cite{DBLP:journals/fmsd/JungesAHJKQV24} for further details on the refinement procedure, including splitting and sampling strategies.
For our example, we (choose to) split $R$ along the value for $p$ into $R_1=[0.3,0.4] \times [0.6,0.7]$ and $R_2 = R \setminus R_1$.
Recursively, GPL verifies  the iMCs
    \(\isub_{R_1}(\D)\) (\cref{fig:eximc2}) and \(\isub_{R_2}(\D)\). 
Checking \(R_1\) yields a maximal probability to reach $\good$ of $0.196<0.2$, implying $\D,R_1 \models \mathbb{P}_{< 0.2}(\Diamond \good)$.
For $R_2$, we get a value of $0.252 \nless 0.2$, resulting in further splitting of $R_2$.
Depending on how the split is performed, at least three more subregions have to be considered to infer that the specification holds in $R_2$.
GPL proves $R$ to be satisfied by checking at least six iMCs in total.

\begin{figure}[t]
    \begin{subfigure}{0.5\linewidth}
        \centering
        \begin{adjustbox}{max height=2.1cm}
        \begin{tikzpicture}[node distance=1.3cm, on grid, auto, initial text =]
            \node[elliptic state, initial] (s0) {wake up};
            \node[elliptic state] (s1) [right of=s0, yshift=1cm] {tailwind};
            \node[elliptic state] (s2) [right of=s0, yshift=-1cm] {headwind};
            \node[state, minimum size=6pt] (bus1) [right of=s1, xshift=1cm] {};
            \node[] (belows1) [right of=s1] {};
            \node[state, minimum size=6pt] (bus2) [right of=s2, xshift=1cm] {};
            \node[] (belows2) [right of=s2] {};
            \node[elliptic state] (good) [right of=belows1, xshift=2cm] {on time \(\good\)};
            \node[elliptic state] (bad) [right of=belows2, xshift=2cm] {too late \(\bad\)};
            \path[->] (s0) edge node[left] {$0.5$} (s1);
            \path[->] (s0) edge node[left] {$0.5$} (s2);
            \path[->] (s1) edge[bend left=20] node[below] {\(p_{\text{bike}}\)} (good);
            \path[->] (s1) edge node[below] {\(1-p_{\text{bike}}\)} (bus1);
            \path[->] (s2) edge[bend right=20] node[above] {\(p_{\text{bike}}\)} (bad);
            \path[->] (s2) edge node[above] {\(1-p_{\text{bike}}\)} (bus2);
            \path[->] (bus1) edge node[above] {$0.4$} (good);
            \path[->] (bus1) edge[bend left=10] node[above left, xshift=-0.5cm] {$0.6$} (bad);
            \path[->] (bus2) edge[bend left=10] node[below left, xshift=-0.3cm] {$0.4$} (good);
            \path[->] (bus2) edge node[above] {$0.6$} (bad);
            \path[->] (good) edge[loop right] node[right] {$1$} (good);
            \path[->] (bad) edge[loop right] node[right] {$1$} (bad);
        \end{tikzpicture}
    \end{adjustbox}
    \caption{Commute choices as pMC \(\D_{c}\)}
    \label{fig:pomdpexamplepmc}
    \end{subfigure}%
    \begin{subfigure}{0.5\linewidth}
        \centering
        \begin{adjustbox}{max height=2.1cm}
        \begin{tikzpicture}[node distance=1.3cm, on grid, auto, initial text =]
            \node[elliptic state, initial] (s0) {wake up};
            \node[elliptic state] (s1) [right of=s0, yshift=1.25cm] {taking bus};
            \node[elliptic state] (s2) [right of=s0, yshift=-1.25cm] {taking bike};
            \node[] (belows1) [right of=s1] {};
            \node[] (belows2) [right of=s2] {};
            \node[elliptic state] (good) [right of=belows1, xshift=1cm] {on time \(\good\)};
            \node[elliptic state] (bad) [right of=belows2, xshift=1cm] {too late \(\bad\)};
            \path[->] (s0) edge node[left] {$1-p_{\text{bike}}$} (s1);
            \path[->] (s0) edge node[left] {$p_{\text{bike}}$} (s2);
            \path[->] (s2) edge[bend left=20] node[above, yshift=0.1cm] {$0.5$} (good);
            \path[->] (s2) edge node[below] {$0.5$} (bad);
            \path[->] (s1) edge node[above] {$0.4$} (good);
            \path[->] (s1) edge[bend right=20] node[below, yshift=-0.1cm] {$0.6$} (bad);
            \path[->] (good) edge[loop right] node[right] {$1$} (good);
            \path[->] (bad) edge[loop right] node[right] {$1$} (bad);
        \end{tikzpicture}
    \end{adjustbox}
    \caption{Reordered commute choices \(\D_{r}\)}
    \label{fig:pomdpexamplepmcreordered}
    \end{subfigure}
    \caption{Reordering commute choices}
    \label{fig:commutechoices}
\end{figure}
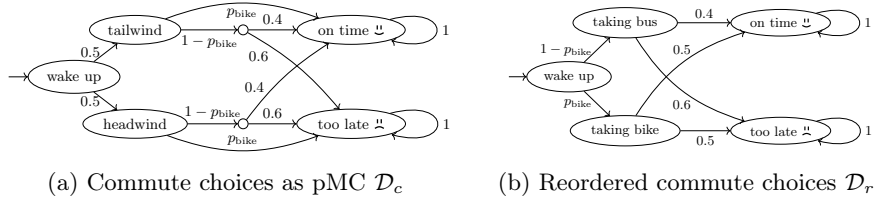

\paragraph{Big-step transformation to require fewer splits.}
The number of iterations, i.e., the number of iMCs that GPL verifies, can be prohibitively large, especially if there are many parameters. Indeed, while~\cite{DBLP:conf/atva/QuatmannD0JK16} evenly splits regions along every parameter, more refined splitting mechanisms were investigated later~\cite{DBLP:journals/fmsd/JungesAHJKQV24}. 
However, the coarse abstraction mechanism is the \emph{root cause} for the required number of iterations.
Here, the novel idea to reduce the number of iterations is to transform the pMC prior to abstraction.
We give two examples to show the effectiveness of this transformation.

\begin{example}
    \label{ex:statelim}
    Consider the pMCs \(\D\) in \cref{fig:expmc} and \(\D'\) in
    \cref{fig:pmcbigstep}. We obtain $\D'$ by applying state elimination~\cite{DBLP:conf/ictac/Daws04}, a transformation that preserves the reachability probability for every parameter value. Verifying only the iMC $\isub_R(\D')$ in \cref{fig:imcbigstep} suffices to verify that the reachability probability in $\D$ in $R$ is below $0.2$.
\end{example}

\begin{example}
    \label{ex:reorder}
    Consider the pMC \(\D_{c}\) in \cref{fig:pomdpexamplepmc} modeling a randomized decision to commute by bus or bike. 
    Depending on the wind direction, taking the bike leads to arriving on time, while the bus is randomly late 60\% of the time.
    Analyzing \(\isub_R(\D_{c})\) for $R=[0,1]$ yields a minimal reachability probability of $0.2$. 
    We can \enquote{reorder} \(\D_{c}\) into pMC \(\D_{r}\) (\cref{fig:pomdpexamplepmcreordered}) without affecting reachability probabilities. Analyzing \(\isub_R(\D_{r})\) for $R=[0,1]$ yields a tight lower bound of $0.4$.
\end{example}%
\noindent
Intuitively, the big-step transformation takes advantage of the fact that pMCs have sub-pMCs in which the same parameter occurs multiple times, and these sub-pMCs can be merged efficiently into one transition. Such a sub-pMC can represent initiating communication between two stations or navigating through a particular room. Shortcuts eliminate sequential occurrences of a parameter within a context, while grouping eliminates parallel occurrences within a context. We refer to \Cref{section:bigstep} for further details.
\begin{figure}[t]
    \centering%
    \begin{minipage}{0.3\linewidth}
        \centering%
        \begin{tikzpicture}[pMC]
            \node[state, initial above] (s0) {$s_0$};
            \node[state] (s1) [right of=s0] {$s_1$};
            \node[state] (s2) [right of=s1] {$\good$};
            \node[state, draw=lightgray, text=lightgray] (bad) [below of=s1, yshift=0.8cm] {$\bad$};
            \path[->, draw=lightgray] (s1) edge (bad);
            \path[->] (s0) edge node[above] {$1-p$} (s1);
            \path[->] (s0) edge[loop below] node[below] {$p$} (s0);
            \path[->] (s1) edge node[above] {$p$} (s2);
            \path[->] (s2) edge[loop below] node[below] {$1$} (s2);
            \path[->, draw=lightgray, text=lightgray] (bad) edge[loop right] node[right] {$1$} (bad);
        \end{tikzpicture}
        \caption{pMC \(\D\)}
        \label{fig:bellmanpmc}
    \end{minipage}%
    \hspace{0.05\linewidth}%
    \begin{minipage}{0.64\linewidth}
    \captionsetup{type=figure} %
    \begin{subfigure}{0.5\linewidth}\centering
        \begin{tikzpicture}[pMC]
            \node[state] (s0) {$s_0$};
            \node[state] (s1) [below of=s0, xshift=-1.5cm, yshift=0.5cm] {$s_1$};
            \node[state] (s2) [below of=s0, xshift=-0.75cm, yshift=0.5cm] {$s_2$};
            \node (dots) [below of=s0, xshift=0.75cm, yshift=0.5cm] {$\ldots$};
            \node[state] (sn) [below of=s0, xshift=1.5cm, yshift=0.5cm] {$s_n$};
            \path[->] (s0) edge node[above left] {$p_1$} (s1);
            \path[->] (s0) edge node[right] {$p_2$} (s2);
            \path[->] (s0) edge node[above right] {$p_n$} (sn);
        \end{tikzpicture}
        \caption{pMC \(\D_n\)}
        \label{fig:pmccontroller}
    \end{subfigure}%
    \hfill%
    \begin{subfigure}{0.5\linewidth}\centering
        \begin{tikzpicture}[pMC]
            \node[state] (s0) {$s_0$};
            \node[state] (s1) [below of=s0, xshift=-1.5cm, yshift=0.5cm] {$s_1$};
            \node[state] (s2) [below of=s0, xshift=-0.75cm, yshift=0.5cm] {$s_2$};
            \node (dots) [below of=s0, xshift=0.75cm, yshift=0.5cm] {$\ldots$};
            \node[state] (sn) [below of=s0, xshift=1.5cm, yshift=0.5cm] {$s_n$};
            \path[->] (s0) edge node[above left=-3pt, pos=0.3] {\scriptsize$[0.1, 0.9]$} (s1);
            \path[->] (s0) edge node[right] {\textquotedbl} (s2);
            \path[->] (s0) edge node[above right=-3pt, pos=0.3] {\scriptsize$[0.1, 0.9]$} (sn);
        \end{tikzpicture}
        \caption{iMC from \(\D_n\)}
        \label{fig:imccontroller}
    \end{subfigure}
    \vspace{-10pt}
    \caption{pMC and corresponding iMC}
    \end{minipage}
\end{figure}

\subsection{Comparing GPL and Standard Parameter Lifting}

The standard parameter lifting (PL) approach~\cite{DBLP:conf/atva/QuatmannD0JK16} considers an abstraction-refinement loop similar to GPL.
In fact, region refinement is performed in an identical way.
The key difference between standard PL and GPL is the abstraction.
While standard PL abstracts possible pMC instantiations using (non-parametric) MDPs, GPL is based on iMCs.
The semantics of iMCs yield various advantages that allow us to lift restrictions (see \cref{tab:comparison}).

\paragraph{Support for regions that are not well defined.}
Consider the pMC \(\D_n\) in \cref{fig:pmccontroller} and \(R = [0.1, 0.9]^n\).
Some points in this region do not induce MCs, e.g.,
for \(n=5\) and the point \(u(p_i) = 0.9\), the probabilities of the
distribution from \(s_0\) add up to \(4.5\).
We call \(R\) not well defined.
Such regions naturally occur, e.g., for controllers that randomly execute some action $a$ with probability $p_a$.
Standard PL does not handle not-well-defined regions, while GPL supports them due to iMC semantics.
\emph{For any region \(R\), GPL will analyze \(wd(R)\), i.e., the Markov chains in R} (\cref{sec:illdefined}).

\paragraph{Support for arbitrary polynomials as transition probabilities.}
The MDP abstraction of standard PL requires transition functions to be monotonic. 
\emph{GPL supports arbitrary polynomials by computing their intervals within each region} to get the iMC (\cref{section:imcs}, \cref{app:largepolys}).
For example, the pMC $\D'$ from \cref{fig:pmcbigstep} is supported by GPL but not by standard PL.
The more general support also enables more elaborate transformations of the pMC. 
In particular, the proposed big-step transformation algorithm (\Cref{section:bigstep}) yields pMCs with non-monotonic transition functions and is thus not applicable for standard PL.

\paragraph{Support for regions that are not graph preserving.} 
\label{sec:overviewnotgraph}
Verifying sets of MCs, where different MCs have different topologies, is at the heart of probabilistic software product line verification~\cite{DBLP:journals/fac/ChrszonDKB18, DBLP:conf/fm/VandinBLL18}. These sets can be represented using pMCs with regions that are not graph preserving and the additional constraint that a parameter is either $0$ or $1$. 
In contrast to standard PL, \emph{GPL supports not-graph-preserving regions via end component analysis} (\cref{section:notgraphperserving}).
In particular, GPL supports the verification of sets of pMCs, i.e., it allows mixing discrete, graph-changing parameters and continuous parameters (see \cref{sec:families}), which is not possible with existing abstraction-refinement techniques for software product line verification~\cite{DBLP:conf/tacas/CeskaJJK19,DBLP:conf/cav/AndriushchenkoC21}.
Region \(R = [0, 1]\) on the pMC \(\D\) in \cref{fig:bellmanpmc} is not graph preserving and yields discontinuous reachability probabilities, as seen when comparing $p=1$ and $p=1-\varepsilon$. Indeed, the verification results for $R'=[\varepsilon, 1-\varepsilon]$ and $R$ are significantly different for almost every threshold \(\varepsilon>0\)!

\section{Verifying Interval Markov Chains}
\nosectionappendix
GPL uses interval Markov chains (iMCs). This section reviews iMC verification and describes a tailored version of end component elimination on iMCs.
\label{section:imcs}
\begin{toappendix}
  \label{app:imcsproofs}
\end{toappendix}

\subsection{Interval Markov Chains}

Interval Markov chains (iMCs) can be seen as simplistic pMCs, where each transition has a unique real-valued parameter, combined with a region that assigns an interval to each such parameter.

\begin{definition}[Interval Markov chain]
    An \emph{interval Markov chain (iMC)} is a tuple \( \I = (S, s_I, \mathcal{P})\) with 
\(S\) and \(s_I\) as in \cref{def:pmc} and transition function \(\mathcal{P}\colon S \times S \rightarrow \intervals\).
The set of MCs induced by iMC \( \I \) is given by
    \(
        \MC(\I) := \{\M = (S, s_I, \mathcal{P}_\M) \mid \M \text{ is an MC s.t.\ }\mathcal{P}_\M(s, s') \in \mathcal{P}(s,s') \text{ for all } s, s'\}.
    \)
\end{definition}%
\noindent
We define the \emph{reachability interval} of iMC \( \I \) as
\[
        {
        \textstyle
        \llangle \I \rrangle :=
        \left[\,
            \min_{\M \in \MC(\I)} {\Pr}^{\M}(s_I \leadsto \good)\,,~
            \max_{\M \in \MC(\I)} {\Pr}^{\M}(s_I \leadsto \good)\,
        \right].
        }
    \]
\noindent    
The reachability interval \( \llangle \I \rrangle \) can be described by a system of Bellman equations.
\begin{definition}[iMC system of equations]
    \label{def:imcsystem}
    Let \(\I = (S, s_I, \mathcal{P})\) be an iMC and \(\opt \in \{\min, \max\}\).
    The system of equations for variables \(x_s = {\Pr}^{\opt}(s \leadsto \good)\) is given by
    \(x_{s} = 1\) for all \(s \in \good\),
    \(x_{s} = 0\) for all \(s \in \bad\),
    and otherwise
    \[
        x_s = \opt \left\{ \textstyle\sum_{t \in S} a_{s, t} \cdot x_{t} \,\middle\vert\, a_{s, t} \in \mathcal{P}(s, t) \text{ for } s, t \in S \text{ such that } \textstyle\sum_{t \in S} a_{s, t} = 1 \right\}.
    \]
\end{definition}
\noindent
The solution of this system of equations is unique if all intervals of the iMC
preserve its graph structure \cite{DBLP:journals/tcs/HaddadM18}.
The solution can be computed via a linear program encoding \cite{DBLP:conf/tacas/BenediktLW13,DBLP:conf/cav/PuggelliLSS13} or a value iteration procedure \cite{DBLP:conf/tacas/SenVA06,
DBLP:journals/ior/NilimG05}.

\subsection{Verifying iMCs With Not-Graph-Preserving Intervals}
\label{section:notgraphperserving}

Most methods for pMCs assume that
regions are graph preserving~\cite{DBLP:conf/birthday/0001JK22}.
Any not-graph-preserving region can be decomposed into exponentially many graph-preserving hyperintervals, which may be open, and are thus not regions suitable for PL \cite{DBLP:journals/fmsd/JungesAHJKQV24}.
We drop the assumption that regions must be graph preserving.
The challenge is that the iMC system of equations may not have a unique solution.

\begin{example}
    Consider the pMC in \cref{fig:bellmanpmc}. The
    probability to reach \(\good\) is \(p\) if \(p < 1\) and zero if \(p = 1\).
    Replacing all parametric transitions with the interval \([0,1]\)
    yields an iMC \(\I\).
    Its minimizing system of equations has a distinct (thus non-unique) solution for each \(r \in [0, 1]\) given by
    \(x_{s_0} = r\), \(x_{\!\good} = 1\), and \(x_{s_1} = x_{\!\bad} = 0\).
\end{example}%
\noindent
By eliminating the end components of the iMC, we can construct an iMC with a unique fixed point~\cite{DBLP:journals/tcs/HaddadM18, DBLP:conf/atva/BrazdilCCFKKPU14, DBLP:journals/corr/abs-2412-10185}.
Eliminating end components on iMCs is all we will need to verify pMCs on not-graph-preserving regions in the next section.

\begin{definition}[iMC end component]
    Let \(\I= (S, s_I, \mathcal{P})\) be an iMC.
    A set \(S'\) of states is an \emph{end component (EC)} if
    \(S' \subseteq S\) is a BSCC for some \( \M \in \MC(\I) \). 
\end{definition}
The union of two overlapping ECs is again an EC \cite{DBLP:journals/tcs/HaddadM18}. Thus, a state belongs to at most one \emph{maximal EC (MEC)}. The states in
\(\good\) and \(\bad\) each form an MEC.
\begin{lemmarep}[\hspace{1sp}{{\cite[Prop.\ 3]{DBLP:journals/tcs/HaddadM18}}}]
    The iMC \(\I\)'s system of equations has a unique solution if the only MECs
    in \(\I\) consist of the states in \(\good \cup \bad\).
    \label{thm:uniquefixedpoint}
\end{lemmarep}
\begin{proof}
    This follows from \textrm{\cite[Prop.\ 3]{DBLP:journals/tcs/HaddadM18}}. Use the fact that such a pMC's min-reduction is the pMC itself.
\end{proof}

\begin{figure}[t]
    \captionsetup{strut=off}
    \centering
    \begin{minipage}{0.5\linewidth}
    \captionsetup[subfigure]{justification=centering,belowskip=-1\baselineskip,strut=off}
    \begin{subfigure}{0.48\linewidth}
    \centering
    \begin{adjustbox}{max width=\linewidth}
    \begin{tikzpicture}[node distance=1.3cm, on grid, auto, initial text =]
        \node[cloud, draw, minimum width=1.5cm] (s0) {$S_i$};
        \node[state] (s1) [right of=s0, xshift=.5cm, yshift=.7cm] {$s_1^i$};
        \node (vdots) [right of=s0, xshift=.5cm,yshift=.2cm] {$\vdots$};
        \node[state] (sn) [right of=s0, xshift=.5cm, yshift=-.5cm] {$s_n^i$};
        \path[->] (s0) edge[loop above] node {$[\cdot, 1]$} (s0);
        \path[->] (s0) edge node[above=.1cm] {$[0, \cdot]$} (s1);
        \path[->] (s0) edge node[below=.1cm] {$[0, \cdot]$} (sn);
    \end{tikzpicture}
    \end{adjustbox}
    \end{subfigure}%
    \vspace{0.04\linewidth}%
    \begin{subfigure}{0.48\linewidth}
    \centering
    \begin{adjustbox}{max width=\linewidth}
    \begin{tikzpicture}[node distance=1.3cm, on grid, auto, initial text =]
        \node[state] (s0) {$s_i$};
        \node[state] (bad) [below of=s0,yshift=.2cm] {$\bad$};
        \node[state] (s1) [right of=s0, xshift=.5cm, yshift=.5cm] {$s_1^i$};
        \node (vdots) [right of=s0, xshift=.5cm] {$\vdots$};
        \node[state] (sn) [right of=s0, xshift=.5cm, yshift=-.8cm] {$s_n^i$};
        \path[->] (s0) edge node[left] {$[0, 1]$} (bad);
        \path[->] (bad) edge[loop left] node {$1$} (bad);
        \path[->] (s0) edge node[above=.1cm] {$[0, 1]$} (s1);
        \path[->] (s0) edge node[below=.1cm] {$[0, 1]$} (sn);
    \end{tikzpicture}
    \end{adjustbox}
   \vspace{-5pt}
    \end{subfigure}%
   \vspace{-10pt}
    \captionof{figure}{Illustration of EC elimination}\label{fig:ecelim}%
    \end{minipage}%
    \begin{minipage}{0.5\linewidth}
    \captionsetup[subfigure]{justification=centering,belowskip=0.5\baselineskip,strut=off}
    \begin{subfigure}{0.5\linewidth}
        \centering
        \begin{adjustbox}{max width=\linewidth}%
        \begin{tikzpicture}[node distance=2cm, on grid, auto, initial text =]
            \node[state] (s0) [initial] {$s_0$};
            \node[state] (s1) [right of=s0] {$s_1$};
            \node[state] (good) [below of=s0, xshift=1cm, yshift=0.6cm] {$\good$};
        
            \path[->] (s0) edge[bend left=20] node {$p$} (s1);
            \path[->] (s1) edge[bend left=20] node {$1-p$} (s0);
            \path[->] (s0) edge[] node[left] {$1-p$} (good);
            \path[->] (s1) edge[] node[right] {$p$} (good);
            \path[->] (good) edge[loop left] node {$1$} (good);
        \end{tikzpicture}
        \end{adjustbox}
    \end{subfigure}%
    \begin{subfigure}{0.5\linewidth}%
        \centering
        \begin{adjustbox}{max width=\linewidth}
        \begin{tikzpicture}[node distance=2cm, on grid, auto, initial text =]
            \node[state] (s0) [initial] {$s_0$};
            \node[state] (s1) [right of=s0] {$s_1$};
            \node[state] (good) [below of=s0, xshift=1cm, yshift=0.6cm] {$\good$};
            \path[->] (s0) edge[bend left=20] node {$[0,1]$} (s1);
            \path[->] (s1) edge[bend left=20] node {$[0,1]$} (s0);
            \path[->] (s0) edge[] node[left] {$[0,1]$} (good);
            \path[->] (s1) edge[] node[right] {$[0,1]$} (good);
            \path[->] (good) edge[loop left] node {$1$} (good);
        \end{tikzpicture}
        \end{adjustbox}
    \end{subfigure}
    \captionof{figure}{Substituting $R=[0,1]$}
    \label{fig:endcomponentcomparison}
    \end{minipage}
\end{figure}

We identify MECs as in \cite[Alg.\ 3]{DBLP:journals/tcs/HaddadM18}
and eliminate them while preserving optimal reachability probabilities.
Our transformation is a variant of the ones in
\cite{DBLP:journals/tcs/HaddadM18}, the difference being that we give a single
transformation instead of two.
Each MEC $S_i$ is collapsed into a single state $s_i$ as sketched in \cref{fig:ecelim}.
To reflect the possibility to never exit the MEC, an additional transition to $\bad$ is added.

\begin{definition}[EC elimination]
    Let iMC \(\I= (S, s_I, \mathcal{P})\) and \(E_\I = \{S_1,
    \ldots, S_n\}\) the set of MECs with  \( S_i \cap (\good \cup \bad) = \emptyset\) for all $1 \le i \le n$.
For $s \in S$, define \( \elimstate{s} = S_i \) if $s \in S_i$ for some $S_i \in E_\I$ and \( \elimstate{s} = s \) otherwise.
The \emph{EC elimination} of \( \I \) is the iMC
    \(\mathrm{elim}(\I) = (\{ \elimstate{s} \mid s \in S \}, \elimstate{s_I}, \hat{\mathcal{P}})\),
where for $s,s' \in S$ \[
\hat{\mathcal{P}}\big(\elimstate{s}, \elimstate{s'}\big) =
\begin{cases}
[0,1] & \text{if } \elimstate{s} \in E_\I \text{, } \elimstate{s} \neq \elimstate{s'} \text{, and } \mathcal{P}\big(\elimstate{s}, \elimstate{s'}\big) \neq [0,0],\\
[0,1] & \text{if } \elimstate{s} \in E_\I \text{ and } \elimstate{s'} \cap \bad \neq \emptyset,\\
\mathcal{P}\big(s, \elimstate{s'}\big) & \text{if } \elimstate{s} \notin E_\I,\\
[0,0] & \text{otherwise.}
\end{cases}
\]
\end{definition}

\begin{theoremrep}
    \label{thm:imcunique}
    For any iMC \(\I\), (a)~\(elim(\I)\)'s system of equations has a unique
    solution and (b)~\(\llangle \I \rrangle = \llangle elim(\I) \rrangle\), i.e., the reachability intervals coincide.
\end{theoremrep}%
\begin{toappendix}
\begin{proofsketch}
    (a): The transformation eliminates all MECs except those in \(\good\)
    and \(\bad\), which, combined with \cref{thm:uniquefixedpoint}, leads to the
    statement.
    (b): We show that all replaced states have the same optimal probabilities and
    thus conclude the same for all other states. Suppose \(S_i\) is a MEC in
    \(\I\) and \(s_i\) its replacement in \(\I'\). We show
    \(\opt_{\M \in \MC(\I)}{\Pr}^{\M}(s' \leadsto
    \good) = \opt_{\M' \in \MC(\I')}{\Pr}^{\M'}(s_i \leadsto
    \good)\) for \(s' \in S_i\).
    For \(\opt = \min\), it is optimal to forever stay in \(S_i\) in \(\I\) and
    choose to go to \(\bad\) in \(\I'\), thus both probabilities are zero.
    For \(\opt = \max\), it is optimal to leave for the state with the largest
    probability to \(\good\) in \(\I\) and \(\I'\), thus both probabilities
    are equal to the probability for that successor state.
\end{proofsketch}
\end{toappendix}

\section{Generalized Parameter Lifting}%
\nosectionappendix%
\label{sec:parameterlifting}%
We have seen how one can verify iMCs with arbitrary intervals. This section introduces GPL. We first introduce Parameter Lifting on iMCs and then show how GPL verifies not-well-defined regions and handles discrete parameters.
\subsection{Computing Region Estimates and Splitting Regions}
\label{sec:estimatesplit}
PL is based on computing \emph{region estimates}, i.e., upper and lower bounds to the
reachability probability within a region. 
\begin{definition}[Region estimate]
   A \emph{region estimate} for pMC \(\D\) in region \(R\)
   is an interval \([a, b] \in \intervals\) such that $a \le {\Pr}^{\D[u]}(s_I \leadsto \good) \le b$ for all $u \in wd(R)$.
\end{definition}
\noindent
To obtain region estimates for pMCs, we replace the transition functions by intervals that cover all instantiations within the region, yielding an iMC. We say an iMC \(\I\) \emph{substitutes} a pMC \(\D\) in region \(R\) if for all \(u \in wd(R)\): \(\D[u] \in \MC(\I)\).

\begin{theoremrep}
    Given a pMC \(\D\), a region \(R\), and an iMC \(\mathcal{I}\) that substitutes \(\D\) in \(R\),
    the reachability interval \(\llangle \mathcal{I} \rrangle\) is a region
    estimate for \(\D\) in \(R\).
    \label{thm:imcsatisfies}
\end{theoremrep}
\begin{proof}
    For all \(u \in wd(R)\): \(\D[u] \in \MC(\I)\). Thus, \(\min_{\M \in
    \MC(\I)} \Pr^{\M}(s_I \leadsto \good) \leq \min_{u \in wd(R)}
    \Pr^{\D[u]}(s_I \leadsto \good)\). There is a symmetric argument for the maximum.
\end{proof}
\noindent
An iMC \(\I\) \emph{refines} another iMC \(\I'\) if both share states \(S\) and for all \(s, s' \in S\): \(\mathcal{P}^\I(s, s') \subseteq \mathcal{P}^{\I'}(s, s')\)
\cite{DBLP:conf/lics/JonssonL91}.
Let the \emph{interval substitution} iMC \(\isub_R(\D)\) be defined as the maximally refined iMC that substitutes \(\D\) in \(R\).
It is obtained by substituting \(\D\)'s parametric transition probabilities with their intervals within \(R\):
\begin{propositionrep}
    For pMC \(\D = (S, s_I, \mathcal{P}, V)\), region \(R\), \(\isub_R(\D)=(S, s_I, \mathcal{P}_{\text{sub}})\):
    \(
        \mathcal{P}_{\text{sub}}(s, s') = \big[\,\min_{u \in wd(R)} \mathcal{P}(s,s')[u]\,,~ \max_{u \in wd(R)} \mathcal{P}(s, s')[u]\,\big]
    \)
    for all \(s, s'\).
\end{propositionrep}%
\begin{proof}
    The iMC \(\isub_R(\D)\) is the maximally refined iMC s.t. for all \(u \in wd(R)\): \(\D[u] \in \MC(\I)\). Thus, its transition probability intervals encompass exactly the transition probabilities of all \(\D[u]\) with \(u \in wd(R)\).
\end{proof}
GPL's abstraction is the interval substitution $\isub_R(\D)$.
Transition intervals in $\isub_R(\D)$ may include 0 as we allow not-graph-preserving regions---unlike standard PL~\cite{DBLP:conf/atva/QuatmannD0JK16}.
Consequently, an EC \(S' \subseteq S\) of $\isub_R(\D)$ might not be a BSCC in any of the instantiations of $\D[u]$.
\emph{Handling such ECs as in \cref{section:notgraphperserving} is the key to providing region estimates for not-graph-preserving regions.}
\begin{example}
The interval substitution $\isub_R(\D)$ for pMC $\D$ and region \(R = [0,1]\) in \cref{fig:endcomponentcomparison} has an EC $\{s_0,s_1\}$ which is no BSCC of any instantiation \(\D[u]\), $u \in R$.
\end{example}
The iMC $\isub_R(\D)$ might induce MCs that do not correspond to any instantiation of the pMC $\D$ due to two reasons.
First, for transition functions over discrete parameters, the (continuous) intervals of $\isub_R(\D)$ potentially contain values not realizable by a discrete parameter assignment.
Second, iMC transition intervals can be instantiated at each state independently, while pMC transition functions with common parameters are coupled. 
If region estimates obtained through interval substitution are not adequate to prove the specification, we may \emph{split} the region into smaller regions which yields \emph{refined} estimates.

\begin{definition}[Region split \cite{DBLP:journals/fmsd/JungesAHJKQV24}]\label{def:regionsplit}
    Let \(R\) be a region and \(R_1, \dots, R_m\) be regions with \(R = \bigcup_{j=1}^{m} R_j\).
    Then we say that \(R\)  \emph{splits} into \(R_1, \dots, R_m\).
\end{definition}

\begin{propositionrep}\label{proposition:regionsplit}
If \(R\) splits into \(R_1, \dots, R_m\)
and \(\I_1, \ldots, \I_m\) are iMCs s.t. \(\I_j\) substitutes \(\D\) in \(R_j\),
then $\bigcup_{j=1}^m \llangle \I_j(\D) \rrangle$ is a region estimate for pMC $\D$ in $R$.
\end{propositionrep}
\begin{proof}
   For each \(u \in R\), there exists a \(k \in \{1, \dots, m\}\) such that \(u
   \in R_k\) and \(\Pr^{\D[u]}(s_0 \leadsto \good) \in \llangle \I_k \rrangle
   \subseteq \bigcup_{j=1}^m \llangle \I_j \rrangle\).
\end{proof}
\begin{example}
For $\D$ and $\isub_R(\D)$ as in \cref{fig:endcomponentcomparison}, we have $\Pr^{\D[u]}(s_0 \leadsto \good) = 1$ for all $u \in R$, but \(\llangle \isub_R(\D) \rrangle = [0,1]\).
Splitting $R$ into $R_1 = [0,0.5]$ and $R_2=[0.5,1]$ yields $\llangle \isub_{R_1}(\D) \rrangle = \llangle \isub_{R_2}(\D) \rrangle = [1,1]$ which results in estimate $[1,1]$ for $R$.
\end{example}
Intuitively, splitting a region $R$ into increasingly smaller subregions $R_j$ yields tighter intervals in the iMCs $\isub_{R_j}(\D)$ and therefore tighter reachability intervals $\llangle \isub_{R_j}(\D) \rrangle$.
This enables obtaining arbitrarily precise region estimates for $R$.

For 
a specification \(\varphi = \mathbb{P}_{\geq \lambda}({\Diamond \good})\),
 a region estimate \([a, b]\) for pMC $\D$ in region $R$ yields three cases:
If $a \geq \lambda$ or $b < \lambda$, \emph{all} well-defined instantiations in $R$ \emph{satisfy} or \emph{violate} $\varphi$, immediately answering our main problem statement.
If $a < \lambda \leq b$, we successively apply region splitting to find an answer by either showing that $\varphi$ holds in all subregions or finding a subregion where $\varphi$ is violated.
This terminates unless the optimum and the threshold coincide.
We refer to \cite{DBLP:journals/fmsd/JungesAHJKQV24} for further details.%

\subsection{Verifying Not-Well-Defined Regions}%
\label{sec:illdefined}

GPL supports the verification of not-well-defined regions, i.e., regions in which some points do not induce a Markov chain as the transition probabilities do not sum up to one.
Such pMCs naturally occur when studying POMDPs~\cite{DBLP:conf/uai/Junges0WQWK018}.
For example, the region \(R_n = [0.1,0.9]^n\) is not well defined on pMC \(\D_n\) in \cref{fig:pmccontroller}.
Reasoning about such regions involves ignoring not-well-defined instantiations.
GPL achieves this by exploiting iMC semantics.
Correctness follows from \cref{thm:imcsatisfies} and the fact that \(\isub_R(\D)\) substitutes \(\D\) in \(R\):
\begin{corollary}
    \label{thm:illdefined}
    Let \(\D\) be a pMC and \(R\) a not-well-defined region. Then for all
    well-defined instantiations \(u \in R\): \(\Pr^{\D[u]}(s_I \leadsto \good)
    \in \llangle \isub_R(\D) \rrangle\).
\end{corollary}
\noindent
A pMC is \emph{simple} if all transitions are constant or of the form $p$ or $1-p$ for $p \in V$.
{\setlength{\intextsep}{0pt}%
\begin{wrapfigure}{r}{0.4\textwidth}
    \begin{adjustbox}{max width=\linewidth}
        \begin{tikzpicture}[node distance=2cm, on grid, auto, initial text =]
            \node[state] (s0) {$s_0'$};
            \node[state] (x1) [right of=s0] {};
            \node (x2) [right of=x1] {$\ldots$};
            \node[state] (sn) [right of=x2] {$s_n'$};
            \node[state] (s1) [below right of=s0, yshift=0.5cm] {$s_1'$};
            \node[state] (s2) [below right of=x1, yshift=0.5cm] {$s_2'$};
            \node[state] (sn1) [below right of=x2, yshift=0.5cm] {$\scriptscriptstyle s_{n-1}'$};
            \path[->] (s0) edge node[below left, yshift=0.1cm] {$p_1'$} (s1);
            \path[->] (s0) edge node[above] {$1-p_1'$} (x1);
            \path[->] (x1) edge node[below left, yshift=0.1cm] {$p_2'$} (s2);
            \path[->] (x1) edge node[above] {$1-p_2'$} (x2);
            \path[->] (x2) edge node[below left, yshift=0.1cm] {$p_{n-1}'$} (sn1);
            \path[->] (x2) edge node[above] {$1-p_{n-1}'$} (sn);
        \end{tikzpicture}
    \end{adjustbox}
    \captionof{figure}{simple pMC \(\D_n'\) from \(\D_n\)}
    \label{fig:pmcchoicesimple}
\end{wrapfigure}
In simple pMCs, all regions $R \subseteq [0,1]^{|V|}$ are well defined.
The pMC $\D_n$ in \cref{fig:pmccontroller} is not simple.
A transformation in \cite{DBLP:conf/uai/Junges0WQWK018} yields the simple pMC $\D'_n$ in \cref{fig:pmcchoicesimple} over new parameters, with a bijection between valuations of $\D_n$ and $\D'_n$. $\D'_n$ can be checked with standard PL over the new parameters.
However, this transformation does not enable us to check $R_n$: \(R_n\) in $\D_n$ has no equivalent hyperinterval region \(R'_n\) in \(\D'_n\) with \(R'_n = wd(R'_n)\). Thus, the region \enquote{go from \(s_0\) to \(s_i\) with probabilities between 0.1 and 0.9} cannot be verified with standard PL. This query only becomes possible with GPL.

} %

\subsection{Reasoning About Families of pMCs Using Discrete Parameters}
\label{sec:families}
Suppose \(\mathfrak{M}\) is a finite family (i.e., a finite set) of Markov chains.
Each such \(\mathfrak{M}\) can be described by a single pMC with additional discrete parameters \(V_D = \{p_1, \ldots, p_n\}\) that take values \(p_i \in \{0, 1\}\)\cite{DBLP:conf/fm/CeskaHJK19}.
For example, consider the pMC in \cref{fig:commutechoices} with \(p_{\text{bike}} \in \{0, 1\}\). This pMC encodes two MCs and models buying either a bus subscription (\(p_{bike}=0\)) or a bike (\(p_{bike}=1\)).
This encoding is used for the analysis of software product lines in, e.g., \cite{DBLP:conf/fm/CeskaHJK19, DBLP:conf/hase/RodriguesANLCSS15}. Previously, these pMCs could not be analyzed with parameter lifting as such regions are not graph preserving.

A similar procedure can be applied to finite families of pMCs \(\mathfrak{D}\)  over parameters \(V_C\), resulting in a single pMC over \(V_D \cup V_C\) that describes all pMCs in~\(\mathfrak{D}\).
With GPL, and given a region \(R_C\) over the parameters \(V_C\), all pMCs can be simultaneously checked by checking the joint pMC over the region \(\{0,1\}^n \times R_C\).
If necessary, GPL splits a parameter over \(\{0,1\}\) into \(\{0\}\) and \(\{1\}\).
To the best of our knowledge, GPL is the first verification method that explicitly supports a mix of discrete and continuous parameters, and thus finite families of pMCs.

\section{Tightening Region Estimates by Transforming pMCs}
\nosectionappendix
\label{section:transformation}
As shown in \cref{ex:statelim,ex:reorder} on page \pageref{ex:statelim}, transforming the pMC before applying interval substitution can improve its region estimates.
In this section, we introduce the requirements for such transformations, present two approaches based on shortcuts and transition grouping, and describe an algorithm combining both ideas.\looseness=-1

\subsection{Tightening Transformations}
\begin{definition}[Tightening transformation]
    An iMC \(\I\) \emph{tightens} iMC \(\I'\) if \(\llangle \I \rrangle
    \subseteq \llangle \I' \rrangle\).
    Let \(\mathfrak{D}_V\) be the set of pMCs with parameters \(V\). A function \(\ttra\colon
    \mathfrak{D}_V \rightarrow \mathfrak{D}_V\) is a \emph{tightening
    transformation} if for all pMCs \(\D\), the pMC $\ttra(\D)$ satisfies for all regions~$R$: \(wd_{\ttra(\D)}(R) = wd_\D(R)\) and
    \(\isub_R(\ttra(\D))\) tightens \(\isub_R(\D)\).
    \label{def:tighteningtransformation}
\end{definition}
A tightening transformation preserves reachability probabilities induced by well-defined pMC instantiations.
Let \(\D \equiv \D'\) denote that two pMCs $\D,\D' \in \mathfrak{D}_V$ have the same reachability probabilities, i.e.,
their well-defined instantiations coincide 
and $\Pr^{\D[u]}(s_0 \leadsto \good) = \Pr^{\D'[u]}(s_0 \leadsto \good)$ for all well-defined $u \in \mathbb{D}^V$.

\begin{lemmarep}\label{lem:tighteningimpliessameprobs}
For tightening transformation \(\ttra\), we have \( \D \equiv \ttra(\D) \) for all pMCs \( \D \).\looseness=-1
\end{lemmarep}
\begin{proof}
A tightening transformation $\ttra$, pMC $\D$, and well-defined  $u \in \mathbb{R}^V$ yield 
\(\{\Pr_{\D[u]}(s_0 \leadsto \good)\} = \llangle \isub_{\{u\}}(\D) \rrangle \subseteq  \llangle \isub_{\{u\}}(\ttra(\D)) \rrangle = \{\Pr_{\ttra(\D)[u]}(s_0 \leadsto \good)\}\).
\end{proof}

Intuitively, the region estimates obtained after applying a tightening transformation shall be at least as tight as the estimates obtained using the original pMC.
The identity $\ttra_\mathrm{id}$ with $\ttra_\mathrm{id}(\D) = \D$ is a (trivial) tightening transformation that does not improve any region estimates.
Another example is the function \(\ttra_{\mathrm{exact}}\) that transforms a pMC \(\D\) into a pMC \(\D'\) over fractions of multivariate polynomials with three states \(\{s_0, \good, \bad\}\) and the single rational transition function that encodes the reachability probabilities in $\D$ as in \cite{DBLP:conf/ictac/Daws04}.
\(\ttra_{\mathrm{exact}}\) is a tightening transformation, since for any region $R$, $\llangle \isub_R(\ttra_{\mathrm{exact}}(\D))\rrangle$ is the tightest possible estimate and thus a subset of $\llangle \isub_{R}(D) \rrangle$ by \Cref{thm:imcsatisfies}.
The result $\ttra_{\mathrm{exact}}(\D)$ has exponentially large fractions of polynomials as transition probabilities \cite{DBLP:journals/iandc/BaierHHJKK20}.

From a practical perspective, neither $\ttra_\mathrm{id}$ nor
$\ttra_{\mathrm{exact}}$ are useful: The transformation \(\ttra_{\mathrm{exact}}\)
yields the tightest region estimates, but is hard to compute and evaluate, the identity \(\ttra_{\mathrm{id}}\) is easy to compute, but
effectless.  Our aim is to find a tightening transformation that (1)~strictly
tightens many region estimates and (2)~is effectively computable, with a fast
evaluation of region estimates. 

\subsection{Shortcuts in pMCs}
\label{sec:shortcut}

Our transformation algorithm is based on two main ideas: \emph{Creating
shortcuts} in single-parameter sub-pMCs as in \cref{ex:statelim} and
\emph{grouping parametric choices} as in \cref{ex:reorder}.  It works on
single-parameter \emph{sub-pMCs} rooted in a state $\hat{s}$.

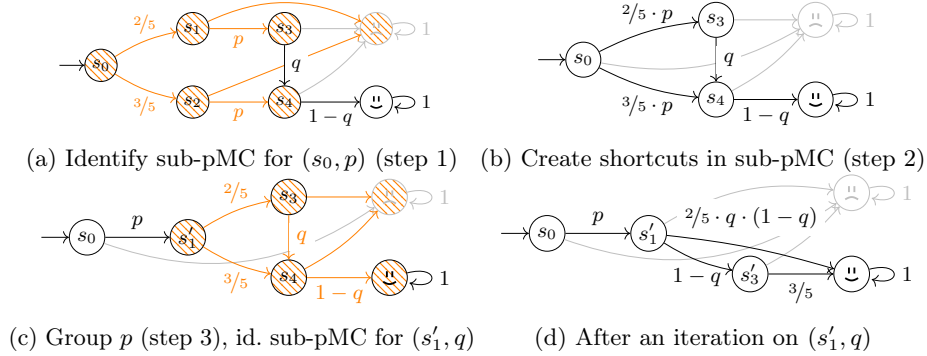
\begin{figure}[t]
\begin{subfigure}{0.5\linewidth}
    \centering%
    \begin{adjustbox}{max height=1.8cm}
        \begin{tikzpicture}[node distance=1.5cm, on grid, auto, initial text =]
            \node[state, initial, pattern=north west lines, pattern color=orange] (s0) {\(s_0\)};
            \node[state, pattern=north west lines, pattern color=orange] (s1) [right of=s0, yshift=0.6cm] {\(s_1\)};
            \node[state, pattern=north west lines, pattern color=orange] (s2) [right of=s0, yshift=-0.6cm] {\(s_2\)};
            \node[state, pattern=north west lines, pattern color=orange] (s3) [right of=s1] {\(s_3\)};
            \node[state, pattern=north west lines, pattern color=orange] (s4) [right of=s2] {\(s_4\)};
            \node[state, draw=lightgray, text=lightgray, pattern=north west lines, pattern color=orange] (s6) [right of=s3] {\(\bad\)};
            \node[state] (s5) [right of=s4] {\(\good\)};

            \path[->, draw=orange] (s0) edge[bend left=10] node[above, text=orange] {$\nicefrac{2}{5}$} (s1);
            \path[->, draw=orange] (s0) edge[bend right=10] node[below, text=orange] {$\nicefrac{3}{5}$} (s2);

            \path[->, draw=orange] (s1) edge[bend left=25] (s6);
            \path[->, draw=orange] (s2) edge[bend right=0] (s6);
            \path[->, draw=lightgray] (s4) edge[bend right=10] (s6);
            \path[->, draw=lightgray] (s3) edge (s6);
            \path[->, draw=lightgray] (s6) edge[loop right] node[right, text=lightgray] {$1$} (s6);

            \path[->, draw=orange] (s1) edge node[below, text=orange] {$p$} (s3);
            \path[->, draw=orange] (s2) edge node[below, text=orange] {$p$} (s4);

            \path[->] (s3) edge[] node[right, fill=white] {$q$} (s4);
            \path[->] (s4) edge[] node[below] {$1-q$} (s5);

            \path[->] (s5) edge[loop right] node[right] {$1$} (s5);
        \end{tikzpicture}
    \end{adjustbox}
    \caption{Identify sub-pMC for \((s_0, p)\) (step 1)}
    \label{fig:bigstep1}
\end{subfigure}%
\begin{subfigure}{0.5\linewidth}
    \centering%
    \begin{adjustbox}{max height=1.8cm}
        \begin{tikzpicture}[node distance=1.5cm, on grid, auto, initial text =]
            \node[state, initial] (s0) {\(s_0\)};
            \node[state] (s3) [right of=s0, xshift=.5cm, yshift=0.6cm] {\(s_3\)};
            \node[state] (s4) [right of=s0, xshift=.5cm, yshift=-0.6cm] {\(s_4\)};
            \node[state, draw=lightgray, text=lightgray] (s6) [right of=s3] {\(\bad\)};
            \node[state] (s5) [right of=s4] {\(\good\)};

            \path[->, draw=lightgray] (s0) edge[bend right=20] (s6);
            \path[->, draw=lightgray] (s3) edge (s6);
            \path[->, draw=lightgray] (s4) edge[bend right=10] (s6);
            \path[->, draw=lightgray] (s6) edge[loop right] node[right, text=lightgray] {$1$} (s6);

            \path[->] (s0) edge[bend left=10] node[above] {$\nicefrac{2}{5} \cdot p$} (s3);
            \path[->] (s0) edge[bend right=10] node[below] {$\nicefrac{3}{5} \cdot p$} (s4);

            \path[->] (s3) edge[] node[right, fill=white] {$q$} (s4);
            \path[->] (s4) edge[] node[below] {$1-q$} (s5);

            \path[->] (s5) edge[loop right] node[right] {$1$} (s5);
        \end{tikzpicture}
    \end{adjustbox}
    \caption{Create shortcuts in sub-pMC (step 2)}
    \label{fig:bigstep2}
\end{subfigure}

\begin{subfigure}{0.5\linewidth}
    \centering%
    \begin{adjustbox}{max height=1.8cm}
        \begin{tikzpicture}[node distance=1.5cm, on grid, auto, initial text =]
            \node[state, initial] (s0) {\(s_0\)};
            \node[state, pattern=north west lines, pattern color=orange] (s1) [right of=s0] {\(s_1'\)};
            \node[state, pattern=north west lines, pattern color=orange] (s3) [right of=s1, yshift=0.6cm] {\(s_3\)};
            \node[state, pattern=north west lines, pattern color=orange] (s4) [right of=s1, yshift=-0.6cm] {\(s_4\)};
            \node[state, draw=lightgray, text=lightgray, pattern=north west lines, pattern color=orange] (s6) [right of=s3] {\(\bad\)};
            \node[state, pattern=north west lines, pattern color=orange] (s5) [right of=s4] {\(\good\)};

            \path[->, draw=lightgray] (s0) edge[bend right=27] (s6);
            \path[->, draw=orange] (s3) edge (s6);
            \path[->, draw=orange] (s4) edge[bend right=10] (s6);
            \path[->, draw=lightgray] (s6) edge[loop right] node[right, text=lightgray] {$1$} (s6);

            \path[->] (s0) edge[] node[above] {$p$} (s1);

            \path[->, draw=orange] (s1) edge[bend left=10] node[above, text=orange] {$\nicefrac{2}{5}$} (s3);
            \path[->, draw=orange] (s1) edge[bend right=10] node[below, text=orange] {$\nicefrac{3}{5}$} (s4);

            \path[->, draw=orange] (s3) edge[] node[right, fill=white, text=orange] {$q$} (s4);
            \path[->, draw=orange] (s4) edge[] node[below, text=orange] {$1-q$} (s5);

            \path[->] (s5) edge[loop right] node[right] {$1$} (s5);
        \end{tikzpicture}
    \end{adjustbox}
    \caption{Group \(p\) (step 3), id.\ sub-pMC for \((s_1', q)\)}
    \label{fig:bigstep3}
\end{subfigure}%
\begin{subfigure}{0.5\linewidth}
    \centering%
    \begin{adjustbox}{max height=1.8cm}
        \begin{tikzpicture}[node distance=1.5cm, on grid, auto, initial text =]
            \node[state, initial] (s0) {\(s_0\)};
            \node[state] (s1) [right of=s0] {\(s_1'\)};
            \node[state] (s4) [right of=s1, yshift=-0.6cm] {\(s_3'\)};
            \node[state, draw=lightgray, text=lightgray] (s6) [right of=s3] {\(\bad\)};
            \node[state] (s5) [right of=s4] {\(\good\)};

            \path[->, draw=lightgray] (s0) edge[bend right=27] (s6);
            \path[->, draw=lightgray] (s4) edge[bend right=10] (s6);
            \path[->, draw=lightgray] (s6) edge[loop right] node[right, text=lightgray] {$1$} (s6);
            \path[->, draw=lightgray] (s1) edge[bend left=20] (s6);

            \path[->] (s0) edge[] node[above] {$p$} (s1);

            \path[->] (s1) edge[bend left=5] node[above, fill=white, yshift=0.17cm] {$\nicefrac{2}{5}\cdot q \cdot (1-q)$} (s5);
            \path[->] (s1) edge[bend right=10] node[below] {$1-q$} (s4);

            \path[->] (s4) edge[] node[below] {$\nicefrac{3}{5}$} (s5);

            \path[->] (s5) edge[loop right] node[right] {$1$} (s5);
        \end{tikzpicture}
    \end{adjustbox}
    \caption{After an iteration on \((s_1', q)\)}
    \label{fig:bigstep4}
\end{subfigure}
\caption{Big-step transformation algorithm on an example pMC}
\label{fig:processingalgorithm}
\end{figure}

\begin{definition}[Sub-pMC rooted in \(\hat{s}\) over \(p\)]
    \label{def:rootedpmc}
    A \emph{sub-pMC of \(\D\) rooted in \(\hat{s} \in S \) over \(p \in V\)} is a pMC \(\D_{\hat{s},p} = (\hat{S}, \hat{s},
    \hat{\mathcal{P}}, \{p\})\) such that  \(\hat{s} \in \hat{S} \subseteq S\) and
\begin{compactitem}
\item the underlying graph $\mathcal{G}_{\hat{s},p} = (\hat{S}, \{(s,t) \in \hat{S}\times\hat{S} \mid \hat{\mathcal{P}}(s,t) \neq 0\})$ is acyclic, i.e., all maximal paths end in $\hat{S}_\mathrm{exit} = \{ s \in \hat{S} \mid \hat{\mathcal{P}}(s,t) = 0 \text{ for all } t \in \hat{S}\}$,
\item every $s \in \hat{S}$ is reachable from $\hat{s}$ in $\mathcal{G}_{\hat{s},p}$, and
\item $s \notin \hat{S}_\mathrm{exit}$ implies $\hat{\mathcal{P}}(s,t) = \mathcal{P}(s,t) \in \mathbb{Q}[\{p\}]$  for all $t \in S$.
\end{compactitem}
\end{definition}
Note that cyclic sub-pMCs can be made acyclic by removing transitions.
A sub-pMC of the pMC in \cref{fig:bigstep1} rooted in \(s_0\) over \(p\) is indicated in orange.
We have $\hat{S}_\mathrm{exit} = \{s_3,s_4,\bad\}$.
Our approach is to take \emph{shortcuts} from $s_0$ directly to $\hat{S}_\mathrm{exit}$---skipping over the intermediate states $s_1$ and $s_2$.
To this end, the outgoing transitions of $s_0$ are replaced in \cref{fig:bigstep2}.
We now fix $\D_{\hat{s},p}$ and $\hat{S}_\mathrm{exit}$ as in \cref{def:rootedpmc}.
\begin{definition}[Shortcut pMC]
    \label{def:shortcutpmc}
The \emph{shortcut pMC} of $\D$ and its sub-pMC \(\D_{\hat{s}, p}\) is the pMC $\ttra_\mathrm{shortcut}(\D, \D_{\hat{s},p}) = (S, s_I, \mathcal{P}_\mathrm{shortcut}, V)$,
with $\mathcal{P}_\mathrm{shortcut}(s,t) = \mathcal{P}(s,t)$ for $s,t \in S$, $s \neq \hat{s}$,
$\mathcal{P}_\mathrm{shortcut}(\hat{s},t) = 0$ for $t \notin \hat{S}_\mathrm{exit}$, and
\[
       \mathcal{P}_\mathrm{shortcut}(\hat{s},t) ~=~ {\Pr}^{\D_{\hat{s},p}}(\hat{s} \leadsto t) ~=~ { \textstyle \sum_{s_0 \dots s_n \in \mathit{Paths}(\hat{s}, t)}~ \prod_{i=1}^{n} \mathcal{P}(s_{i-1},s_i) }   \]
for $t \in \hat{S}_\mathrm{exit}$, where $\mathit{Paths}(\hat{s}, t)$ denotes the set of paths from $\hat{s}$ to $t$.
\end{definition}%
\noindent
The set $\mathit{Paths}(\hat{s}, t)$ for $t \in \hat{S}$ is finite as \cref{def:rootedpmc} requires $\D_{\hat{s},p}$ to be acyclic.
It follows that $\mathcal{P}_\mathrm{shortcut}(\hat{s},t) = {\Pr}^{\D_{\hat{s},p}}(\hat{s} \leadsto t)$ is a univariate polynomial over parameter $p$.
The polynomials ${\Pr}^{\D_{\hat{s},p}}(\hat{s} \leadsto t)$ for all $t \in \hat{S}$ can effectively be computed in a dynamic programming fashion by traversing the states of $\D_{\hat{s},p}$ in a topological order. 
Our implementation uses a factorized representation, cf. \cref{app:shortcutprobs}.

\begin{lemmarep}\label{lem:shortcuttightens}
$\isub_R(\ttra_\mathrm{shortcut}(\D, \D_{\hat{s},p}))$ tightens $\isub_R(\D)$ for any region $R$.
\end{lemmarep}
\begin{proof}
The two iMCs only differ at state $\hat{s}$.
For $t \in S_\mathrm{exit}$, we have $\mathcal{P}_\mathrm{shortcut}(\hat{s}, t) = \{ {\Pr}^{\D}(\hat{s} \leadsto t)[u] \mid u \in wd_\D(R)\}$.
 The claim follows as $\isub_R(\D)$ substitutes $\D$.%
\end{proof}

\subsection{Grouping Transitions}

Our approach is to iteratively apply transformations using shortcuts.
The following example suggests interleaving shortcuts with a grouping of transitions.

\begin{example}
After adding an intermediate state $s_1'$ as in \cref{fig:bigstep3}, we obtain a sub-pMC rooted in $s_1'$ over $q$.
This is a larger sub-pMC than the candidates in \cref{fig:bigstep2}.
\cref{fig:bigstep4} shows the corresponding shortcut pMC.
\end{example}

\begin{definition}[Grouped pMC]\label{def:grouping}
    Suppose we have \(\hat{s} \in S\), \(S' = \{s_1, \ldots, s_k\} \subseteq S\)
    s.t. $\mathcal{P}(\hat{s}, s_i) = g_i + c_i \cdot f$
    for some polynomials $f, g_i \in \mathbb{Q}[V]$ and factors $c_i \in \mathbb{Q}$
    with $c = \sum_{j=1}^k c_j$.
    Then for the \emph{grouped pMC}
    $\ttra_\mathrm{group}(\D,\hat{s},f) = (S \uplus \{s'\}, s_I,
    \mathcal{P}_\mathrm{group}, V)$ we have
    $\mathcal{P}_\mathrm{group}(\hat{s}, s') = c \cdot f$,
    $\mathcal{P}_\mathrm{group}(\hat{s}, s_i) = g_i$,
    $\mathcal{P}_\mathrm{group}(s', s_i) = \nicefrac{c_i}{c}$,
    $\mathcal{P}_\mathrm{group}(s', t') = 0$ for all \(t' \notin S'\), and
    $\mathcal{P}_\mathrm{group}(t, t') = \mathcal{P}(t, t')$ in all other cases.
\end{definition}

\begin{lemmarep}\label{lem:groupingtightens}
    $\isub_R(\ttra_\mathrm{group}(\D,\hat{s},f))$ tightens $\isub_R(\D)$ for any region $R$, \(f \in \mathbb{Q}[V]\).
\end{lemmarep}
\begin{proof}
Let \(\I_1 = \isub_R(\D)\) and
\(\I_2 = \isub_R(\ttra_\mathrm{group}(\D,\hat{s},f))\).
We have to prove that
$\llangle \I_2 \rrangle \subseteq \llangle \I_1 \rrangle$,
i.e.,
$\{\Pr^{\M}(s_I \leadsto \good) \mid \M \in \MC(\I_2)\}
\subseteq \{\Pr^{\M}(s_I \leadsto \good) \mid \M \in \MC(\I_1)\}$.
We consider an MC induced by \(\I_2\) and show that we can (re)construct another MC induced by \(\I_1\) with the same reachability probability.
Let \(c\), \(c_1, \ldots, c_k\) and \(y_1, \ldots, y_k\) be defined as in \cref{def:grouping}.
For \(\M \in \MC(\I_2)\),
let \(x = \mathcal{P}^\M(\hat{s}, s')\)
and \(y_i = \mathcal{P}^\M(\hat{s}, s_i)\),
then (re)construct \(\M' \in \MC(\I_1)\)
s.t. \(\mathcal{P}(\hat{s}, s_i) = y_i + c_i \cdot \nicefrac{x}{c}\).
Then, \(\Pr^{\M}(s_I \leadsto \good) = \Pr^{\M'}(s_I \leadsto \good)\).
\end{proof}
\begin{example}
    Creating shortcuts and grouping together \emph{reorders}
    parametric transitions to come before constant transitions. Consider the
    pMCs \(\D_a\) in \cref{fig:bigstep1} and \(\D_c\) in \cref{fig:bigstep3}.
    \(\D_c\) takes the \(p\)-transition before taking the constant transitions.
\end{example}
Our algorithm never changes the order in which parameters occur along a path,
because reordering parameters tends to lead to non-tightening transitions.
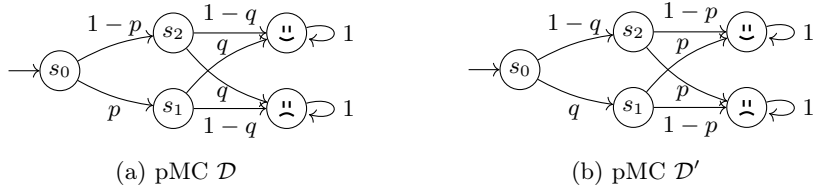
\begin{figure}[t]
\begin{subfigure}{0.49\linewidth}
    \centering
    \begin{adjustbox}{max width=\linewidth}
        \begin{tikzpicture}[node distance=1.5cm, on grid, auto, initial text =]
            \node[state, initial] (s0) {\(s_0\)};
            \node[state] (s1) [right of=s0, yshift=.5cm] {\(s_2\)};
            \node[state] (s2) [right of=s0, yshift=-.5cm] {\(s_1\)};
            \node[state] (good) [right of=s1] {\(\good\)};
            \node[state] (bad) [right of=s2] {\(\bad\)};
            \path[->] (s0) edge[bend left=10] node[above] {$1-p$} (s1);
            \path[->] (s0) edge[bend right=10] node[below] {$p$} (s2);
            \path[->] (s2) edge[bend left=16] node[above, yshift=-.05cm] {$q$} (good);
            \path[->] (s2) edge[] node[below] {$1-q$} (bad);
            \path[->] (s1) edge[] node[above] {$1-q$} (good);
            \path[->] (s1) edge[bend right=16] node[below, yshift=.05cm] {$q$} (bad);
            \path[->] (good) edge[loop right] node[right] {$1$} (good);
            \path[->] (bad) edge[loop right] node[right] {$1$} (bad);
        \end{tikzpicture}
    \end{adjustbox}
    \caption{pMC \(\D\)}
\label{fig:commutechoicesparams}
\end{subfigure}
\begin{subfigure}{0.49\linewidth}
    \centering
    \begin{adjustbox}{max width=\linewidth}
        \begin{tikzpicture}[node distance=1.5cm, on grid, auto, initial text =]
            \node[state, initial] (s0) {\(s_0\)};
            \node[state] (s1) [right of=s0, yshift=.5cm] {\(s_2\)};
            \node[state] (s2) [right of=s0, yshift=-.5cm] {\(s_1\)};
            \node[state] (good) [right of=s1] {\(\good\)};
            \node[state] (bad) [right of=s2] {\(\bad\)};
            \path[->] (s0) edge[bend left=10] node[above] {$1-q$} (s1);
            \path[->] (s0) edge[bend right=10] node[below] {$q$} (s2);
            \path[->] (s2) edge[bend left=16] node[above, yshift=-.05cm] {$p$} (good);
            \path[->] (s2) edge[] node[below] {$1-p$} (bad);
            \path[->] (s1) edge[] node[above] {$1-p$} (good);
            \path[->] (s1) edge[bend right=16] node[below, yshift=.05cm] {$p$} (bad);
            \path[->] (good) edge[loop right] node[right] {$1$} (good);
            \path[->] (bad) edge[loop right] node[right] {$1$} (bad);
        \end{tikzpicture}
    \end{adjustbox}
    \caption{pMC \(\D'\)}
\end{subfigure}
\caption{Two pMCs over parameters $V=\{p,q\}$ with different orderings}
\label{fig:reorderingorder}
\end{figure}
\begin{example}\label{ex:reorderinggonewrong}
    Consider \(\D\) and \(\D'\) with \(\D \equiv \D'\) in \cref{fig:reorderingorder}.
    Regions \(R = [0.1,0.5] \times [0.6,0.7]\), \(R' = [0.6,0.7] \times [0.1,0.5]\)
    yield \(\llangle \isub_{R}(\D) \rrangle = \llangle \isub_{R'}(\D') \rrangle = [0.22,0.66]\)
    and \(\llangle \isub_{R}(\D') \rrangle = \llangle \isub_{R'}(\D) \rrangle = [0.33,0.55]\).
    Consequently, any transformation \(\ttra\) with either $\ttra(\D) = \D'$ or $\ttra(\D') = \D$ is \emph{not} tightening.
    A similar observation appears in flip-hoisting~\cite{DBLP:journals/corr/abs-2110-10284}.
    Note that one could intersect these region estimates and always tighten estimates with the result. We leave this as future work.
\end{example}

\subsection{Big-step Transformation Algorithm for pMCs}
\label{section:bigstep}

We combine shortcuts and grouping into the big-step algorithm. Its steps are:
\begin{description}
    \item[Step 1:] Find a suitable sub-pMC rooted in some $\hat{s} \in S$ over \(p \in V\) (or terminate).
    \item[Step 2:] Construct the shortcut pMC \(\ttra_\mathrm{shortcut}(\D, \D_{\hat{s},p})\).
    \item[Step 3:] If possible, construct grouping pMCs \(\ttra_\mathrm{group}(\D',\hat{s},\cdot)\). Go to step 1.
\end{description}

\paragraph{Step 1: Picking transformations over \((\hat{s}, p)\).}
States \(\hat{s} \in S\) are selected through a stack, which is initially a
topological ordering from the initial state, and all parameters \(p \in V\) are
selected such that each \((\hat{s}, p)\) is visited once. Applying
transformations only makes sense if \(\D_{\hat{s}, p}\) has more than one
occurrence of \(p\).  To check this efficiently, we define a map \(\gamma\colon
S \times V \rightarrow 2^{S}\), such that for all \(s \in \gamma(\hat{s}, p)\), \(s\) is reachable from
\(\hat{s}\) by constant transitions and \(s\) has a \(p\)-transition.
The mapping $\gamma$ is computable by a standard graph search.
 We pick \((\hat{s}, p)\) if
\[
|\gamma(\hat{s}, p)| \geq 2 \quad \text{or} \quad
 \Big( \gamma(\hat{s}, p) = \{s\} \text{ and  } \exists s'\in S: \mathcal{P}(s,s') \neq 0 \wedge  \gamma(s', p) \neq \emptyset \Big).
\]
The above condition implies the existence of a suitable sub-pMC \(\D_{\hat{s}, p}\) with more than one occurrence of \(p\).
Additionally, checking that we will make at least one state from \(\D\)
unreachable in an iteration on \((\hat{s}, p)\) makes the algorithm terminate.

\paragraph{Step 2: Applying \(\ttra_\mathrm{shortcut}\).}
We compute \(\D_{\hat{s}, p}\) using a DFS, where we add reachable states \(s\)
if they conform to \cref{def:rootedpmc} and if \(|\gamma(s, p)| \neq
\emptyset\). We then compute \(\ttra_\mathrm{shortcut}(\D, \D_{\hat{s}, p})\)
as discussed in \cref{app:shortcutprobs}.

\paragraph{Step 3: Applying \(\ttra_\mathrm{group}\).}
With \(\D'\) starting as the shortcut pMC,
we compute \(\D' \leftarrow \ttra_\mathrm{group}(\D',\hat{s},f)\) if we find at least two shortcuts
with a common factor \(f\). This is done repeatedly until no more common factors are found.
Assuming a factorized representation of polynomials, their common factors can be identified by a syntactical comparison.
The new states \(s'\) are pushed to the top of the stack. Grouping changes the map
\(\gamma\), which has to be recomputed locally.

\begin{theoremrep}
    The big-step transformation is tightening in the sense of
    \cref{def:tighteningtransformation}.
    \label{thm:improvesregionestimates}
\end{theoremrep}
\begin{toappendix}
\begin{proof}

The big-step transformation is a composition of a series of shortcut and grouping transformations. It satisfies all properties of a tightening transformation:
\begin{itemize}
    \item Tightens region estimates: The shortcut and grouping transformations tighten region estimates, as proven in \Cref{lem:shortcuttightens,lem:groupingtightens}, so the big-step transformation itself tightens region estimates by transitivity.
    \item Preserves set of parameters: The big-step transformation preserves parameters as the set of parameters $V$ stays the same in both transformations.
    \item Preserves well-defined instantiations: Suppose $u$ is a well-defined instantiation. The shortcut transformation replaces the transitions involving a parameter $p \in V$ by polynomials over $p$, which aggregate the total probability mass to each exit of the shortcut pMC. These new polynomials are non-negative and sum to one. For the grouping transformation, the sum of the outgoing probabilities from $\hat{s}$ and the newly introduced $s'$ remains one, as the probabilities of the grouped transitions are summed into $s'$, then distributed according to normalized ratios.
\end{itemize}
\end{proof}
\end{toappendix}
\noindent
\Cref{thm:improvesregionestimates} follows from \Cref{lem:shortcuttightens,lem:groupingtightens}.
The big-step transformation may result in iMCs with large polynomial transitions. We use a Newton method to compute an iMC that substitutes \(\D\) in \(R\), cf.\  \cref{app:largepolys}.

\section{Experiments}
\nosectionappendix%
\label{section:experiments}%
\paragraph{Research questions and methodology.}
We evaluate the performance of GPL and the big-step transformation (Q1\&2) and the wider applicability of GPL (Q3\&4):
\begin{compactitem}
	\item (Q1) What is the effect of the big-step transformation (\cref{section:bigstep})?%
	\item (Q2) How does GPL's performance compare against standard PL \cite{DBLP:conf/atva/QuatmannD0JK16}? Can GPL compete with standard PL on benchmarks supported by both?%
	\item (Q3) Is GPL efficient on regions that standard PL cannot handle?
	\item (Q4) Can GPL efficiently analyze a family of pMCs using discrete parameters?
\end{compactitem}
We implemented GPL and the big-step transformation in
Storm~\cite{DBLP:journals/sttt/HenselJKQV22}, improving upon its
implementation of robust value iteration (VI) on iMCs~\cite{DBLP:journals/ior/NilimG05}. The implementation is released as an open-source component of Storm.
The experiments ran on a
single core of an AMD~Ryzen~TRP~5965WX with 60 minutes timeout and 32GB available memory.
We use  VI with default precision of \(10^{-6}\). 
For region refinement (\cref{sec:parameterlifting}), we split on four parameters. Preliminary experiments indicated that the
results are not sensitive to this hyperparameter.
Our benchmarks consist of a model $\D$, a region $R$, and a specification $\varphi$.
\begin{figure}[t]
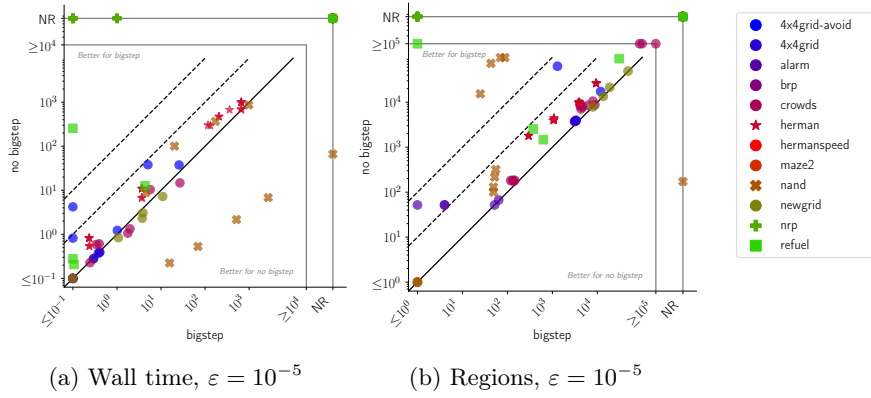

    \centering
    \begin{subfigure}{0.38\linewidth}
    \begin{adjustbox}{max width=\linewidth}
        \input{misc/benchmarks/time-1e-05.pgf}
    \end{adjustbox}
    \caption{Wall time, \(\varepsilon=10^{-5}\)}
    \label{fig:bigsteptime}
    \end{subfigure}%
    \begin{subfigure}{0.38\linewidth}
    \begin{adjustbox}{max width=\linewidth}
        \input{misc/benchmarks/regions-1e-05.pgf}
    \end{adjustbox}
    \caption{Regions, \(\varepsilon=10^{-5}\)}
    \label{fig:bigstepregions}
    \end{subfigure}%
    \begin{subfigure}{0.2\linewidth}
    \begin{adjustbox}{max width=\linewidth, clip, trim=0cm 9.5cm 0cm 0cm}
        \input{misc/benchmarks/time-1e-05-legend.pgf}
    \end{adjustbox}
    \end{subfigure}
    \caption{Effectiveness of the big-step transformation (Q1)}
\end{figure}%

\subsubsection{Q1: What is the effect of the big-step transformation?}
\label{section:experimentbigstep}%
We compare number of regions and runtime for GPL with and without big-step transformation.
 
\paragraph{Setup.}
We consider simple pMCs (\texttt{4x4grid}, \texttt{evade}, \texttt{maze2},
\texttt{nrp}, \texttt{refuel}) synthesized from POMDPs~\cite{DBLP:journals/corr/abs-2405-13583,DBLP:conf/uai/Junges0WQWK018}, pMCs
from~\cite{DBLP:conf/atva/QuatmannD0JK16,DBLP:journals/dc/VolkBKA22} 
(\texttt{brp}, \texttt{crowds}, \texttt{nand}, \texttt{herman}, \texttt{hermanspeed}),
and a pMC generated from a Bayesian network (\texttt{alarm})
from~\cite{DBLP:journals/jair/SalmaniK23}.
Of those pMCs, we choose multiple instances that
(a)~require at least one refinement step and (b)~are solvable in 60 minutes in at least one case. 
We verify five different regions on each benchmark: 
$[0.2,1]^{|V|}$, $[0,0.8]^{|V|}$, and $[\delta, 1-\delta]^{|V|}$ for \(\delta \in \{0, 10^{-6}, 0.1\}\).
Note that most of these regions are not graph preserving.
We obtain challenging probability thresholds for the specifications using gradient descent (GD) \cite{DBLP:conf/vmcai/HeckSJMK22} by running GD for at least ten converging iterations and using the best value that it has found.
We add an \(\varepsilon
\in \{10^{-1}, 10^{-2}, 10^{-4}, 10^{-5}\}\) away from the optimum and ask GPL to
verify it. 
The task thus is to prove the
\(\varepsilon\)-optimality of the bound found by GD. 
We do not run experiments on trivial probability thresholds like $\lambda=1.01$. Instances where the specification does not hold are excluded from our evaluation, as they can be efficiently solved by GD.
Here, we enabled the standard splitting heuristic in PL. In \cref{app:roundrobin}, we confirm that our observations also hold when regions are split by round-robin.

\paragraph{Results.}
\Cref{fig:bigsteptime,fig:bigstepregions} compare the performance of GPL with and without the big-step transformation for \(\varepsilon = 10^{-5}\).  The (log-scale!) plots show the wall time of the
entire Storm execution and the number of regions needed to prove the specification \(\varphi\).
A point \((x, y)\) indicates that GPL needed \(x\) seconds
(regions) to prove \(\varphi\) with the big-step transformation and \(y\) seconds (regions) without. Points above the diagonal mean that the big-step transformation is
beneficial, the two dashed lines indicate an improvement of factor 10 and 100
respectively. Detailed results, also for other values of \(\varepsilon\), are in \cref{app:exbigstep}. Smaller \(\varepsilon\) constitute
more difficult benchmarks that require more region refinements, as they imply the statement for all larger \(\varepsilon\).
In \cref{appendix:exsimple}, we compare against standard PL.

\paragraph{Discussion.} 
The big-step transformation reduces the number of regions on almost all models.
GPL with big-step solves \texttt{nrp} within one region and two seconds, even on an
instance with 100 parameters, while GPL without big-step already times out on the
instance with five parameters.
We illustrate what happens on \texttt{nrp} in \cref{app:nrp}.
Big-step also helps tremendously in other cases, such as \texttt{refuel} (34 parameters) and
some \texttt{nand} (2 parameters).  While \texttt{nrp} has many
parameters that big-step reorders, \texttt{nand} has many
parameters from which big-step creates shortcuts. While proving the
bound on some regions on \texttt{nand} is much faster with the big-step
algorithm enabled, the algorithm without is faster on other regions,
outcompeting the transformation time. 
The big-step overhead usually pays off, as it is rarely the case that the added
transformation time outweighs the time saved while running GPL.
Further experiments show that the transformation scales to many states, but
handling large shortcuts, as in \texttt{nand}, is expensive in our
implementation, which can be improved in the future.

\subsubsection{Q2: How does GPL's performance compare against standard PL?}

\paragraph{Setup.}
We now compare GPL without big-step transformation to standard PL.
We use the benchmarks
from~Q1 with graph-preserving regions, as the
others cannot be handled by standard PL.
We drop the non-monotonic \texttt{hermanspeed} benchmark as it is not supported by standard PL.
We measure wall-clock time on the benchmarks where the execution took more than
one second.

\begin{figure}[!t]
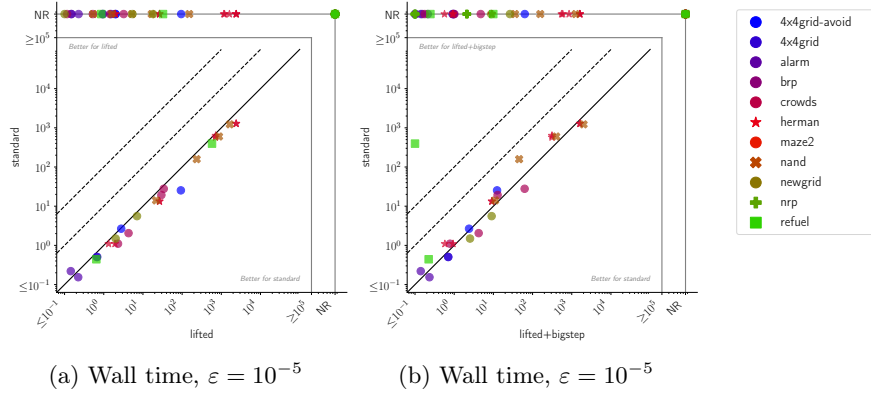

    \centering
    \begin{subfigure}{0.38\linewidth}
    \begin{adjustbox}{max width=\linewidth}
        \input{misc/benchmarks/time-slowdown.pgf}
    \end{adjustbox}
    \caption{Wall time, \(\varepsilon=10^{-5}\)}
    \label{fig:standardbench}
    \end{subfigure}%
    \begin{subfigure}{0.38\linewidth}
    \begin{adjustbox}{max width=\linewidth}
        \input{misc/benchmarks/time-st-big.pgf}
    \end{adjustbox}
    \caption{Wall time, \(\varepsilon=10^{-5}\)}
    \label{fig:standardbigstep}
    \end{subfigure}%
    \begin{subfigure}{0.2\linewidth}
    \begin{adjustbox}{max width=\linewidth, clip, trim=0cm 12cm 0cm 0cm}
        \input{misc/benchmarks/time-slowdown-legend.pgf}
    \end{adjustbox}
    \end{subfigure}
    \caption{Comparison of generalized and standard PL (Q2). Many no-result points for standard PL indicate that the benchmark is not supported as the region is not graph preserving}
    \label{fig:genvsstandard}
\end{figure}

\paragraph{Results.}
On average, GPL needs 1.46x the runtime of standard PL on these benchmarks, with a median of 1.37x.
In \cref{fig:standardbench}, we
compare wall-time between generalized and standard PL on simple pMCs.  In
\cref{fig:standardbigstep}, we show the same with the big-step transformation
enabled on generalized PL.
The runtimes of the algorithms scale equally on harder benchmarks.
The \texttt{4x4grid-avoid} benchmark is comparatively much slower for generalized PL
than for standard PL, we have not investigated why this is. It becomes faster
than standard PL with big-step enabled.
We present more detailed results in \cref{appendix:exsimple}.

\paragraph{Discussion.}
The runtime overhead of GPL is mostly due to performing VI on iMCs which takes slightly more time per iteration compared to value iteration on MDPs, which is used by PL.
A hybrid iMC/MDP approach could speed up GPL.

\subsubsection{Q3: Is GPL efficient on regions that standard PL cannot handle?}
\label{section:experimentnonsimple}

\paragraph{Setup.}
We have already seen in Q1 that GPL can handle not-graph-preserving regions.
We further evaluate performance on a handcrafted, parameterized pMCs $\D_n = ( \{s_0, \dots, s_n, \good, \bad\}, \{p_1, \dots, p_{n-1}\}, s_0, \mathcal{P})$, where $1\le n\le32$, $\mathcal{P}(s_0,s_i)=p_i$ ($1\le i < n$), $\mathcal{P}(s_0,s_n)=1 - \sum_{1 \leq i < n} p_i$, and $\mathcal{P}(s_i,\good)=\nicefrac{1}{i}=1-\mathcal{P}(s_i,\bad)$ ($1\le i \le n$).
$\D_n$ reflects a multi-parameter distribution coming out of the initial state $s_0$---a worst-case scenario for standard PL as the used MDP abstraction requires $2^{n-1}$ distinct actions. See \cref{app:notwelldefined}.
We consider the specification \(\varphi=\Pr(s_0 \leadsto \good) \geq
0.01\) with regions \(R_1 = [10^{-6}, \nicefrac{1}{n}]^{n-1}\), \(R_2 = [0, \nicefrac{1}{n}]^{n-1}\), and \(R_3 = [0, \nicefrac{2}{n}]^{n-1}\).
Only $R_1$ is supported by standard PL. $R_2$ and $R_3$ are not graph preserving and $R_3$ is also not well defined.

\paragraph{Results.}
Standard PL takes 126.0s to verify \(R_1\) for \(n=23\) and has a mem-out (>32GB) for \(n \geq 24\) allocating \(2^{n-1}\) MDP actions.
$R_2$ and $R_3$ are not supported.

Our proposed GPL proves \(\varphi\) on \(R_1\), \(R_2\) and \(R_3\) without region refinement for all \(1\le n\le 32\) in under 1s.
Detailed results  are in \cref{app:notwelldefined}.

\paragraph{Discussion.}
GPL can efficiently verify the scaling benchmark on not-well-defined and not-graph-preserving regions.
Verifying properties on pMCs with many parameters in a single state's
distribution, even on graph-preserving and well-defined regions like \(R_1\),
only becomes feasible with generalized PL. %
As we discuss in \cref{sec:illdefined}, there is no simple way around
this limitation in standard PL.

\subsubsection{Q4: Can GPL efficiently analyze a family of pMCs?}%
\label{section:experimentnongraphpreserving}%

\paragraph{Setup.}
We run an experiment on a pMC generated from a family of pMCs as discussed in \cref{sec:families}.
We use a variant of \emph{Dynamic Power
Management}~\cite{DBLP:journals/tcad/BeniniBPM99, DBLP:conf/fm/CeskaHJK19} with
16 discrete and two continuous parameters.
We choose the region \(\{0, 1\}^{16} \times [0.4, 0.6] \times [0.7, 0.9]\).
The discrete parameters describe the topology of DPM's controller, while the continuous parameters describe probabilities to start and continue sending packets.
The bound we use is the one
found by gradient descent\footnote{Ignoring the discreteness of some parameters. The bound is correct as GPL proves it.}
minus \(\varepsilon=10^{-5}\). We compare against enumerating all $2^{16}$ possible discrete parameter valuations and verifying each resulting pMC using (standard) PL.

\paragraph{Results and discussion.}
GPL proves the property with a refinement into 128 subregions within 0.62s.
For 90 regions, verifying a \emph{single} iMC implies the specification for \emph{multiple} family members.
GPL thus reasons effectively about the pMC family. 
Enumerating and solving all family members with PL takes 698.51s.

\section{Related Work}
Closest to our work are abstraction-refinement loops for verifying pMCs~\cite{DBLP:conf/atva/QuatmannD0JK16}, discussed in \Cref{sec:intro}, and for related models in~\cite{DBLP:journals/acta/CeskaDPKB17,DBLP:conf/tacas/CeskaJJK19,DBLP:conf/cav/AndriushchenkoBCJKM23,DBLP:conf/atva/GiroR12}.
Crucially, the abstraction in these approaches ignores parameter dependencies between different states.  
Global monotonicity of certain parameters~\cite{DBLP:conf/atva/SpelJK19} allows avoiding useless region splits~\cite{DBLP:conf/tacas/SpelJK21}.
An application of PL to distributed protocols~\cite{DBLP:journals/dc/VolkBKA22} overcomes the necessity for monotonic transition functions by splitting the region a-priori.

We compute solution functions in our shortcut transformation. Their computation is heavily studied, originally in~\cite{DBLP:conf/ictac/Daws04,DBLP:journals/fac/LanotteMT07,DBLP:journals/sttt/HahnHZ11,DBLP:journals/fmsd/JungesAHJKQV24}.
A polynomial-time algorithm for a fixed number of parameters is given in
\cite{DBLP:journals/iandc/BaierHHJKK20}.
Improvements of state-elimination include exploiting similarities between multiple models~\cite{DBLP:conf/qest/GainerHS18}
and achieving speed-ups with a graph-like function representations~\cite{DBLP:conf/atva/GainerHS18}.
Similarly to our pMC transformation, solution function computation that
first considers fragments is investigated in \cite{DBLP:journals/tse/FangCGA23}.
Other computational problems on pMCs have gained quite some attention: For feasibility, dual to verification, incomplete approaches are popular~\cite{DBLP:conf/tase/ChenHHKQ013,DBLP:journals/tac/CubuktepeJJKT22,
DBLP:conf/vmcai/HeckSJMK22} and scale to thousands of parameters.
Discrete and continuous parameters are mixed in \cite{DBLP:conf/icsa/CalinescuCGKP17} to find locally Pareto-optimal designs.
More work considers verification of pMDPs
\cite{DBLP:conf/qest/PolgreenWHA17, DBLP:conf/l4dc/RickardAM24}, pCTMCs
\cite{DBLP:conf/tacas/BortolussiS18}, and MDPs with latent parameters
\cite{DBLP:conf/aaai/CostenRLH23}.

Our pMC transformations have parallels in probabilistic programming.
Flip-hoisting \cite{DBLP:journals/corr/abs-2110-10284} merges parallel
equivalent flip statements, while we merge parallel parameter
transitions on the same parameter.
Big-step semantics \cite[p.\ 24]{DBLP:books/daglib/0070910} join sequential
statements, while we join sequential parametric transitions.
Our use on Newton's method to compute iMCs from pMCs given regions is taken from
\cite[p.\ 105]{DBLP:books/daglib/0022249}.  Specialized variations of Newton's
method have been used to verify recursive MCs
\cite{DBLP:journals/jacm/EtessamiY09} and recursive stochastic games
\cite{DBLP:conf/icalp/EsparzaGKS08}.

\section{Conclusion and Outlook}
This paper presents generalized parameter lifting (GPL), an abstraction-refinement loop for pMC verification. 
GPL enhances the state of the art by its ability to solve a wider class of practically motivated pMCs on a wider class of parameter regions. 
In particular, in contrast to standard PL, GPL can prove specifications for every induced Markov chain of any given pMC.
GPL also allows for a novel big-step transformation of pMCs that yields finer abstractions.
Future work includes exploring new application areas of pMCs enabled by GPL and investigating pMC transformations that reorder parameters.

\newpage

\nosectionappendix
\bibliography{paper}

\newpage
\begin{toappendix}
\section{Pseudocode for Full Algorithm}
\label{app:pseudocode}

We give a pseudocode description of the full big-step algorithm in \cref{alg:bigstep}.
\begin{algorithm}
\begin{algorithmic}
    \Procedure{GPLWithBigStep}{$\D, R, \varphi$}
    \State $\D \gets \Call{BigStep}{\D}$
    \Comment{\cref{section:transformation}}
    \State $q \gets \Call{Queue}{}()$
    \State $\Call{Push}{q, R}$
    \While{$\lnot \Call{Empty}{q}$}
        \State $R' \gets \Call{Pop}{q}$
        \State $\mathcal{I} \gets \Call{ConstructiMC}{\D, R'}$ \Comment{\cref{sec:estimatesplit}}
        \State $\mathcal{I}' \gets \Call{EliminateECs}{\mathcal{I}}$ \Comment{\cref{section:notgraphperserving}}
        \If{$\mathcal{I} \not\vDash \varphi$}
        \Comment{\cref{sec:estimatesplit}, not-well-defined $R$: \cref{sec:illdefined}}
            \If{$\mathcal{I} \vDash \lnot \varphi$}
                \State \textbf{return false}
            \Else
                \State $R_1, \ldots, R_n \gets \Call{Split}{R}$ \Comment{\cref{sec:estimatesplit}, discrete params: \cref{sec:families}}
                \State $\Call{Push}{q, R_1, \ldots, R_n}$
            \EndIf
        \EndIf
    \EndWhile
    \State \textbf{return true}
\EndProcedure
\end{algorithmic}
\caption{GPL with big-step transformation}
\label{alg:bigstep}
\end{algorithm}

\section{Full Big-Step Transformation Example}
\label{app:bigstep}

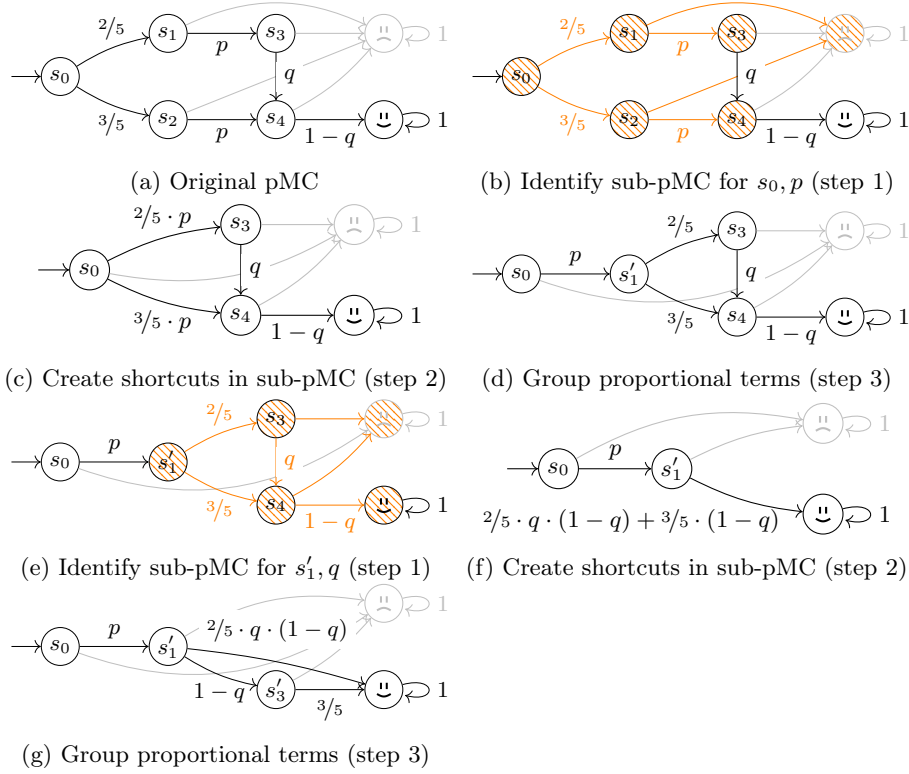
\begin{figure}[t]
\begin{subfigure}{0.5\linewidth}
    \centering%
    \begin{adjustbox}{max width=\linewidth}
        \begin{tikzpicture}[node distance=1.5cm, on grid, auto, initial text =]
            \node[state, initial] (s0) {\(s_0\)};
            \node[state] (s1) [right of=s0, yshift=0.6cm] {\(s_1\)};
            \node[state] (s2) [right of=s0, yshift=-0.6cm] {\(s_2\)};
            \node[state] (s3) [right of=s1] {\(s_3\)};
            \node[state] (s4) [right of=s2] {\(s_4\)};
            \node[state, draw=lightgray, text=lightgray] (s6) [right of=s3] {\(\bad\)};
            \node[state] (s5) [right of=s4] {\(\good\)};

            \path[->] (s0) edge[bend left=10] node[above] {$\nicefrac{2}{5}$} (s1);
            \path[->] (s0) edge[bend right=10] node[below] {$\nicefrac{3}{5}$} (s2);

            \path[->, draw=lightgray] (s1) edge[bend left=25] (s6);
            \path[->, draw=lightgray] (s2) edge[bend right=0] (s6);
            \path[->, draw=lightgray] (s4) edge[bend right=10] (s6);
            \path[->, draw=lightgray] (s3) edge (s6);
            \path[->, draw=lightgray] (s6) edge[loop right] node[right, text=lightgray] {$1$} (s6);

            \path[->] (s1) edge node[below] {$p$} (s3);
            \path[->] (s2) edge node[below] {$p$} (s4);

            \path[->] (s3) edge[] node[right, fill=white] {$q$} (s4);
            \path[->] (s4) edge[] node[below] {$1-q$} (s5);

            \path[->] (s5) edge[loop right] node[right] {$1$} (s5);
        \end{tikzpicture}
    \end{adjustbox}
    \caption{Original pMC}
\end{subfigure}%
\begin{subfigure}{0.5\linewidth}
    \centering%
    \begin{adjustbox}{max width=\linewidth}
        \begin{tikzpicture}[node distance=1.5cm, on grid, auto, initial text =]
            \node[state, initial, pattern=north west lines, pattern color=orange] (s0) {\(s_0\)};
            \node[state, pattern=north west lines, pattern color=orange] (s1) [right of=s0, yshift=0.6cm] {\(s_1\)};
            \node[state, pattern=north west lines, pattern color=orange] (s2) [right of=s0, yshift=-0.6cm] {\(s_2\)};
            \node[state, pattern=north west lines, pattern color=orange] (s3) [right of=s1] {\(s_3\)};
            \node[state, pattern=north west lines, pattern color=orange] (s4) [right of=s2] {\(s_4\)};
            \node[state, draw=lightgray, text=lightgray, pattern=north west lines, pattern color=orange] (s6) [right of=s3] {\(\bad\)};
            \node[state] (s5) [right of=s4] {\(\good\)};

            \path[->, draw=orange] (s0) edge[bend left=10] node[above, text=orange] {$\nicefrac{2}{5}$} (s1);
            \path[->, draw=orange] (s0) edge[bend right=10] node[below, text=orange] {$\nicefrac{3}{5}$} (s2);

            \path[->, draw=orange] (s1) edge[bend left=25] (s6);
            \path[->, draw=orange] (s2) edge[bend right=0] (s6);
            \path[->, draw=lightgray] (s4) edge[bend right=10] (s6);
            \path[->, draw=lightgray] (s3) edge (s6);
            \path[->, draw=lightgray] (s6) edge[loop right] node[right, text=lightgray] {$1$} (s6);

            \path[->, draw=orange] (s1) edge node[below, text=orange] {$p$} (s3);
            \path[->, draw=orange] (s2) edge node[below, text=orange] {$p$} (s4);

            \path[->] (s3) edge[] node[right, fill=white] {$q$} (s4);
            \path[->] (s4) edge[] node[below] {$1-q$} (s5);

            \path[->] (s5) edge[loop right] node[right] {$1$} (s5);
        \end{tikzpicture}
    \end{adjustbox}
    \caption{Identify sub-pMC for \(s_0, p\) (step 1)}
\end{subfigure}

\begin{subfigure}{0.5\linewidth}
    \centering%
    \begin{adjustbox}{max width=\linewidth}
        \begin{tikzpicture}[node distance=1.5cm, on grid, auto, initial text =]
            \node[state, initial] (s0) {\(s_0\)};
            \node[state] (s3) [right of=s0, xshift=.5cm, yshift=0.6cm] {\(s_3\)};
            \node[state] (s4) [right of=s0, xshift=.5cm, yshift=-0.6cm] {\(s_4\)};
            \node[state, draw=lightgray, text=lightgray] (s6) [right of=s3] {\(\bad\)};
            \node[state] (s5) [right of=s4] {\(\good\)};

            \path[->, draw=lightgray] (s0) edge[bend right=20] (s6);
            \path[->, draw=lightgray] (s3) edge (s6);
            \path[->, draw=lightgray] (s4) edge[bend right=10] (s6);
            \path[->, draw=lightgray] (s6) edge[loop right] node[right, text=lightgray] {$1$} (s6);

            \path[->] (s0) edge[bend left=10] node[above] {$\nicefrac{2}{5} \cdot p$} (s3);
            \path[->] (s0) edge[bend right=10] node[below] {$\nicefrac{3}{5} \cdot p$} (s4);

            \path[->] (s3) edge[] node[right, fill=white] {$q$} (s4);
            \path[->] (s4) edge[] node[below] {$1-q$} (s5);

            \path[->] (s5) edge[loop right] node[right] {$1$} (s5);
        \end{tikzpicture}
    \end{adjustbox}
    \caption{Create shortcuts in sub-pMC (step 2)}
\end{subfigure}%
\begin{subfigure}{0.5\linewidth}
    \centering%
    \begin{adjustbox}{max width=\linewidth}
        \begin{tikzpicture}[node distance=1.5cm, on grid, auto, initial text =]
            \node[state, initial] (s0) {\(s_0\)};
            \node[state] (s1) [right of=s0] {\(s_1'\)};
            \node[state] (s3) [right of=s1, yshift=0.6cm] {\(s_3\)};
            \node[state] (s4) [right of=s1, yshift=-0.6cm] {\(s_4\)};
            \node[state, draw=lightgray, text=lightgray] (s6) [right of=s3] {\(\bad\)};
            \node[state] (s5) [right of=s4] {\(\good\)};

            \path[->, draw=lightgray] (s0) edge[bend right=27] (s6);
            \path[->, draw=lightgray] (s3) edge (s6);
            \path[->, draw=lightgray] (s4) edge[bend right=10] (s6);
            \path[->, draw=lightgray] (s6) edge[loop right] node[right, text=lightgray] {$1$} (s6);

            \path[->] (s0) edge[] node[above] {$p$} (s1);

            \path[->] (s1) edge[bend left=10] node[above] {$\nicefrac{2}{5}$} (s3);
            \path[->] (s1) edge[bend right=10] node[below] {$\nicefrac{3}{5}$} (s4);

            \path[->] (s3) edge[] node[right, fill=white] {$q$} (s4);
            \path[->] (s4) edge[] node[below] {$1-q$} (s5);

            \path[->] (s5) edge[loop right] node[right] {$1$} (s5);
        \end{tikzpicture}
    \end{adjustbox}
    \caption{Group proportional terms (step 3)}
\end{subfigure}

\begin{subfigure}{0.5\linewidth}
    \centering%
    \begin{adjustbox}{max width=\linewidth}
        \begin{tikzpicture}[node distance=1.5cm, on grid, auto, initial text =]
            \node[state, initial] (s0) {\(s_0\)};
            \node[state, pattern=north west lines, pattern color=orange] (s1) [right of=s0] {\(s_1'\)};
            \node[state, pattern=north west lines, pattern color=orange] (s3) [right of=s1, yshift=0.6cm] {\(s_3\)};
            \node[state, pattern=north west lines, pattern color=orange] (s4) [right of=s1, yshift=-0.6cm] {\(s_4\)};
            \node[state, draw=lightgray, text=lightgray, pattern=north west lines, pattern color=orange] (s6) [right of=s3] {\(\bad\)};
            \node[state, pattern=north west lines, pattern color=orange] (s5) [right of=s4] {\(\good\)};

            \path[->, draw=lightgray] (s0) edge[bend right=27] (s6);
            \path[->, draw=orange] (s3) edge (s6);
            \path[->, draw=orange] (s4) edge[bend right=10] (s6);
            \path[->, draw=lightgray] (s6) edge[loop right] node[right, text=lightgray] {$1$} (s6);

            \path[->] (s0) edge[] node[above] {$p$} (s1);

            \path[->, draw=orange] (s1) edge[bend left=10] node[above, text=orange] {$\nicefrac{2}{5}$} (s3);
            \path[->, draw=orange] (s1) edge[bend right=10] node[below, text=orange] {$\nicefrac{3}{5}$} (s4);

            \path[->, draw=orange] (s3) edge[] node[right, fill=white, text=orange] {$q$} (s4);
            \path[->, draw=orange] (s4) edge[] node[below, text=orange] {$1-q$} (s5);

            \path[->] (s5) edge[loop right] node[right] {$1$} (s5);
        \end{tikzpicture}
    \end{adjustbox}
    \caption{Identify sub-pMC for \(s_1', q\) (step 1)}
\end{subfigure}%
\begin{subfigure}{0.5\linewidth}
    \centering%
    \begin{adjustbox}{max width=\linewidth}
        \begin{tikzpicture}[node distance=1.5cm, on grid, auto, initial text =]
            \node[state, initial] (s0) {\(s_0\)};
            \node[state] (s1) [right of=s0] {\(s_1'\)};
            \node[state, draw=lightgray, text=lightgray] (s6) [right of=s1, yshift=0.6cm, xshift=0.5cm] {\(\bad\)};
            \node[state] (s5) [right of=s1, yshift=-0.6cm, xshift=0.5cm] {\(\good\)};

            \path[->, draw=lightgray] (s0) edge[bend left=20] (s6);
            \path[->, draw=lightgray] (s6) edge[loop right] node[right, text=lightgray] {$1$} (s6);
            \path[->, draw=lightgray] (s1) edge[bend left=10] (s6);

            \path[->] (s0) edge[] node[above] {$p$} (s1);
            \path[->] (s1) edge[bend right=10] node[below, xshift=-1.5cm] {$\nicefrac{2}{5} \cdot q \cdot (1-q) + \nicefrac{3}{5} \cdot (1-q)$} (s5);

            \path[->] (s5) edge[loop right] node[right] {$1$} (s5);
        \end{tikzpicture}
    \end{adjustbox}
    \caption{Create shortcuts in sub-pMC (step 2)}
\end{subfigure}

\begin{subfigure}{0.5\linewidth}
    \centering%
    \begin{adjustbox}{max width=\linewidth}
        \begin{tikzpicture}[node distance=1.5cm, on grid, auto, initial text =]
            \node[state, initial] (s0) {\(s_0\)};
            \node[state] (s1) [right of=s0] {\(s_1'\)};
            \node[state] (s4) [right of=s1, yshift=-0.6cm] {\(s_3'\)};
            \node[state, draw=lightgray, text=lightgray] (s6) [right of=s3] {\(\bad\)};
            \node[state] (s5) [right of=s4] {\(\good\)};

            \path[->, draw=lightgray] (s0) edge[bend right=27] (s6);
            \path[->, draw=lightgray] (s4) edge[bend right=10] (s6);
            \path[->, draw=lightgray] (s6) edge[loop right] node[right, text=lightgray] {$1$} (s6);
            \path[->, draw=lightgray] (s1) edge[bend left=20] (s6);

            \path[->] (s0) edge[] node[above] {$p$} (s1);

            \path[->] (s1) edge[bend left=5] node[above, fill=white, yshift=0.17cm] {$\nicefrac{2}{5}\cdot q \cdot (1-q)$} (s5);
            \path[->] (s1) edge[bend right=10] node[below] {$1-q$} (s4);
            \path[->] (s4) edge[] node[below] {$\nicefrac{3}{5}$} (s5);
            \path[->] (s5) edge[loop right] node[right] {$1$} (s5);
        \end{tikzpicture}
    \end{adjustbox}
    \caption{Group proportional terms (step 3)}
\end{subfigure}
\caption{Big-step transformation algorithm exemplified}
\label{fig:processingalgorithmfull}
\end{figure}

The full version of \cref{fig:processingalgorithm} can be seen in
\cref{fig:processingalgorithmfull}.

\section{Computing Shortcut Probabilities}\label{app:shortcutprobs}

Let $\D_{\hat{s},p}$ be a sub-pMC with state space  $\hat{S}$ as in \cref{def:rootedpmc} in \cref{sec:shortcut}.

We outline our algorithm to efficiently compute the shortcut probabilities
\[
{\Pr}^{\D_{\hat{s},p}}(\hat{s} \leadsto t) ~=~ \sum_{s_0 \dots s_n \in \mathit{Paths}(\hat{s}, t)}~ \prod_{i=1}^{n} \mathcal{P}(s_{i-1},s_i)
\]
for all $t \in \hat{S}$.

Let $\hat{S} = \{ s_0, \dots, s_n\}$, where the states $s_0, \dots, s_n$ are in a topological order according to the underlying (acyclic) graph of $\D_{\hat{s},p}$, i.e., $\hat{s} = s_0$ and for all  $s_i,s_j \in \hat{S}$ with $\hat{\mathcal{P}}(s_i,s_j) \neq 0$ we have $i < j$.

We inductively define polynomials $f^\mathrm{reach}_i$ over $p$ for $i=0, \dots, n$ as follows:
\[
f^\mathrm{reach}_i ~=~
\begin{cases}
1 & \text{if } i=0\\
\sum_{j=0}^{i-1} \mathcal{P}(s_j,s_i) \cdot f^\mathrm{reach}_j & \text{if } i > 0.
\end{cases}
\]

\begin{lemma}
For all $0 \le i \le n$: $f^\mathrm{reach}_i = {\Pr}^{\D_{\hat{s},p}}(\hat{s} \leadsto s_i)$.
\end{lemma}
\begin{proof}
Follows from the topological ordering of $s_0, \dots, s_n$ and a simple induction over the length of the longest path to $t$.
\end{proof}

Our algorithm computes the polynomials $f^\mathrm{reach}_0, f^\mathrm{reach}_1, \dots, f^\mathrm{reach}_n$ which---by the above lemma---coincide with the desired shortcut probabilities.

To allow for an efficient computation, we syntactically represent transition probabilities as a sum of factorized monomials:
Let $\mathfrak{F} = \{f_1, \dots, f_m\}$ be the set of non-constant polynomials over $p$ occurring in $\D_{\hat{s},p}$.
In our implementation, we represent the polynomials $f^\mathrm{reach}_i$ as
\[
    f^\mathrm{reach}_i = \sum_{k=1}^\ell c_k \prod_{j=1^m} (f_j)^{b_{k,j}}. \tag{$\star$} \label{eq:transition}
\]

\section{Big-Step Transformation on NRP}
\label{app:nrp}
The big-step transformation on NRP is illustrated in \cref{fig:bigstepnrp}.

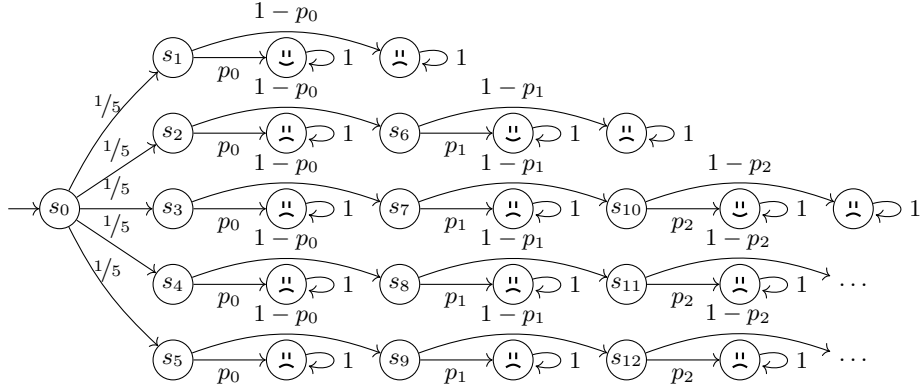
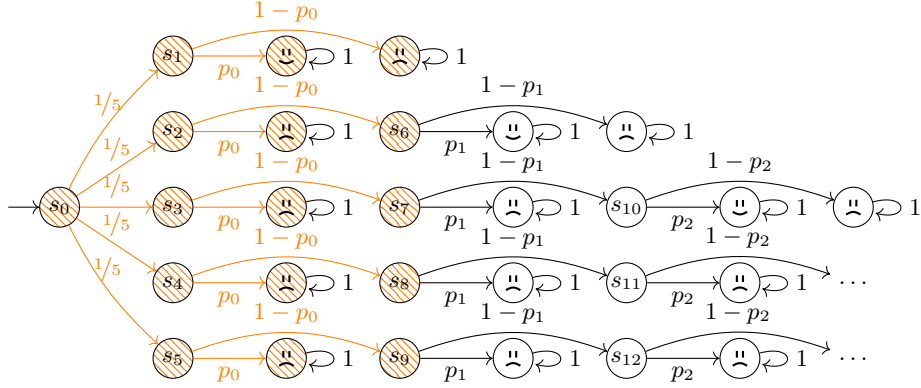
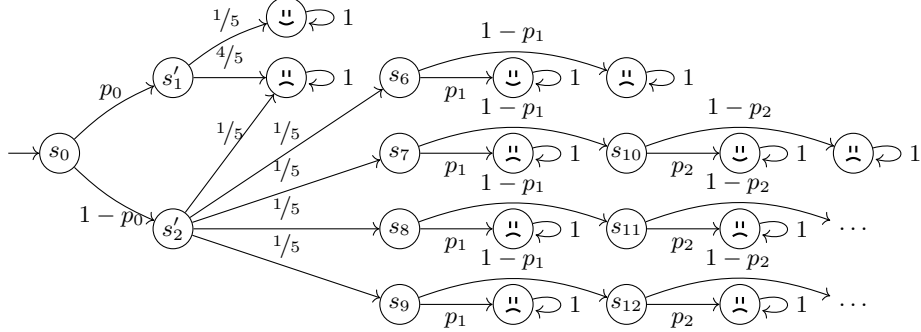
\begin{figure}
    \begin{subfigure}{1\linewidth}%
    \centering
    \begin{tikzpicture}[node distance=1.5cm, on grid, auto, initial text =]
        \node[state, initial] (s0) {\(s_0\)};
        \node[state] (s1) [right of=s0, yshift=2cm] {\(s_1\)};
        \node[state] (s2) [right of=s0, yshift=1cm] {\(s_2\)};
        \node[state] (s3) [right of=s0, yshift=0cm] {\(s_3\)};
        \node[state] (s4) [right of=s0, yshift=-1cm] {\(s_4\)};
        \node[state] (s5) [right of=s0, yshift=-2cm] {\(s_5\)};
        \node[state] (good1) [right of=s1]  {\(\good\)};
        \node[state] (bad0) [right of=good1]  {\(\bad\)};
        \node[state] (bad1) [right of=s2]  {\(\bad\)};
        \node[state] (s6) [right of=bad1]  {\(s_6\)};
        \node[state] (bad2) [right of=s3]  {\(\bad\)};
        \node[state] (s7) [right of=bad2]  {\(s_7\)};
        \node[state] (bad3) [right of=s4]  {\(\bad\)};
        \node[state] (s8) [right of=bad3]  {\(s_8\)};
        \node[state] (bad4) [right of=s5]  {\(\bad\)};
        \node[state] (s9) [right of=bad4]  {\(s_{9}\)};
        \path[->] (s0) edge[bend left=10] node[above] {$\nicefrac{1}{5}$} (s1);
        \path[->] (s0) edge[] node[above] {$\nicefrac{1}{5}$} (s2);
        \path[->] (s0) edge[] node[above] {$\nicefrac{1}{5}$} (s3);
        \path[->] (s0) edge[] node[above] {$\nicefrac{1}{5}$} (s4);
        \path[->] (s0) edge[bend right=10] node[above] {$\nicefrac{1}{5}$} (s5);
        \path[->] (s1) edge[] node[below] {$p_0$} (good1);
        \path[->] (s2) edge[] node[below] {$p_0$} (bad1);
        \path[->] (s3) edge[] node[below] {$p_0$} (bad2);
        \path[->] (s4) edge[] node[below] {$p_0$} (bad3);
        \path[->] (s5) edge[] node[below] {$p_0$} (bad4);
        \path[->] (s1) edge[bend left=20] node[above] {$1-p_0$} (bad0);
        \path[->] (s2) edge[bend left=20] node[above] {$1-p_0$} (s6);
        \path[->] (s3) edge[bend left=20] node[above] {$1-p_0$} (s7);
        \path[->] (s4) edge[bend left=20] node[above] {$1-p_0$} (s8);
        \path[->] (s5) edge[bend left=20] node[above] {$1-p_0$} (s9);
        \node[state] (bad5) [right of=s6] {\(\good\)};
        \node[state] (s10) [right of=bad5] {\(\bad\)};
        \node[state] (bad6) [right of=s7] {\(\bad\)};
        \node[state] (s11) [right of=bad6] {\(s_{10}\)};
        \node[state] (bad7) [right of=s8] {\(\bad\)};
        \node[state] (s12) [right of=bad7] {\(s_{11}\)};
        \node[state] (bad8) [right of=s9] {\(\bad\)};
        \node[state] (s13) [right of=bad8] {\(s_{12}\)};
        \path[->] (s6) edge[] node[below] {$p_1$} (bad5);
        \path[->] (s6) edge[bend left=20] node[above] {$1-p_1$} (s10);
        \path[->] (s7) edge[] node[below] {$p_1$} (bad6);
        \path[->] (s7) edge[bend left=20] node[above] {$1-p_1$} (s11);
        \path[->] (s8) edge[] node[below] {$p_1$} (bad7);
        \path[->] (s8) edge[bend left=20] node[above] {$1-p_1$} (s12);
        \path[->] (s9) edge[] node[below] {$p_1$} (bad8);
        \path[->] (s9) edge[bend left=20] node[above] {$1-p_1$} (s13);
        
        \node[state] (bad10) [right of=s11] {\(\good\)};
        \node[state] (bad11) [right of=s12] {\(\bad\)};
        \node[state] (bad12) [right of=s13] {\(\bad\)};
        
        \node[state] (bad13) [right of=bad10] {\(\bad\)};
        \node (dots2) [right of=bad11] {\(\ldots\)};
        \node (dots3) [right of=bad12] {\(\ldots\)};
        
        \path[->] (s11) edge[] node[below] {$p_2$} (bad10);
        \path[->] (s12) edge[] node[below] {$p_2$} (bad11);
        \path[->] (s13) edge[] node[below] {$p_2$} (bad12);
        
        \path[->] (s11) edge[bend left=20] node[above] {$1-p_2$} (bad13);
        \path[->] (s12) edge[bend left=20] node[above] {$1-p_2$} (dots2);
        \path[->] (s13) edge[bend left=20] node[above] {$1-p_2$} (dots3);
        \path[->] (good1) edge[loop right] node[right] {$1$} (good1);
        \path[->] (bad0) edge[loop right] node[right] {$1$} (bad0);
        \path[->] (bad1) edge[loop right] node[right] {$1$} (bad1);
        \path[->] (bad2) edge[loop right] node[right] {$1$} (bad2);
        \path[->] (bad3) edge[loop right] node[right] {$1$} (bad3);
        \path[->] (bad4) edge[loop right] node[right] {$1$} (bad4);
        \path[->] (bad5) edge[loop right] node[right] {$1$} (bad5);
        \path[->] (bad6) edge[loop right] node[right] {$1$} (bad6);
        \path[->] (bad7) edge[loop right] node[right] {$1$} (bad7);
        \path[->] (bad8) edge[loop right] node[right] {$1$} (bad8);
        \path[->] (s10) edge[loop right] node[right] {$1$} (s10);
        \path[->] (bad10) edge[loop right] node[right] {$1$} (bad10);
        \path[->] (bad11) edge[loop right] node[right] {$1$} (bad11);
        \path[->] (bad12) edge[loop right] node[right] {$1$} (bad12);
        \path[->] (bad13) edge[loop right] node[right] {$1$} (bad13);
    \end{tikzpicture}
    \caption{Fragment of NRP with \(N=5\). The \(\good\) and \(\bad\) states are not merged for clarity, they would usually already be merged into a single state with many incoming transitions}
    \end{subfigure}
    \begin{subfigure}{1\linewidth}%
    \centering
    \begin{tikzpicture}[node distance=1.5cm, on grid, auto, initial text =]
        \node[state, initial, pattern=north west lines, pattern color=orange] (s0) {\(s_0\)};
        \node[state, pattern=north west lines, pattern color=orange] (s1) [right of=s0, yshift=2cm] {\(s_1\)};
        \node[state, pattern=north west lines, pattern color=orange] (s2) [right of=s0, yshift=1cm] {\(s_2\)};
        \node[state, pattern=north west lines, pattern color=orange] (s3) [right of=s0, yshift=0cm] {\(s_3\)};
        \node[state, pattern=north west lines, pattern color=orange] (s4) [right of=s0, yshift=-1cm] {\(s_4\)};
        \node[state, pattern=north west lines, pattern color=orange] (s5) [right of=s0, yshift=-2cm] {\(s_5\)};
        \node[state, pattern=north west lines, pattern color=orange] (good1) [right of=s1]  {\(\good\)};
        \node[state, pattern=north west lines, pattern color=orange] (bad0) [right of=good1]  {\(\bad\)};
        \node[state, pattern=north west lines, pattern color=orange] (bad1) [right of=s2]  {\(\bad\)};
        \node[state, pattern=north west lines, pattern color=orange] (s6) [right of=bad1]  {\(s_6\)};
        \node[state, pattern=north west lines, pattern color=orange] (bad2) [right of=s3]  {\(\bad\)};
        \node[state, pattern=north west lines, pattern color=orange] (s7) [right of=bad2]  {\(s_7\)};
        \node[state, pattern=north west lines, pattern color=orange] (bad3) [right of=s4]  {\(\bad\)};
        \node[state, pattern=north west lines, pattern color=orange] (s8) [right of=bad3]  {\(s_8\)};
        \node[state, pattern=north west lines, pattern color=orange] (bad4) [right of=s5]  {\(\bad\)};
        \node[state, pattern=north west lines, pattern color=orange] (s9) [right of=bad4]  {\(s_{9}\)};
        \path[->, draw=orange] (s0) edge[bend left=10] node[above, text=orange] {$\nicefrac{1}{5}$} (s1);
        \path[->, draw=orange] (s0) edge[] node[above, text=orange] {$\nicefrac{1}{5}$} (s2);
        \path[->, draw=orange] (s0) edge[] node[above, text=orange] {$\nicefrac{1}{5}$} (s3);
        \path[->, draw=orange] (s0) edge[] node[above, text=orange] {$\nicefrac{1}{5}$} (s4);
        \path[->, draw=orange] (s0) edge[bend right=10, text=orange] node[above] {$\nicefrac{1}{5}$} (s5);
        \path[->, draw=orange] (s1) edge[] node[below, text=orange] {$p_0$} (good1);
        \path[->, draw=orange] (s2) edge[] node[below, text=orange] {$p_0$} (bad1);
        \path[->, draw=orange] (s3) edge[] node[below, text=orange] {$p_0$} (bad2);
        \path[->, draw=orange] (s4) edge[] node[below, text=orange] {$p_0$} (bad3);
        \path[->, draw=orange] (s5) edge[] node[below, text=orange] {$p_0$} (bad4);
        \path[->, draw=orange] (s1) edge[bend left=20] node[above, text=orange] {$1-p_0$} (bad0);
        \path[->, draw=orange] (s2) edge[bend left=20] node[above, text=orange] {$1-p_0$} (s6);
        \path[->, draw=orange] (s3) edge[bend left=20] node[above, text=orange] {$1-p_0$} (s7);
        \path[->, draw=orange] (s4) edge[bend left=20] node[above, text=orange] {$1-p_0$} (s8);
        \path[->, draw=orange] (s5) edge[bend left=20] node[above, text=orange] {$1-p_0$} (s9);
        \node[state] (bad5) [right of=s6] {\(\good\)};
        \node[state] (s10) [right of=bad5] {\(\bad\)};
        \node[state] (bad6) [right of=s7] {\(\bad\)};
        \node[state] (s11) [right of=bad6] {\(s_{10}\)};
        \node[state] (bad7) [right of=s8] {\(\bad\)};
        \node[state] (s12) [right of=bad7] {\(s_{11}\)};
        \node[state] (bad8) [right of=s9] {\(\bad\)};
        \node[state] (s13) [right of=bad8] {\(s_{12}\)};
        \path[->] (s6) edge[] node[below] {$p_1$} (bad5);
        \path[->] (s6) edge[bend left=20] node[above] {$1-p_1$} (s10);
        \path[->] (s7) edge[] node[below] {$p_1$} (bad6);
        \path[->] (s7) edge[bend left=20] node[above] {$1-p_1$} (s11);
        \path[->] (s8) edge[] node[below] {$p_1$} (bad7);
        \path[->] (s8) edge[bend left=20] node[above] {$1-p_1$} (s12);
        \path[->] (s9) edge[] node[below] {$p_1$} (bad8);
        \path[->] (s9) edge[bend left=20] node[above] {$1-p_1$} (s13);
        
        \node[state] (bad10) [right of=s11] {\(\good\)};
        \node[state] (bad11) [right of=s12] {\(\bad\)};
        \node[state] (bad12) [right of=s13] {\(\bad\)};
        
        \node[state] (bad13) [right of=bad10] {\(\bad\)};
        \node (dots2) [right of=bad11] {\(\ldots\)};
        \node (dots3) [right of=bad12] {\(\ldots\)};
        
        \path[->] (s11) edge[] node[below] {$p_2$} (bad10);
        \path[->] (s12) edge[] node[below] {$p_2$} (bad11);
        \path[->] (s13) edge[] node[below] {$p_2$} (bad12);
        
        \path[->] (s11) edge[bend left=20] node[above] {$1-p_2$} (bad13);
        \path[->] (s12) edge[bend left=20] node[above] {$1-p_2$} (dots2);
        \path[->] (s13) edge[bend left=20] node[above] {$1-p_2$} (dots3);
        \path[->] (good1) edge[loop right] node[right] {$1$} (good1);
        \path[->] (bad0) edge[loop right] node[right] {$1$} (bad0);
        \path[->] (bad1) edge[loop right] node[right] {$1$} (bad1);
        \path[->] (bad2) edge[loop right] node[right] {$1$} (bad2);
        \path[->] (bad3) edge[loop right] node[right] {$1$} (bad3);
        \path[->] (bad4) edge[loop right] node[right] {$1$} (bad4);
        \path[->] (bad5) edge[loop right] node[right] {$1$} (bad5);
        \path[->] (bad6) edge[loop right] node[right] {$1$} (bad6);
        \path[->] (bad7) edge[loop right] node[right] {$1$} (bad7);
        \path[->] (bad8) edge[loop right] node[right] {$1$} (bad8);
        \path[->] (s10) edge[loop right] node[right] {$1$} (s10);
        \path[->] (bad10) edge[loop right] node[right] {$1$} (bad10);
        \path[->] (bad11) edge[loop right] node[right] {$1$} (bad11);
        \path[->] (bad12) edge[loop right] node[right] {$1$} (bad12);
        \path[->] (bad13) edge[loop right] node[right] {$1$} (bad13);
    \end{tikzpicture}
    \caption{Identify sub-pMC over \(p_0\) rooted in \(s_0\)}
    \end{subfigure}
        \begin{subfigure}{1\linewidth}%
    \centering
    \begin{tikzpicture}[node distance=1.5cm, on grid, auto, initial text =]
        \node[state, initial] (s0) {\(s_0\)};
        \node[state] (s1n) [right of=s0, yshift=1cm] {\(s_1'\)};
        \node[state] (s2n) [right of=s0, yshift=-1cm] {\(s_2'\)};
        \node[state] (good1) [right of=s1n, yshift=0.8cm]  {\(\good\)};
        \node[state] (bad1) [right of=s1n]  {\(\bad\)};
        \node[state] (s6) [right of=bad1]  {\(s_6\)};
        \node[state] (s7) [right of=bad1, yshift=-1cm]  {\(s_7\)};
        \node[state] (s8) [right of=bad1, yshift=-2cm]  {\(s_8\)};
        \node[state] (s9) [right of=bad1, yshift=-3cm]  {\(s_{9}\)};
        \path[->] (s0) edge[bend left=10] node[above] {$p_0$} (s1n);
        \path[->] (s0) edge[bend right=10] node[below] {$1-p_0$} (s2n);
        \path[->] (s1n) edge[bend left=10] node[above] {$\nicefrac{1}{5}$} (good1);
        \path[->] (s1n) edge[] node[above] {$\nicefrac{4}{5}$} (bad1);
        \path[->] (s2n) edge[] node[above] {$\nicefrac{1}{5}$} (bad1);
        \path[->] (s2n) edge[] node[above] {$\nicefrac{1}{5}$} (s6);
        \path[->] (s2n) edge[] node[above] {$\nicefrac{1}{5}$} (s7);
        \path[->] (s2n) edge[] node[above] {$\nicefrac{1}{5}$} (s8);
        \path[->] (s2n) edge[] node[above] {$\nicefrac{1}{5}$} (s9);
        \node[state] (bad5) [right of=s6] {\(\good\)};
        \node[state] (s10) [right of=bad5] {\(\bad\)};
        \node[state] (bad6) [right of=s7] {\(\bad\)};
        \node[state] (s11) [right of=bad6] {\(s_{10}\)};
        \node[state] (bad7) [right of=s8] {\(\bad\)};
        \node[state] (s12) [right of=bad7] {\(s_{11}\)};
        \node[state] (bad8) [right of=s9] {\(\bad\)};
        \node[state] (s13) [right of=bad8] {\(s_{12}\)};
        \path[->] (s6) edge[] node[below] {$p_1$} (bad5);
        \path[->] (s6) edge[bend left=20] node[above] {$1-p_1$} (s10);
        \path[->] (s7) edge[] node[below] {$p_1$} (bad6);
        \path[->] (s7) edge[bend left=20] node[above] {$1-p_1$} (s11);
        \path[->] (s8) edge[] node[below] {$p_1$} (bad7);
        \path[->] (s8) edge[bend left=20] node[above] {$1-p_1$} (s12);
        \path[->] (s9) edge[] node[below] {$p_1$} (bad8);
        \path[->] (s9) edge[bend left=20] node[above] {$1-p_1$} (s13);
        
        \node[state] (bad10) [right of=s11] {\(\good\)};
        \node[state] (bad11) [right of=s12] {\(\bad\)};
        \node[state] (bad12) [right of=s13] {\(\bad\)};
        
        \node[state] (bad13) [right of=bad10] {\(\bad\)};
        \node (dots2) [right of=bad11] {\(\ldots\)};
        \node (dots3) [right of=bad12] {\(\ldots\)};
        
        \path[->] (s11) edge[] node[below] {$p_2$} (bad10);
        \path[->] (s12) edge[] node[below] {$p_2$} (bad11);
        \path[->] (s13) edge[] node[below] {$p_2$} (bad12);
        
        \path[->] (s11) edge[bend left=20] node[above] {$1-p_2$} (bad13);
        \path[->] (s12) edge[bend left=20] node[above] {$1-p_2$} (dots2);
        \path[->] (s13) edge[bend left=20] node[above] {$1-p_2$} (dots3);
        \path[->] (good1) edge[loop right] node[right] {$1$} (good1);
        \path[->] (bad1) edge[loop right] node[right] {$1$} (bad1);
        \path[->] (bad5) edge[loop right] node[right] {$1$} (bad5);
        \path[->] (bad6) edge[loop right] node[right] {$1$} (bad6);
        \path[->] (bad7) edge[loop right] node[right] {$1$} (bad7);
        \path[->] (bad8) edge[loop right] node[right] {$1$} (bad8);
        \path[->] (s10) edge[loop right] node[right] {$1$} (s10);
        \path[->] (bad10) edge[loop right] node[right] {$1$} (bad10);
        \path[->] (bad11) edge[loop right] node[right] {$1$} (bad11);
        \path[->] (bad12) edge[loop right] node[right] {$1$} (bad12);
        \path[->] (bad13) edge[loop right] node[right] {$1$} (bad13);
    \end{tikzpicture}
    \caption{Big-step and group (\(\bad\) states in sub-pMC are merged now for clarity, in this example the state \(s_1'\) would have four transitions with \(\nicefrac{1}{5}\) each going to a separate \(\bad\) state). Note that we merged all occurences of \(p_0\) into a single occurence. Continue with the sub-pMC rooted in $s_2'$ over $p_1$}
    \end{subfigure}
    \caption{Illustration of the big-step transformation on NRP}
    \label{fig:bigstepnrp}
\end{figure}

\section{Substituting Regions Into pMCs with Large Polynomials}\label{app:largepolys}

\begin{figure}[t]
    \begin{center}
    \begin{adjustbox}{max width=0.8\linewidth}
        \input{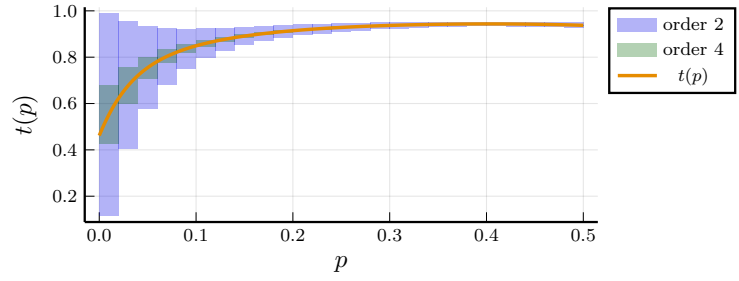}
    \end{adjustbox}
    \end{center}
    \caption{Intervals computed using the interval Newton method from a
    polynomial with 397 terms that appears in \texttt{nand}, with intervals of
    width 0.02.}
    \label{fig:midpoint}
\end{figure}

Having completed the transformation and given a region \(R\), we compute the
substituted intervals. For polynomials and smaller monomials, we ask an SMT
solver for the roots of the derivatives. For bigger monomials, as
they are produced in the transformation algorithm, we instead use the
\emph{interval Newton method} \cite[p.\ 105]{DBLP:books/daglib/0022249} combined
with interval arithmetic to arrive at an overapproximation of the desired
interval.

We now have a pMC with transitions of the form \cref{eq:transition} where we
know the factorizations to \(f_j(p)\) but not the factorization of \(t(p)\). To
be able to lift the pMC to an iMC given a region, we need to efficiently compute
a reasonably tight interval around \(t(p)\).

The core of our method is an evaluation of the polynomial through interval arithmetic.
We improve bounds by using the interval Newton method
\cite[p.\ 105]{DBLP:books/daglib/0022249}.  Suppose we have a bound \([\underline{t'},
\overline{t'}]\) on the derivative of \(t\) in the interval \(i_p = [\underline{p}, \overline{p}]\)
and \(t = \max(|\underline{t'}|, |\overline{t'}|)\).
Let \(w(i_p) = \overline{p} - \underline{p}\) be the width of said interval and \(m(i_p) =
\nicefrac{(\underline{p} + \overline{p})}2\) its midpoint. Then, by the midpoint method, we can bound
\(t\) by \[
    t(m(i_p)) - \frac{w(i_p)}{2 \cdot t} \leq t(p) \leq t(m(i_p)) + \frac{w(i_p)}{2 \cdot t}
    \text{ for } p \in [\underline{p}, \overline{p}].
\]
With our representation of \(t(p)\) as a sum of products, we can efficiently
compute the \(k\)th derivative of \(t(p)\) if there are few \(f_j\).  We bound
the \(k\)th derivative directly by interval arithmetic. Then we go up through
the derivatives and compute the bounds through the midpoint method. In
\cref{fig:midpoint}, we have computed bounds on a large function through the
interval Newton method, going down two and four orders. The interval Newton
method converges quadratically
\cite[p.\ 107]{DBLP:books/daglib/0022249}.

\newpage
\section{Experiments: Big-Step Transformation}
\label{app:exbigstep}

We provide a table of all results.
Note that some pMCs have different numbers of parameters for graph-preserving
and non-graph-preserving regions because bisimulation can be more aggressive if
it does not need to preserve graph-preserving instantiations.

\begin{scriptsize}
\subsubsection{Table for \(\varepsilon=10^{-5}\)}
\begin{longtable}{llrrllrrrr}

        \toprule
        Model & Const & $|S|$ & $|V|$ & $\delta$ & Prop & \multicolumn{2}{c}{Time (s)} & \multicolumn{2}{c}{Regions} \\
        \cmidrule(ll){7-8} \cmidrule(ll){9-10}
        & & & & & & nobig & big & nobig & big \\
        \midrule
        
 4x4grid       &          &     	47 &   3 & 0     & R$\geq$5.04  & 0.37     & 0.374    & 4281    & 3585   \\
 4x4grid       &          &     	47 &   3 & 0,0.8 & R$\geq$6.69  & 0.396    & 0.406    & 4553    & 3841   \\
 4x4grid       &          &     	47 &   3 & 0.1   & R$\geq$6.42  & 0.28     & 0.294    & 4433    & 3753   \\
 4x4grid       &          &     	47 &   3 & 0.2,1 & R$\geq$6.69  & 0.388    & 0.405    & 4433    & 3841   \\
 4x4grid       &          &     	49 &   3 & 1e-06 & R$\geq$5.04  & 0.28     & 0.294    & 4281    & 3561   \\
 4x4grid-avoid &          &     	45 &   3 & 0     & P$\leq$0.93  & 4.216    & 0.08     & 929     & 137    \\
 4x4grid-avoid &          &     	47 &   3 & 0,0.8 & P$\leq$0.86  & 37.546   & 25.709   & 6721    & 3289   \\
 4x4grid-avoid &          &     	47 &   3 & 0.1   & P$\leq$0.85  & 1.23     & 1.02     & 19297   & 13329  \\
 4x4grid-avoid &          &     	45 &   3 & 0.2,1 & P$\leq$0.93  & 0.822    & 0.045    & 449     & 65     \\
 4x4grid-avoid &          &     	47 &   3 & 1e-06 & P$\leq$0.93  & 38.022   & 5.049    & 49569   & 1497   \\
 alarm         &          &     	15 &   2 & 0     & P$\geq$0.04  & 0.066    & 0.062    & 69      & 5      \\
 alarm         &          &     	15 &   2 & 0,0.8 & P$\geq$0.04  & 0.066    & 0.062    & 69      & 1      \\
 alarm         &          &     	31 &   2 & 0.1   & P$\geq$0.24  & 0.065    & 0.067    & 89      & 85     \\
 alarm         &          &     	60 &   2 & 0.2,1 & P$\geq$0.12  & 0.065    & 0.068    & 69      & 69     \\
 alarm         &          &     	31 &   2 & 1e-06 & P$\geq$0.04  & 0.064    & 0.062    & 69      & 5      \\
 brp           & (16, 2)  &    	183 &   4 & 0     & P$\leq$0.02  & MO       & MO       & MO      & MO     \\
 brp           & (16, 2)  &    	330 &   4 & 0,0.8 & P$\leq$0.02  & MO       & MO       & MO      & MO     \\
 brp           & (16, 2)  &    	324 &   4 & 0.1   & P$\leq$0.02  & MO       & MO       & MO      & MO     \\
 brp           & (16, 2)  &    	183 &   4 & 0.2,1 & P$\leq$0.02  & MO       & MO       & MO      & MO     \\
 brp           & (16, 2)  &    	177 &   4 & 1e-06 & P$\leq$0.02  & MO       & MO       & MO      & MO     \\
 brp           & (32, 4)  &    	551 &   4 & 0     & P$\leq$0.01  & TO       & TO       & TO      & TO     \\
 brp           & (32, 4)  &    	551 &   4 & 0,0.8 & P$\leq$0.01  & TO       & TO       & TO      & TO     \\
 brp           & (32, 4)  &    	545 &   4 & 0.1   & P$\leq$0.01  & TO       & TO       & TO      & TO     \\
 brp           & (32, 4)  &   	1100 &   4 & 0.2,1 & P$\leq$0.01  & TO       & TO       & TO      & TO     \\
 brp           & (32, 4)  &    	545 &   4 & 1e-06 & P$\leq$0.01  & TO       & TO       & TO      & TO     \\
 brp           & (64, 8)  &   	3984 &   4 & 0     & P$\leq$0.01  & TO       & TO       & TO      & TO     \\
 brp           & (64, 8)  &   	3984 &   4 & 0,0.8 & P$\leq$0.01  & TO       & TO       & TO      & TO     \\
 brp           & (64, 8)  &   	1857 &   4 & 0.1   & P$\leq$0.01  & TO       & TO       & TO      & TO     \\
 brp           & (64, 8)  &   	3984 &   4 & 0.2,1 & P$\leq$0.01  & TO       & TO       & TO      & TO     \\
 brp           & (64, 8)  &   	1857 &   4 & 1e-06 & P$\leq$0.01  & TO       & TO       & TO      & TO     \\
 crowds        & (10, 2)  &     	32 &   2 & 0     & P$\leq$0.02  & MO       & MO       & MO      & MO     \\
 crowds        & (10, 2)  &     	32 &   2 & 0,0.8 & P$\leq$0.0   & 0.609    & 0.396    & 10429   & 6241   \\
 crowds        & (10, 2)  &     	11 &   2 & 0.1   & P$\leq$0.01  & 0.588    & 0.338    & 12557   & 6909   \\
 crowds        & (10, 2)  &     	32 &   2 & 0.2,1 & P$\leq$0.01  & 0.058    & 0.049    & 241     & 157    \\
 crowds        & (10, 2)  &     	20 &   2 & 1e-06 & P$\leq$0.02  & 10.376   & 5.689    & 221205  & 114265 \\
 crowds        & (10, 4)  &     	52 &   2 & 0,0.8 & P$\leq$0.02  & 1.334    & 1.984    & 10345   & 8697   \\
 crowds        & (10, 4)  &     	84 &   2 & 0.1   & P$\leq$0.03  & 1.062    & 1.765    & 11745   & 9761   \\
 crowds        & (10, 4)  &     	96 &   2 & 0.2,1 & P$\leq$0.07  & 0.227    & 0.244    & 233     & 185    \\
 crowds        & (10, 4)  &     	42 &   2 & 1e-06 & P$\leq$0.1   & 14.706   & 26.914   & 187169  & 144597 \\
 herman        & 1        &  	40690 &   1 & 0     & R$\geq$12.1  & 974.535  & 655.365  & 19239   & 7965   \\
 herman        & 1        &  	40690 &   1 & 0,0.8 & R$\geq$12.1  & 1011.977 & 644.335  & 20233   & 7985   \\
 herman        & 1        &  	40690 &   1 & 0.1   & R$\geq$12.1  & 697.86   & 311.2    & 20245   & 8041   \\
 herman        & 1        &  	18872 &   1 & 0.2,1 & R$\geq$12.1  & 1012.064 & 662.52   & 20201   & 7965   \\
 herman        & 1        &  	18872 &   1 & 1e-06 & R$\geq$12.1  & 674.756  & 320.971  & 19225   & 7983   \\
 herman        & 5        &    	196 &   1 & 0     & R$\geq$1.93  & 0.818    & 0.237    & 3781    & 631    \\
 herman        & 5        &    	196 &   1 & 0,0.8 & R$\geq$1.93  & 0.819    & 0.232    & 3827    & 605    \\
 herman        & 5        &     	89 &   1 & 0.1   & R$\geq$1.93  & 0.545    & 0.236    & 3993    & 823    \\
 herman        & 5        &    	196 &   1 & 0.2,1 & R$\geq$1.93  & 0.829    & 0.231    & 3881    & 611    \\
 herman        & 5        &     	89 &   1 & 1e-06 & R$\geq$1.93  & 0.54     & 0.27     & 3795    & 823    \\
 herman        & 7        &    	680 &   1 & 0     & R$\geq$4.49  & 10.941   & 3.84     & 7719    & 2083   \\
 herman        & 7        &    	680 &   1 & 0,0.8 & R$\geq$4.49  & 10.88    & 3.676    & 7767    & 2009   \\
 herman        & 7        &   	1422 &   1 & 0.1   & R$\geq$4.49  & 6.702    & 2.487    & 7767    & 2055   \\
 herman        & 7        &   	1422 &   1 & 0.2,1 & R$\geq$4.49  & 10.921   & 3.676    & 7783    & 2009   \\
 herman        & 7        &    	680 &   1 & 1e-06 & R$\geq$4.49  & 6.815    & 2.964    & 7709    & 2167   \\
 herman        & 9        &   	8008 &   1 & 0     & R$\geq$7.92  & 674.245  & 361.1    & 80383   & 32725  \\
 herman        & 9        &   	3707 &   1 & 0,0.8 & R$\geq$7.92  & 471.581  & 207.915  & 55877   & 19175  \\
 herman        & 9        &   	8008 &   1 & 0.1   & R$\geq$7.92  & 302.109  & 116.856  & 54551   & 18971  \\
 herman        & 9        &   	8008 &   1 & 0.2,1 & R$\geq$7.92  & 471.028  & 206.876  & 54849   & 19025  \\
 herman        & 9        &   	8008 &   1 & 1e-06 & R$\geq$7.92  & 294.046  & 131.597  & 52893   & 20937  \\
 hermanspeed   & 3        &     	22 &   3 & 0     & R$\geq$0.58  & MO       & MO       & MO      & MO     \\
 hermanspeed   & 3        &     	22 &   3 & 0,0.8 & R$\geq$0.58  & MO       & MO       & MO      & MO     \\
 hermanspeed   & 3        &     	22 &   3 & 0.1   & R$\geq$0.58  & MO       & MO       & MO      & MO     \\
 hermanspeed   & 3        &     	22 &   3 & 0.2,1 & R$\geq$0.58  & MO       & MO       & MO      & MO     \\
 hermanspeed   & 3        &     	22 &   3 & 1e-06 & R$\geq$0.58  & MO       & MO       & MO      & MO     \\
 hermanspeed   & 5        &    	673 &   1 & 0     & R$\geq$2.5   & TO       & TO       & TO      & TO     \\
 hermanspeed   & 5        &    	673 &   1 & 0,0.8 & R$\geq$2.5   & TO       & TO       & TO      & TO     \\
 hermanspeed   & 5        &    	673 &   1 & 0.1   & R$\geq$2.52  & TO       & TO       & TO      & TO     \\
 hermanspeed   & 5        &    	673 &   1 & 0.2,1 & R$\geq$2.55  & TO       & TO       & TO      & TO     \\
 hermanspeed   & 5        &    	673 &   1 & 1e-06 & R$\geq$2.5   & TO       & TO       & TO      & TO     \\
 maze2         &          &     	41 &  15 & 0     & R$\geq$14.32 & MO       & MO       & MO      & MO     \\
 maze2         &          &     	50 &  15 & 0,0.8 & R$\geq$16.14 & MO       & MO       & MO      & MO     \\
 maze2         &          &     	50 &  15 & 0.1   & R$\geq$17.52 & MO       & MO       & MO      & MO     \\
 maze2         &          &     	50 &  15 & 0.2,1 & R$\geq$19.64 & MO       & MO       & MO      & MO     \\
 maze2         &          &     	41 &  15 & 1e-06 & R$\geq$14.32 & MO       & MO       & MO      & MO     \\
 nand          & (5, 5)   &    	112 &   2 & 0,0.8 & P$\leq$0.76  & 0.224    & 15.713   & 133     & 65     \\
 nand          & (5, 5)   &   	2687 &   2 & 0.1   & P$\leq$0.5   & 8.886    & 4.55     & 20501   & 33     \\
 nand          & (5, 10)  &   	5448 &   2 & 0,0.8 & P$\leq$0.81  & 0.529    & 69.137   & 173     & 65     \\
 nand          & (5, 10)  &   	5447 &   2 & 0.1   & P$\leq$0.5   & 100.543  & 20.228   & 97753   & 57     \\
 nand          & (5, 25)  &  	13728 &   2 & 0,0.8 & P$\leq$0.82  & 2.167    & 519.333  & 289     & 69     \\
 nand          & (5, 25)  &  	13727 &   2 & 0.1   & P$\leq$0.5   & 366.85   & 172.618  & 131069  & 93     \\
 nand          & (5, 50)  &    	562 &   2 & 0,0.8 & P$\leq$0.82  & 6.818    & 2716.399 & 425     & 73     \\
 nand          & (5, 50)  &    	562 &   2 & 0.1   & P$\leq$0.5   & 864.776  & 981.843  & 131069  & 117    \\
 nand          & (10, 50) & 	250403 &   2 & 0,0.8 & P$\leq$0.91  & 66.399   & TO       & 229     & TO     \\
 nand          & (10, 50) & 	346962 &   2 & 0.1   & P$\leq$0.28  & TO       & TO       & TO      & TO     \\
 newgrid       & 2        &     	47 &   3 & 0,0.8 & P$\leq$0.79  & 2.311    & 3.747    & 25689   & 23025  \\
 newgrid       & 2        &     	32 &   3 & 0.1   & P$\leq$0.89  & 3.071    & 3.886    & 56553   & 55041  \\
 newgrid       & 2        &     	37 &   3 & 0.2,1 & P$\leq$0.99  & 0.018    & 0.02     & 1       & 1      \\
 newgrid       & 4        &     	76 &   3 & 0,0.8 & P$\leq$0.92  & 7.252    & 10.983   & 15697   & 15817  \\
 newgrid       & 4        &     	72 &   3 & 0.1   & P$\leq$0.97  & 0.839    & 1.077    & 10713   & 10665  \\
 newgrid       & 4        &    	110 &   3 & 0.2,1 & P$\leq$0.99  & MO       & MO       & MO      & MO     \\
 nrp           & 5        &     	34 &   5 & 0     & P$\leq$0.2   & MO       & 0.016    & MO      & 1      \\
 nrp           & 5        &     	12 &   5 & 0,0.8 & P$\leq$0.2   & MO       & MO       & MO      & MO     \\
 nrp           & 5        &     	33 &   5 & 0.1   & P$\leq$0.2   & MO       & MO       & MO      & MO     \\
 nrp           & 5        &     	34 &   5 & 0.2,1 & P$\leq$0.2   & MO       & 0.017    & MO      & 1      \\
 nrp           & 5        &     	33 &   5 & 1e-06 & P$\leq$0.2   & MO       & 0.018    & MO      & 1      \\
 nrp           & 10       &     	22 &  10 & 0     & P$\leq$0.1   & MO       & 0.018    & MO      & 1      \\
 nrp           & 10       &    	114 &  10 & 0,0.8 & P$\leq$0.1   & MO       & 0.018    & MO      & 1      \\
 nrp           & 10       &    	113 &  10 & 0.1   & P$\leq$0.1   & MO       & 0.018    & MO      & 1      \\
 nrp           & 10       &     	22 &  10 & 0.2,1 & P$\leq$0.1   & MO       & 0.018    & MO      & 1      \\
 nrp           & 10       &     	22 &  10 & 1e-06 & P$\leq$0.1   & MO       & 0.018    & MO      & 1      \\
 nrp           & 100      &  	10104 & 100 & 0     & P$\leq$0.01  & MO       & 1.003    & MO      & 1      \\
 nrp           & 100      &    	202 & 100 & 0,0.8 & P$\leq$0.01  & MO       & 0.997    & MO      & 1      \\
 nrp           & 100      &    	202 & 100 & 0.1   & P$\leq$0.01  & MO       & 1.036    & MO      & 1      \\
 nrp           & 100      &  	10104 & 100 & 0.2,1 & P$\leq$0.01  & MO       & 1.0      & MO      & 1      \\
 nrp           & 100      &  	10103 & 100 & 1e-06 & P$\leq$0.01  & MO       & 0.949    & MO      & 1      \\
 refuel        & 3        &     	51 &  18 & 0     & P$\leq$0.09  & MO       & MO       & MO      & MO     \\
 refuel        & 3        &     	47 &  18 & 0,0.8 & P$\leq$0.06  & 0.206    & 0.106    & 1601    & 673    \\
 refuel        & 3        &     	34 &  18 & 0.1   & P$\leq$0.06  & 0.278    & 0.06     & 2641    & 401    \\
 refuel        & 3        &     	51 &  18 & 0.2,1 & P$\leq$0.07  & 12.693   & 4.338    & 99553   & 32753  \\
 refuel        & 3        &     	32 &  18 & 1e-06 & P$\leq$0.09  & 255.516  & 0.022    & 2288977 & 1      \\
\bottomrule
\end{longtable}
\subsubsection{Table for \(\varepsilon=10{-4}\)}
\begin{longtable}{llrrllrrrr}

        \toprule
        Model & Const & $|S|$ & $|V|$ & $\delta$ & Prop & \multicolumn{2}{c}{Time (s)} & \multicolumn{2}{c}{Regions} \\
        \cmidrule(ll){7-8} \cmidrule(ll){9-10}
        & & & & & & nobig & big & nobig & big \\
        \midrule
        
 4x4grid       &          &     	49 &   3 & 0     & R$\geq$5.04  & 0.121    & 0.127    & 1257    & 1097    \\
 4x4grid       &          &     	49 &   3 & 0,0.8 & R$\geq$6.69  & 0.134    & 0.149    & 1385    & 1297    \\
 4x4grid       &          &     	47 &   3 & 0.1   & R$\geq$6.42  & 0.099    & 0.179    & 1369    & 1273    \\
 4x4grid       &          &     	49 &   3 & 0.2,1 & R$\geq$6.69  & 0.133    & 0.149    & 1385    & 1297    \\
 4x4grid       &          &     	47 &   3 & 1e-06 & R$\geq$5.04  & 0.093    & 0.1      & 1257    & 1097    \\
 4x4grid-avoid &          &     	47 &   3 & 0     & P$\leq$0.93  & 4.223    & 0.08     & 929     & 137     \\
 4x4grid-avoid &          &     	47 &   3 & 0,0.8 & P$\leq$0.86  & 16.661   & 13.111   & 4377    & 2617    \\
 4x4grid-avoid &          &     	45 &   3 & 0.1   & P$\leq$0.85  & 0.409    & 0.363    & 6305    & 4489    \\
 4x4grid-avoid &          &     	47 &   3 & 0.2,1 & P$\leq$0.93  & 0.823    & 0.045    & 449     & 65      \\
 4x4grid-avoid &          &     	47 &   3 & 1e-06 & P$\leq$0.93  & 26.304   & 0.641    & 9833    & 481     \\
 alarm         &          &     	15 &   2 & 0     & P$\geq$0.04  & 0.065    & 0.062    & 57      & 5       \\
 alarm         &          &     	15 &   2 & 0,0.8 & P$\geq$0.04  & 0.065    & 0.062    & 57      & 1       \\
 alarm         &          &     	15 &   2 & 0.1   & P$\geq$0.24  & 0.064    & 0.067    & 73      & 73      \\
 alarm         &          &     	15 &   2 & 0.2,1 & P$\geq$0.12  & 0.064    & 0.067    & 53      & 53      \\
 alarm         &          &     	31 &   2 & 1e-06 & P$\geq$0.04  & 0.064    & 0.062    & 57      & 5       \\
 brp           & (16, 2)  &    	330 &   4 & 0     & P$\leq$0.02  & MO       & MO       & MO      & MO      \\
 brp           & (16, 2)  &    	330 &   4 & 0,0.8 & P$\leq$0.02  & 1831.332 & 1774.264 & 7034485 & 7034485 \\
 brp           & (16, 2)  &    	177 &   4 & 0.1   & P$\leq$0.02  & MO       & MO       & MO      & MO      \\
 brp           & (16, 2)  &    	330 &   4 & 0.2,1 & P$\leq$0.02  & MO       & MO       & MO      & MO      \\
 brp           & (16, 2)  &    	177 &   4 & 1e-06 & P$\leq$0.02  & MO       & MO       & MO      & MO      \\
 brp           & (32, 4)  &    	551 &   4 & 0     & P$\leq$0.01  & TO       & TO       & TO      & TO      \\
 brp           & (32, 4)  &   	1100 &   4 & 0,0.8 & P$\leq$0.01  & TO       & TO       & TO      & TO      \\
 brp           & (32, 4)  &   	1094 &   4 & 0.1   & P$\leq$0.01  & TO       & TO       & TO      & TO      \\
 brp           & (32, 4)  &   	1100 &   4 & 0.2,1 & P$\leq$0.01  & TO       & TO       & TO      & TO      \\
 brp           & (32, 4)  &    	545 &   4 & 1e-06 & P$\leq$0.01  & TO       & TO       & TO      & TO      \\
 brp           & (64, 8)  &   	1863 &   4 & 0     & P$\leq$0.01  & TO       & TO       & TO      & TO      \\
 brp           & (64, 8)  &   	3984 &   4 & 0,0.8 & P$\leq$0.01  & TO       & TO       & TO      & TO      \\
 brp           & (64, 8)  &   	3978 &   4 & 0.1   & P$\leq$0.01  & TO       & TO       & TO      & TO      \\
 brp           & (64, 8)  &   	3984 &   4 & 0.2,1 & P$\leq$0.01  & TO       & TO       & TO      & TO      \\
 brp           & (64, 8)  &   	3978 &   4 & 1e-06 & P$\leq$0.01  & TO       & TO       & TO      & TO      \\
 crowds        & (10, 2)  &     	32 &   2 & 0     & P$\leq$0.02  & MO       & MO       & MO      & MO      \\
 crowds        & (10, 2)  &     	19 &   2 & 0,0.8 & P$\leq$0.0   & 0.238    & 0.164    & 3697    & 2173    \\
 crowds        & (10, 2)  &     	20 &   2 & 0.1   & P$\leq$0.01  & 0.208    & 0.123    & 4109    & 1969    \\
 crowds        & (10, 2)  &     	32 &   2 & 0.2,1 & P$\leq$0.01  & 0.05     & 0.048    & 189     & 113     \\
 crowds        & (10, 2)  &     	20 &   2 & 1e-06 & P$\leq$0.02  & 3.048    & 1.416    & 65673   & 28449   \\
 crowds        & (10, 4)  &     	96 &   2 & 0,0.8 & P$\leq$0.02  & 0.585    & 0.762    & 3597    & 2745    \\
 crowds        & (10, 4)  &     	84 &   2 & 0.1   & P$\leq$0.03  & 0.469    & 0.718    & 3953    & 3249    \\
 crowds        & (10, 4)  &     	52 &   2 & 0.2,1 & P$\leq$0.07  & 0.218    & 0.234    & 181     & 145     \\
 crowds        & (10, 4)  &     	42 &   2 & 1e-06 & P$\leq$0.1   & 4.495    & 8.524    & 56641   & 45745   \\
 herman        & 1        &  	40690 &   1 & 0     & R$\geq$12.1  & 367.63   & 239.613  & 6457    & 2559    \\
 herman        & 1        &  	18872 &   1 & 0,0.8 & R$\geq$12.1  & 372.977  & 222.642  & 6529    & 2383    \\
 herman        & 1        &  	18872 &   1 & 0.1   & R$\geq$12.1  & 272.214  & 112.287  & 6531    & 2387    \\
 herman        & 1        &  	40690 &   1 & 0.2,1 & R$\geq$12.1  & 378.213  & 224.697  & 6525    & 2377    \\
 herman        & 1        &  	18872 &   1 & 1e-06 & R$\geq$12.1  & 269.012  & 122.283  & 6455    & 2583    \\
 herman        & 5        &    	196 &   1 & 0     & R$\geq$1.93  & 0.288    & 0.119    & 1165    & 195     \\
 herman        & 5        &    	196 &   1 & 0,0.8 & R$\geq$1.93  & 0.27     & 0.117    & 1071    & 195     \\
 herman        & 5        &     	89 &   1 & 0.1   & R$\geq$1.93  & 0.185    & 0.117    & 1075    & 237     \\
 herman        & 5        &    	196 &   1 & 0.2,1 & R$\geq$1.93  & 0.269    & 0.121    & 1071    & 195     \\
 herman        & 5        &    	196 &   1 & 1e-06 & R$\geq$1.93  & 0.204    & 0.134    & 1177    & 235     \\
 herman        & 7        &    	680 &   1 & 0     & R$\geq$4.49  & 3.777    & 1.327    & 2423    & 639     \\
 herman        & 7        &   	1422 &   1 & 0,0.8 & R$\geq$4.49  & 3.898    & 1.38     & 2555    & 673     \\
 herman        & 7        &    	680 &   1 & 0.1   & R$\geq$4.49  & 2.552    & 0.979    & 2551    & 691     \\
 herman        & 7        &   	1422 &   1 & 0.2,1 & R$\geq$4.49  & 3.9      & 1.383    & 2555    & 673     \\
 herman        & 7        &    	680 &   1 & 1e-06 & R$\geq$4.49  & 2.455    & 1.015    & 2423    & 655     \\
 herman        & 9        &   	8008 &   1 & 0     & R$\geq$7.92  & 121.885  & 54.8     & 13589   & 4841    \\
 herman        & 9        &   	3707 &   1 & 0,0.8 & R$\geq$7.92  & 109.567  & 54.128   & 12173   & 4829    \\
 herman        & 9        &   	8008 &   1 & 0.1   & R$\geq$7.92  & 74.651   & 30.371   & 12175   & 4827    \\
 herman        & 9        &   	8008 &   1 & 0.2,1 & R$\geq$7.92  & 109.678  & 57.134   & 12175   & 4827    \\
 herman        & 9        &   	8008 &   1 & 1e-06 & R$\geq$7.92  & 81.727   & 32.532   & 13101   & 4819    \\
 hermanspeed   & 3        &     	22 &   3 & 0     & R$\geq$0.58  & MO       & MO       & MO      & MO      \\
 hermanspeed   & 3        &     	22 &   3 & 0,0.8 & R$\geq$0.58  & MO       & MO       & MO      & MO      \\
 hermanspeed   & 3        &     	22 &   3 & 0.1   & R$\geq$0.58  & MO       & MO       & MO      & MO      \\
 hermanspeed   & 3        &     	22 &   3 & 0.2,1 & R$\geq$0.58  & MO       & MO       & MO      & MO      \\
 hermanspeed   & 3        &     	22 &   3 & 1e-06 & R$\geq$0.58  & MO       & MO       & MO      & MO      \\
 hermanspeed   & 5        &    	673 &   1 & 0     & R$\geq$2.5   & TO       & TO       & TO      & TO      \\
 hermanspeed   & 5        &    	673 &   1 & 0,0.8 & R$\geq$2.5   & TO       & TO       & TO      & TO      \\
 hermanspeed   & 5        &    	673 &   1 & 0.1   & R$\geq$2.52  & TO       & TO       & TO      & TO      \\
 hermanspeed   & 5        &    	673 &   1 & 0.2,1 & R$\geq$2.55  & TO       & TO       & TO      & TO      \\
 hermanspeed   & 5        &    	673 &   1 & 1e-06 & R$\geq$2.5   & TO       & TO       & TO      & TO      \\
 maze2         &          &     	50 &  15 & 0     & R$\geq$14.31 & MO       & MO       & MO      & MO      \\
 maze2         &          &     	41 &  15 & 0,0.8 & R$\geq$16.14 & 547.256  & 629.841  & 4319969 & 4319889 \\
 maze2         &          &     	50 &  15 & 0.1   & R$\geq$17.52 & MO       & MO       & MO      & MO      \\
 maze2         &          &     	50 &  15 & 0.2,1 & R$\geq$19.64 & MO       & MO       & MO      & MO      \\
 maze2         &          &     	50 &  15 & 1e-06 & R$\geq$14.31 & 718.85   & 815.799  & 5919249 & 5919521 \\
 nand          & (5, 5)   &    	112 &   2 & 0,0.8 & P$\leq$0.76  & 0.194    & 12.876   & 109     & 53      \\
 nand          & (5, 5)   &   	2687 &   2 & 0.1   & P$\leq$0.5   & 2.364    & 3.604    & 5609    & 25      \\
 nand          & (5, 10)  &   	5448 &   2 & 0,0.8 & P$\leq$0.81  & 0.46     & 56.707   & 145     & 53      \\
 nand          & (5, 10)  &    	162 &   2 & 0.1   & P$\leq$0.5   & 12.811   & 16.078   & 14929   & 29      \\
 nand          & (5, 25)  &    	312 &   2 & 0,0.8 & P$\leq$0.82  & 1.953    & 485.882  & 253     & 57      \\
 nand          & (5, 25)  &    	312 &   2 & 0.1   & P$\leq$0.5   & 34.232   & 143.464  & 16381   & 33      \\
 nand          & (5, 50)  &    	562 &   2 & 0,0.8 & P$\leq$0.82  & 6.02     & 2150.894 & 361     & 57      \\
 nand          & (5, 50)  &    	562 &   2 & 0.1   & P$\leq$0.5   & 66.014   & 768.139  & 16381   & 33      \\
 nand          & (10, 50) & 	250403 &   2 & 0,0.8 & P$\leq$0.91  & 58.666   & TO       & 205     & TO      \\
 nand          & (10, 50) & 	250402 &   2 & 0.1   & P$\leq$0.28  & 1079.708 & TO       & 16381   & TO      \\
 newgrid       & 2        &     	37 &   3 & 0,0.8 & P$\leq$0.79  & 1.065    & 1.76     & 8193    & 7385    \\
 newgrid       & 2        &     	32 &   3 & 0.1   & P$\leq$0.89  & 0.886    & 1.115    & 16393   & 15937   \\
 newgrid       & 2        &     	47 &   3 & 0.2,1 & P$\leq$0.99  & 0.018    & 0.019    & 1       & 1       \\
 newgrid       & 4        &     	76 &   3 & 0,0.8 & P$\leq$0.92  & 6.633    & 9.228    & 9105    & 9369    \\
 newgrid       & 4        &     	74 &   3 & 0.1   & P$\leq$0.97  & 0.427    & 0.542    & 5209    & 5217    \\
 newgrid       & 4        &     	76 &   3 & 0.2,1 & P$\leq$0.99  & MO       & MO       & MO      & MO      \\
 nrp           & 5        &     	12 &   5 & 0     & P$\leq$0.2   & MO       & 0.017    & MO      & 1       \\
 nrp           & 5        &     	34 &   5 & 0,0.8 & P$\leq$0.2   & MO       & MO       & MO      & MO      \\
 nrp           & 5        &     	12 &   5 & 0.1   & P$\leq$0.2   & MO       & 0.015    & MO      & 1       \\
 nrp           & 5        &     	34 &   5 & 0.2,1 & P$\leq$0.2   & MO       & 0.016    & MO      & 1       \\
 nrp           & 5        &     	33 &   5 & 1e-06 & P$\leq$0.2   & MO       & 0.017    & MO      & 1       \\
 nrp           & 10       &     	22 &  10 & 0     & P$\leq$0.1   & MO       & 0.018    & MO      & 1       \\
 nrp           & 10       &    	114 &  10 & 0,0.8 & P$\leq$0.1   & MO       & 0.019    & MO      & 1       \\
 nrp           & 10       &    	113 &  10 & 0.1   & P$\leq$0.1   & MO       & 0.019    & MO      & 1       \\
 nrp           & 10       &    	114 &  10 & 0.2,1 & P$\leq$0.1   & MO       & 0.019    & MO      & 1       \\
 nrp           & 10       &     	22 &  10 & 1e-06 & P$\leq$0.1   & MO       & 0.018    & MO      & 1       \\
 nrp           & 100      &    	202 & 100 & 0     & P$\leq$0.01  & MO       & 0.99     & MO      & 1       \\
 nrp           & 100      &  	10104 & 100 & 0,0.8 & P$\leq$0.01  & MO       & 0.999    & MO      & 1       \\
 nrp           & 100      &  	10103 & 100 & 0.1   & P$\leq$0.01  & MO       & 1.049    & MO      & 1       \\
 nrp           & 100      &  	10104 & 100 & 0.2,1 & P$\leq$0.01  & MO       & 0.994    & MO      & 1       \\
 nrp           & 100      &  	10103 & 100 & 1e-06 & P$\leq$0.01  & MO       & 0.956    & MO      & 1       \\
 refuel        & 3        &     	51 &  18 & 0     & P$\leq$0.09  & MO       & MO       & MO      & MO      \\
 refuel        & 3        &     	47 &  18 & 0,0.8 & P$\leq$0.06  & 0.146    & 0.082    & 1073    & 481     \\
 refuel        & 3        &     	32 &  18 & 0.1   & P$\leq$0.06  & 0.2      & 0.051    & 1841    & 305     \\
 refuel        & 3        &     	51 &  18 & 0.2,1 & P$\leq$0.07  & 1.188    & 0.272    & 9601    & 2033    \\
 refuel        & 3        &     	34 &  18 & 1e-06 & P$\leq$0.09  & 18.138   & 0.022    & 169761  & 1       \\
\bottomrule
\end{longtable}
\subsubsection{Table for \(\varepsilon=0.01\)}
\begin{longtable}{llrrllrrrr}

        \toprule
        Model & Const & $|S|$ & $|V|$ & $\delta$ & Prop & \multicolumn{2}{c}{Time (s)} & \multicolumn{2}{c}{Regions} \\
        \cmidrule(ll){7-8} \cmidrule(ll){9-10}
        & & & & & & nobig & big & nobig & big \\
        \midrule
        
 4x4grid       &          &     	47 &   3 & 0     & R$\geq$4.99  & 0.029   & 0.029    & 153    & 121    \\
 4x4grid       &          &     	47 &   3 & 0,0.8 & R$\geq$6.62  & 0.028   & 0.028    & 129    & 105    \\
 4x4grid       &          &     	47 &   3 & 0.1   & R$\geq$6.36  & 0.025   & 0.027    & 137    & 121    \\
 4x4grid       &          &     	49 &   3 & 0.2,1 & R$\geq$6.62  & 0.028   & 0.029    & 121    & 105    \\
 4x4grid       &          &     	47 &   3 & 1e-06 & R$\geq$4.99  & 0.026   & 0.026    & 153    & 121    \\
 4x4grid-avoid &          &     	45 &   3 & 0     & P$\leq$0.94  & 0.072   & 0.043    & 137    & 73     \\
 4x4grid-avoid &          &     	47 &   3 & 0,0.8 & P$\leq$0.87  & 0.555   & 0.074    & 369    & 177    \\
 4x4grid-avoid &          &     	45 &   3 & 0.1   & P$\leq$0.85  & 0.06    & 0.057    & 489    & 361    \\
 4x4grid-avoid &          &     	47 &   3 & 0.2,1 & P$\leq$0.94  & 0.043   & 0.023    & 65     & 25     \\
 4x4grid-avoid &          &     	45 &   3 & 1e-06 & P$\leq$0.94  & 0.261   & 0.037    & 265    & 73     \\
 alarm         &          &     	60 &   2 & 0     & P$\geq$0.04  & 0.063   & 0.062    & 29     & 5      \\
 alarm         &          &     	15 &   2 & 0,0.8 & P$\geq$0.04  & 0.063   & 0.062    & 29     & 1      \\
 alarm         &          &     	15 &   2 & 0.1   & P$\geq$0.23  & 0.063   & 0.065    & 49     & 41     \\
 alarm         &          &     	15 &   2 & 0.2,1 & P$\geq$0.12  & 0.063   & 0.064    & 29     & 29     \\
 alarm         &          &     	15 &   2 & 1e-06 & P$\geq$0.04  & 0.064   & 0.062    & 29     & 5      \\
 brp           & (16, 2)  &    	183 &   4 & 0     & P$\leq$0.02  & 13.488  & 16.266   & 67669  & 67669  \\
 brp           & (16, 2)  &    	330 &   4 & 0,0.8 & P$\leq$0.02  & 1.097   & 1.245    & 5685   & 5685   \\
 brp           & (16, 2)  &    	324 &   4 & 0.1   & P$\leq$0.02  & 2.944   & 4.365    & 27709  & 27709  \\
 brp           & (16, 2)  &    	330 &   4 & 0.2,1 & P$\leq$0.02  & 9.224   & 11.264   & 47741  & 47741  \\
 brp           & (16, 2)  &    	177 &   4 & 1e-06 & P$\leq$0.02  & 7.315   & 12.102   & 67677  & 67677  \\
 brp           & (32, 4)  &    	551 &   4 & 0     & P$\leq$0.01  & 59.761  & 79.855   & 96681  & 96681  \\
 brp           & (32, 4)  &    	551 &   4 & 0,0.8 & P$\leq$0.01  & 21.03   & 28.877   & 36013  & 36013  \\
 brp           & (32, 4)  &   	1094 &   4 & 0.1   & P$\leq$0.01  & 21.744  & 40.187   & 71397  & 71397  \\
 brp           & (32, 4)  &   	1100 &   4 & 0.2,1 & P$\leq$0.01  & 70.382  & 86.994   & 104897 & 104897 \\
 brp           & (32, 4)  &   	1094 &   4 & 1e-06 & P$\leq$0.01  & 29.844  & 60.657   & 96681  & 96681  \\
 brp           & (64, 8)  &   	3984 &   4 & 0     & P$\leq$0.01  & 276.095 & 468.664  & 118281 & 118281 \\
 brp           & (64, 8)  &   	1863 &   4 & 0,0.8 & P$\leq$0.01  & 181.508 & 322.027  & 81741  & 81741  \\
 brp           & (64, 8)  &   	3978 &   4 & 0.1   & P$\leq$0.01  & 146.612 & 354.7    & 118741 & 118741 \\
 brp           & (64, 8)  &   	3984 &   4 & 0.2,1 & P$\leq$0.01  & 383.961 & 611.236  & 153465 & 153465 \\
 brp           & (64, 8)  &   	3978 &   4 & 1e-06 & P$\leq$0.01  & 154.456 & 389.596  & 124681 & 124681 \\
 crowds        & (10, 2)  &     	32 &   2 & 0     & P$\leq$0.02  & MO      & MO       & MO     & MO     \\
 crowds        & (10, 2)  &     	32 &   2 & 0,0.8 & P$\leq$0.0   & 0.056   & 0.054    & 309    & 237    \\
 crowds        & (10, 2)  &     	20 &   2 & 0.1   & P$\leq$0.01  & 0.056   & 0.05     & 397    & 201    \\
 crowds        & (10, 2)  &     	32 &   2 & 0.2,1 & P$\leq$0.01  & 0.045   & 0.044    & 81     & 57     \\
 crowds        & (10, 2)  &     	20 &   2 & 1e-06 & P$\leq$0.02  & 0.22    & 0.127    & 4237   & 2009   \\
 crowds        & (10, 4)  &     	96 &   2 & 0,0.8 & P$\leq$0.02  & 0.234   & 0.256    & 309    & 249    \\
 crowds        & (10, 4)  &     	42 &   2 & 0.1   & P$\leq$0.03  & 0.214   & 0.238    & 369    & 277    \\
 crowds        & (10, 4)  &     	96 &   2 & 0.2,1 & P$\leq$0.07  & 0.209   & 0.221    & 81     & 61     \\
 crowds        & (10, 4)  &     	84 &   2 & 1e-06 & P$\leq$0.1   & 0.491   & 0.685    & 3873   & 2905   \\
 herman        & 1        &  	40690 &   1 & 0     & R$\geq$11.98 & 47.168  & 43.8     & 467    & 193    \\
 herman        & 1        &  	18872 &   1 & 0,0.8 & R$\geq$11.98 & 46.361  & 44.758   & 463    & 195    \\
 herman        & 1        &  	40690 &   1 & 0.1   & R$\geq$11.98 & 38.975  & 29.168   & 467    & 195    \\
 herman        & 1        &  	18872 &   1 & 0.2,1 & R$\geq$11.98 & 45.938  & 43.336   & 463    & 195    \\
 herman        & 1        &  	40690 &   1 & 1e-06 & R$\geq$11.98 & 40.66   & 30.221   & 467    & 195    \\
 herman        & 5        &     	89 &   1 & 0     & R$\geq$1.91  & 0.062   & 0.072    & 103    & 19     \\
 herman        & 5        &     	89 &   1 & 0,0.8 & R$\geq$1.91  & 0.061   & 0.073    & 101    & 19     \\
 herman        & 5        &    	196 &   1 & 0.1   & R$\geq$1.91  & 0.054   & 0.067    & 107    & 21     \\
 herman        & 5        &    	196 &   1 & 0.2,1 & R$\geq$1.91  & 0.061   & 0.074    & 101    & 19     \\
 herman        & 5        &     	89 &   1 & 1e-06 & R$\geq$1.91  & 0.054   & 0.072    & 103    & 25     \\
 herman        & 7        &    	680 &   1 & 0     & R$\geq$4.45  & 0.5     & 0.318    & 203    & 59     \\
 herman        & 7        &    	680 &   1 & 0,0.8 & R$\geq$4.45  & 0.453   & 0.301    & 179    & 49     \\
 herman        & 7        &   	1422 &   1 & 0.1   & R$\geq$4.45  & 0.34    & 0.261    & 179    & 51     \\
 herman        & 7        &   	1422 &   1 & 0.2,1 & R$\geq$4.45  & 0.47    & 0.299    & 179    & 49     \\
 herman        & 7        &   	1422 &   1 & 1e-06 & R$\geq$4.45  & 0.382   & 0.279    & 203    & 59     \\
 herman        & 9        &   	8008 &   1 & 0     & R$\geq$7.84  & 5.689   & 3.576    & 363    & 159    \\
 herman        & 9        &   	3707 &   1 & 0,0.8 & R$\geq$7.84  & 6.047   & 3.137    & 401    & 121    \\
 herman        & 9        &   	3707 &   1 & 0.1   & R$\geq$7.84  & 4.694   & 2.172    & 403    & 123    \\
 herman        & 9        &   	8008 &   1 & 0.2,1 & R$\geq$7.84  & 6.084   & 3.037    & 401    & 121    \\
 herman        & 9        &   	8008 &   1 & 1e-06 & R$\geq$7.84  & 4.662   & 2.504    & 365    & 159    \\
 hermanspeed   & 3        &     	22 &   3 & 0     & R$\geq$0.58  & 1.593   & 1.588    & 10825  & 10825  \\
 hermanspeed   & 3        &     	22 &   3 & 0,0.8 & R$\geq$0.58  & 1.346   & 1.334    & 9057   & 9057   \\
 hermanspeed   & 3        &     	22 &   3 & 0.1   & R$\geq$0.58  & 0.801   & 0.813    & 5953   & 6025   \\
 hermanspeed   & 3        &     	22 &   3 & 0.2,1 & R$\geq$0.58  & 0.415   & 0.417    & 2665   & 2665   \\
 hermanspeed   & 3        &     	22 &   3 & 1e-06 & R$\geq$0.58  & 1.682   & 1.694    & 10809  & 11049  \\
 hermanspeed   & 5        &    	673 &   1 & 0     & R$\geq$2.48  & 150.732 & 152.301  & 19393  & 19393  \\
 hermanspeed   & 5        &    	673 &   1 & 0,0.8 & R$\geq$2.48  & 145.492 & 146.416  & 18641  & 18641  \\
 hermanspeed   & 5        &    	673 &   1 & 0.1   & R$\geq$2.5   & 109.93  & 111.545  & 14857  & 14753  \\
 hermanspeed   & 5        &    	673 &   1 & 0.2,1 & R$\geq$2.52  & 60.591  & 61.446   & 7737   & 7737   \\
 hermanspeed   & 5        &    	673 &   1 & 1e-06 & R$\geq$2.48  & 162.995 & 164.563  & 19569  & 19393  \\
 maze2         &          &     	50 &  15 & 0     & R$\geq$14.17 & 11.539  & 13.297   & 101601 & 101601 \\
 maze2         &          &     	50 &  15 & 0,0.8 & R$\geq$15.98 & 3.533   & 4.023    & 28673  & 28673  \\
 maze2         &          &     	50 &  15 & 0.1   & R$\geq$17.34 & 8.434   & 9.526    & 74049  & 74049  \\
 maze2         &          &     	41 &  15 & 0.2,1 & R$\geq$19.45 & 15.616  & 17.765   & 120801 & 120801 \\
 maze2         &          &     	50 &  15 & 1e-06 & R$\geq$14.17 & 4.34    & 4.959    & 33937  & 33937  \\
 nand          & (5, 5)   &   	2688 &   2 & 0,0.8 & P$\leq$0.76  & 0.131   & 6.372    & 57     & 25     \\
 nand          & (5, 5)   &   	2687 &   2 & 0.1   & P$\leq$0.5   & 0.107   & 0.798    & 125    & 1      \\
 nand          & (5, 10)  &   	5448 &   2 & 0,0.8 & P$\leq$0.82  & 0.325   & 28.456   & 89     & 25     \\
 nand          & (5, 10)  &   	5447 &   2 & 0.1   & P$\leq$0.5   & 0.188   & 3.661    & 125    & 1      \\
 nand          & (5, 25)  &  	13728 &   2 & 0,0.8 & P$\leq$0.82  & 1.427   & 271.543  & 169    & 29     \\
 nand          & (5, 25)  &    	312 &   2 & 0.1   & P$\leq$0.5   & 0.489   & 33.256   & 125    & 1      \\
 nand          & (5, 50)  &    	562 &   2 & 0,0.8 & P$\leq$0.82  & 4.515   & 1274.192 & 241    & 33     \\
 nand          & (5, 50)  &  	27527 &   2 & 0.1   & P$\leq$0.5   & 1.131   & 231.017  & 125    & 1      \\
 nand          & (10, 50) & 	346962 &   2 & 0,0.8 & P$\leq$0.91  & 49.766  & TO       & 149    & TO     \\
 nand          & (10, 50) & 	346962 &   2 & 0.1   & P$\leq$0.29  & 19.779  & TO       & 125    & TO     \\
 newgrid       & 2        &     	37 &   3 & 0,0.8 & P$\leq$0.8   & 0.512   & 0.855    & 1145   & 1105   \\
 newgrid       & 2        &     	32 &   3 & 0.1   & P$\leq$0.9   & 0.05    & 0.059    & 609    & 601    \\
 newgrid       & 2        &     	47 &   3 & 0.2,1 & P$\leq$1.0   & 0.018   & 0.02     & 1      & 1      \\
 newgrid       & 4        &    	110 &   3 & 0,0.8 & P$\leq$0.93  & 4.305   & 4.736    & 1097   & 1121   \\
 newgrid       & 4        &     	74 &   3 & 0.1   & P$\leq$0.98  & 0.041   & 0.045    & 177    & 177    \\
 newgrid       & 4        &    	110 &   3 & 0.2,1 & P$\leq$1.0   & 0.023   & 0.024    & 1      & 1      \\
 nrp           & 5        &     	12 &   5 & 0     & P$\leq$0.2   & MO      & 0.016    & MO     & 1      \\
 nrp           & 5        &     	34 &   5 & 0,0.8 & P$\leq$0.2   & MO      & 0.017    & MO     & 1      \\
 nrp           & 5        &     	33 &   5 & 0.1   & P$\leq$0.2   & MO      & 0.015    & MO     & 1      \\
 nrp           & 5        &     	34 &   5 & 0.2,1 & P$\leq$0.2   & MO      & 0.017    & MO     & 1      \\
 nrp           & 5        &     	12 &   5 & 1e-06 & P$\leq$0.2   & MO      & 0.017    & MO     & 1      \\
 nrp           & 10       &    	114 &  10 & 0     & P$\leq$0.1   & MO      & 0.018    & MO     & 1      \\
 nrp           & 10       &    	114 &  10 & 0,0.8 & P$\leq$0.1   & MO      & 0.018    & MO     & 1      \\
 nrp           & 10       &     	22 &  10 & 0.1   & P$\leq$0.1   & MO      & 0.019    & MO     & 1      \\
 nrp           & 10       &     	22 &  10 & 0.2,1 & P$\leq$0.1   & MO      & 0.019    & MO     & 1      \\
 nrp           & 10       &    	113 &  10 & 1e-06 & P$\leq$0.1   & MO      & 0.018    & MO     & 1      \\
 nrp           & 100      &  	10104 & 100 & 0     & P$\leq$0.01  & MO      & 1.0      & MO     & 1      \\
 nrp           & 100      &  	10104 & 100 & 0,0.8 & P$\leq$0.01  & MO      & 0.984    & MO     & 1      \\
 nrp           & 100      &    	202 & 100 & 0.1   & P$\leq$0.01  & MO      & 1.046    & MO     & 1      \\
 nrp           & 100      &  	10104 & 100 & 0.2,1 & P$\leq$0.01  & MO      & 0.994    & MO     & 1      \\
 nrp           & 100      &    	202 & 100 & 1e-06 & P$\leq$0.01  & MO      & 0.967    & MO     & 1      \\
 refuel        & 3        &     	47 &  18 & 0     & P$\leq$0.09  & 0.382   & 0.347    & 3281   & 2705   \\
 refuel        & 3        &     	51 &  18 & 0,0.8 & P$\leq$0.06  & 0.042   & 0.035    & 177    & 97     \\
 refuel        & 3        &     	34 &  18 & 0.1   & P$\leq$0.06  & 0.053   & 0.03     & 337    & 81     \\
 refuel        & 3        &     	47 &  18 & 0.2,1 & P$\leq$0.07  & 0.033   & 0.024    & 97     & 17     \\
 refuel        & 3        &     	34 &  18 & 1e-06 & P$\leq$0.09  & 0.15    & 0.022    & 1233   & 1      \\
\bottomrule
\end{longtable}
\subsubsection{Table for \(\varepsilon=0.1\)}
\begin{longtable}{llrrllrrrr}

        \toprule
        Model & Const & $|S|$ & $|V|$ & $\delta$ & Prop & \multicolumn{2}{c}{Time (s)} & \multicolumn{2}{c}{Regions} \\
        \cmidrule(ll){7-8} \cmidrule(ll){9-10}
        & & & & & & nobig & big & nobig & big \\
        \midrule
        
 4x4grid       &          &     	47 &   3 & 0     & R$\geq$4.54  & 0.02   & 0.021   & 41      & 41   \\
 4x4grid       &          &     	47 &   3 & 0,0.8 & R$\geq$6.02  & 0.019  & 0.02    & 25      & 25   \\
 4x4grid       &          &     	49 &   3 & 0.1   & R$\geq$5.78  & 0.019  & 0.022   & 41      & 41   \\
 4x4grid       &          &     	47 &   3 & 0.2,1 & R$\geq$6.02  & 0.019  & 0.02    & 25      & 25   \\
 4x4grid       &          &     	49 &   3 & 1e-06 & R$\geq$4.54  & 0.019  & 0.02    & 41      & 41   \\
 4x4grid-avoid &          &     	45 &   3 & 0,0.8 & P$\leq$0.94  & 0.026  & 0.029   & 41      & 9    \\
 4x4grid-avoid &          &     	47 &   3 & 0.1   & P$\leq$0.93  & 0.022  & 0.023   & 33      & 25   \\
 alarm         &          &     	60 &   2 & 0     & P$\geq$0.03  & 0.063  & 0.062   & 17      & 5    \\
 alarm         &          &     	15 &   2 & 0,0.8 & P$\geq$0.03  & 0.062  & 0.062   & 17      & 1    \\
 alarm         &          &     	15 &   2 & 0.1   & P$\geq$0.21  & 0.062  & 0.063   & 21      & 17   \\
 alarm         &          &     	60 &   2 & 0.2,1 & P$\geq$0.11  & 0.062  & 0.063   & 13      & 13   \\
 alarm         &          &     	31 &   2 & 1e-06 & P$\geq$0.03  & 0.062  & 0.062   & 17      & 5    \\
 brp           & (16, 2)  &    	183 &   4 & 0     & P$\leq$0.03  & 0.329  & 0.375   & 1569    & 1569 \\
 brp           & (16, 2)  &    	183 &   4 & 0,0.8 & P$\leq$0.03  & 0.091  & 0.102   & 305     & 305  \\
 brp           & (16, 2)  &    	324 &   4 & 0.1   & P$\leq$0.03  & 0.154  & 0.215   & 1185    & 1185 \\
 brp           & (16, 2)  &    	183 &   4 & 0.2,1 & P$\leq$0.03  & 0.407  & 0.46    & 1985    & 1985 \\
 brp           & (16, 2)  &    	177 &   4 & 1e-06 & P$\leq$0.03  & 0.192  & 0.282   & 1569    & 1569 \\
 brp           & (32, 4)  &    	551 &   4 & 0     & P$\leq$0.01  & 2.09   & 2.781   & 3569    & 3569 \\
 brp           & (32, 4)  &    	551 &   4 & 0,0.8 & P$\leq$0.01  & 0.546  & 0.728   & 905     & 905  \\
 brp           & (32, 4)  &    	545 &   4 & 0.1   & P$\leq$0.01  & 0.542  & 0.996   & 1741    & 1741 \\
 brp           & (32, 4)  &   	1100 &   4 & 0.2,1 & P$\leq$0.01  & 1.44   & 1.94    & 2485    & 2485 \\
 brp           & (32, 4)  &    	545 &   4 & 1e-06 & P$\leq$0.01  & 1.073  & 2.05    & 3569    & 3569 \\
 brp           & (64, 8)  &   	1863 &   4 & 0     & P$\leq$0.01  & 12.926 & 19.221  & 5033    & 5033 \\
 brp           & (64, 8)  &   	3984 &   4 & 0,0.8 & P$\leq$0.01  & 5.924  & 10.992  & 2989    & 2989 \\
 brp           & (64, 8)  &   	3978 &   4 & 0.1   & P$\leq$0.01  & 4.361  & 12.092  & 4357    & 4357 \\
 brp           & (64, 8)  &   	1863 &   4 & 0.2,1 & P$\leq$0.01  & 11.401 & 21.011  & 5673    & 5673 \\
 brp           & (64, 8)  &   	1857 &   4 & 1e-06 & P$\leq$0.01  & 5.868  & 14.52   & 5057    & 5057 \\
 crowds        & (10, 2)  &     	19 &   2 & 0     & P$\leq$0.02  & MO     & MO      & MO      & MO   \\
 crowds        & (10, 2)  &     	19 &   2 & 0,0.8 & P$\leq$0.0   & 0.045  & 0.046   & 89      & 65   \\
 crowds        & (10, 2)  &     	20 &   2 & 0.1   & P$\leq$0.01  & 0.045  & 0.044   & 97      & 53   \\
 crowds        & (10, 2)  &     	19 &   2 & 0.2,1 & P$\leq$0.01  & 0.043  & 0.043   & 41      & 25   \\
 crowds        & (10, 2)  &     	20 &   2 & 1e-06 & P$\leq$0.02  & 0.073  & 0.059   & 809     & 381  \\
 crowds        & (10, 4)  &     	52 &   2 & 0,0.8 & P$\leq$0.02  & 0.212  & 0.221   & 89      & 77   \\
 crowds        & (10, 4)  &     	84 &   2 & 0.1   & P$\leq$0.03  & 0.196  & 0.206   & 89      & 77   \\
 crowds        & (10, 4)  &     	96 &   2 & 0.2,1 & P$\leq$0.08  & 0.205  & 0.211   & 37      & 25   \\
 crowds        & (10, 4)  &     	42 &   2 & 1e-06 & P$\leq$0.11  & 0.266  & 0.289   & 733     & 561  \\
 herman        & 1        &  	40690 &   1 & 0     & R$\geq$10.89 & 18.915 & 27.802  & 87      & 39   \\
 herman        & 1        &  	18872 &   1 & 0,0.8 & R$\geq$10.89 & 16.988 & 26.403  & 85      & 25   \\
 herman        & 1        &  	18872 &   1 & 0.1   & R$\geq$10.89 & 15.647 & 20.853  & 87      & 27   \\
 herman        & 1        &  	40690 &   1 & 0.2,1 & R$\geq$10.89 & 17.273 & 26.517  & 85      & 25   \\
 herman        & 1        &  	40690 &   1 & 1e-06 & R$\geq$10.89 & 17.455 & 21.497  & 87      & 39   \\
 herman        & 5        &    	196 &   1 & 0     & R$\geq$1.74  & 0.043  & 0.069   & 23      & 5    \\
 herman        & 5        &     	89 &   1 & 0,0.8 & R$\geq$1.74  & 0.049  & 0.069   & 21      & 3    \\
 herman        & 5        &     	89 &   1 & 0.1   & R$\geq$1.74  & 0.04   & 0.067   & 23      & 5    \\
 herman        & 5        &    	196 &   1 & 0.2,1 & R$\geq$1.74  & 0.041  & 0.069   & 21      & 3    \\
 herman        & 5        &     	89 &   1 & 1e-06 & R$\geq$1.74  & 0.041  & 0.068   & 25      & 7    \\
 herman        & 7        &    	680 &   1 & 0     & R$\geq$4.04  & 0.19   & 0.229   & 39      & 11   \\
 herman        & 7        &    	680 &   1 & 0,0.8 & R$\geq$4.04  & 0.188  & 0.225   & 39      & 9    \\
 herman        & 7        &   	1422 &   1 & 0.1   & R$\geq$4.04  & 0.159  & 0.202   & 39      & 11   \\
 herman        & 7        &   	1422 &   1 & 0.2,1 & R$\geq$4.04  & 0.194  & 0.225   & 39      & 9    \\
 herman        & 7        &   	1422 &   1 & 1e-06 & R$\geq$4.04  & 0.169  & 0.213   & 39      & 11   \\
 herman        & 9        &   	3707 &   1 & 0     & R$\geq$7.13  & 2.013  & 1.9     & 67      & 23   \\
 herman        & 9        &   	8008 &   1 & 0,0.8 & R$\geq$7.13  & 1.683  & 1.825   & 49      & 21   \\
 herman        & 9        &   	3707 &   1 & 0.1   & R$\geq$7.13  & 1.473  & 1.553   & 51      & 23   \\
 herman        & 9        &   	8008 &   1 & 0.2,1 & R$\geq$7.13  & 1.707  & 1.826   & 49      & 21   \\
 herman        & 9        &   	3707 &   1 & 1e-06 & R$\geq$7.13  & 1.824  & 1.586   & 67      & 23   \\
 hermanspeed   & 3        &     	22 &   3 & 0     & R$\geq$0.53  & 0.038  & 0.04    & 33      & 33   \\
 hermanspeed   & 3        &     	22 &   3 & 0,0.8 & R$\geq$0.53  & 0.039  & 0.039   & 25      & 25   \\
 hermanspeed   & 3        &     	22 &   3 & 0.1   & R$\geq$0.53  & 0.037  & 0.038   & 25      & 25   \\
 hermanspeed   & 3        &     	22 &   3 & 0.2,1 & R$\geq$0.53  & 0.036  & 0.035   & 1       & 1    \\
 hermanspeed   & 3        &     	22 &   3 & 1e-06 & R$\geq$0.53  & 0.041  & 0.042   & 33      & 33   \\
 hermanspeed   & 5        &    	673 &   1 & 0     & R$\geq$2.25  & 4.649  & 4.665   & 481     & 481  \\
 hermanspeed   & 5        &    	673 &   1 & 0,0.8 & R$\geq$2.25  & 3.918  & 3.946   & 393     & 393  \\
 hermanspeed   & 5        &    	673 &   1 & 0.1   & R$\geq$2.27  & 2.947  & 2.959   & 289     & 289  \\
 hermanspeed   & 5        &    	673 &   1 & 0.2,1 & R$\geq$2.3   & 1.935  & 1.946   & 137     & 137  \\
 hermanspeed   & 5        &    	673 &   1 & 1e-06 & R$\geq$2.25  & 5.015  & 5.007   & 489     & 481  \\
 maze2         &          &     	50 &  15 & 0     & R$\geq$12.88 & 0.214  & 0.244   & 1825    & 1825 \\
 maze2         &          &     	50 &  15 & 0,0.8 & R$\geq$14.52 & 0.13   & 0.147   & 961     & 961  \\
 maze2         &          &     	50 &  15 & 0.1   & R$\geq$15.77 & 0.19   & 0.208   & 1537    & 1537 \\
 maze2         &          &     	41 &  15 & 0.2,1 & R$\geq$17.68 & 0.384  & 0.434   & 2881    & 2881 \\
 maze2         &          &     	50 &  15 & 1e-06 & R$\geq$12.88 & 0.566  & 0.651   & 1297    & 1297 \\
 nand          & (5, 5)   &    	112 &   2 & 0,0.8 & P$\leq$0.83  & 0.112  & 2.679   & 41      & 9    \\
 nand          & (5, 5)   &   	2687 &   2 & 0.1   & P$\leq$0.55  & 0.063  & 0.806   & 13      & 1    \\
 nand          & (5, 10)  &    	162 &   2 & 0,0.8 & P$\leq$0.89  & 0.228  & 11.815  & 49      & 9    \\
 nand          & (5, 10)  &    	162 &   2 & 0.1   & P$\leq$0.55  & 0.103  & 3.676   & 13      & 1    \\
 nand          & (5, 25)  &  	13728 &   2 & 0,0.8 & P$\leq$0.9   & 0.954  & 88.449  & 93      & 9    \\
 nand          & (5, 25)  &    	312 &   2 & 0.1   & P$\leq$0.55  & 0.27   & 33.298  & 13      & 1    \\
 nand          & (5, 50)  &  	27528 &   2 & 0,0.8 & P$\leq$0.9   & 3.472  & 522.869 & 157     & 9    \\
 nand          & (5, 50)  &    	562 &   2 & 0.1   & P$\leq$0.55  & 0.708  & 229.969 & 13      & 1    \\
 nand          & (10, 50) & 	250403 &   2 & 0,0.8 & P$\leq$1.0   & 36.756 & TO      & 65      & TO   \\
 nand          & (10, 50) & 	346962 &   2 & 0.1   & P$\leq$0.31  & 11.63  & TO      & 13      & TO   \\
 newgrid       & 2        &     	37 &   3 & 0,0.8 & P$\leq$0.87  & 0.135  & 0.169   & 273     & 281  \\
 newgrid       & 2        &     	32 &   3 & 0.1   & P$\leq$0.98  & 0.019  & 0.02    & 17      & 17   \\
 nrp           & 5        &     	12 &   5 & 0     & P$\leq$0.22  & 73.998 & 0.016   & 1194257 & 1    \\
 nrp           & 5        &     	12 &   5 & 0,0.8 & P$\leq$0.22  & 17.764 & 0.017   & 278993  & 1    \\
 nrp           & 5        &     	33 &   5 & 0.1   & P$\leq$0.22  & 19.798 & 0.016   & 388673  & 1    \\
 nrp           & 5        &     	34 &   5 & 0.2,1 & P$\leq$0.22  & 19.724 & 0.017   & 309009  & 1    \\
 nrp           & 5        &     	12 &   5 & 1e-06 & P$\leq$0.22  & 65.133 & 0.017   & 1194257 & 1    \\
 nrp           & 10       &    	114 &  10 & 0     & P$\leq$0.11  & MO     & 0.018   & MO      & 1    \\
 nrp           & 10       &    	114 &  10 & 0,0.8 & P$\leq$0.11  & MO     & 0.018   & MO      & 1    \\
 nrp           & 10       &    	113 &  10 & 0.1   & P$\leq$0.11  & MO     & 0.019   & MO      & 1    \\
 nrp           & 10       &    	114 &  10 & 0.2,1 & P$\leq$0.11  & MO     & 0.018   & MO      & 1    \\
 nrp           & 10       &    	113 &  10 & 1e-06 & P$\leq$0.11  & MO     & 0.019   & MO      & 1    \\
 nrp           & 100      &    	202 & 100 & 0     & P$\leq$0.01  & MO     & 0.996   & MO      & 1    \\
 nrp           & 100      &  	10104 & 100 & 0,0.8 & P$\leq$0.01  & MO     & 0.998   & MO      & 1    \\
 nrp           & 100      &  	10103 & 100 & 0.1   & P$\leq$0.01  & MO     & 1.043   & MO      & 1    \\
 nrp           & 100      &    	202 & 100 & 0.2,1 & P$\leq$0.01  & MO     & 0.99    & MO      & 1    \\
 nrp           & 100      &  	10103 & 100 & 1e-06 & P$\leq$0.01  & MO     & 0.955   & MO      & 1    \\
 refuel        & 3        &     	47 &  18 & 0     & P$\leq$0.1   & 0.021  & 0.022   & 1       & 1    \\
 refuel        & 3        &     	51 &  18 & 0,0.8 & P$\leq$0.06  & 0.021  & 0.022   & 1       & 1    \\
 refuel        & 3        &     	34 &  18 & 0.1   & P$\leq$0.06  & 0.023  & 0.022   & 17      & 1    \\
 refuel        & 3        &     	51 &  18 & 0.2,1 & P$\leq$0.08  & 0.021  & 0.022   & 1       & 1    \\
 refuel        & 3        &     	34 &  18 & 1e-06 & P$\leq$0.1   & 0.022  & 0.022   & 1       & 1    \\
\bottomrule
\end{longtable}

\end{scriptsize}

\newpage
\section{Experiments: Big-Step Transformation with Round-Robin Splitting}
\label{app:roundrobin}

\begin{figure}[!t]
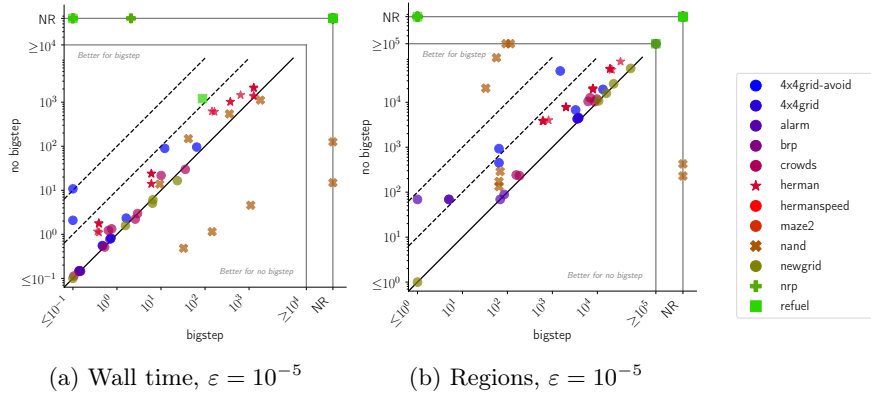

    \centering
    \begin{subfigure}{0.38\linewidth}
    \begin{adjustbox}{max width=\linewidth}
        \input{misc/benchmarks/time-1e-05-rr.pgf}
    \end{adjustbox}
    \caption{Wall time, \(\varepsilon=10^{-5}\)}
    \end{subfigure}%
    \begin{subfigure}{0.38\linewidth}
    \begin{adjustbox}{max width=\linewidth}
        \input{misc/benchmarks/regions-1e-05-rr.pgf}
    \end{adjustbox}
    \caption{Regions, \(\varepsilon=10^{-5}\)}
    \end{subfigure}%
    \begin{subfigure}{0.2\linewidth}
    \begin{adjustbox}{max width=\linewidth, clip, trim=0cm 12cm 0cm 0cm}
        \centering
        \input{misc/benchmarks/time-1e-05-legend-rr.pgf}
    \end{adjustbox}
    \end{subfigure}
    \caption{Effectivity of Big-Step with Round-Robin Splitting}
    \label{fig:bigsteprr}
\end{figure}

The experiments in (Q1) in \cref{section:experiments} enable the standard splitting heuristic in PL. In these experiments, we have replaced this heuristic by a simple round-robin method, we plot regions and wall time in \cref{fig:bigsteprr}. Overall, the effect of big-step is unchanged and the statements in the discussion of (Q1) still hold true.

\begin{scriptsize}
\subsubsection{Table for \(\varepsilon=10^{-5}\)}
\begin{longtable}{llrrllrrrr}

        \toprule
        Model & Const & $|S|$ & $|V|$ & $\delta$ & Prop & \multicolumn{2}{c}{Time (s)} & \multicolumn{2}{c}{Regions} \\
        \cmidrule(ll){7-8} \cmidrule(ll){9-10}
        & & & & & & nobig & big & nobig & big \\
        \midrule
        
 4x4grid       &          &     	49 &   3 & 0     & R$\geq$5.04  & 0.774    & 0.689    & 4281    & 3585   \\
 4x4grid       &          &     	47 &   3 & 0,0.8 & R$\geq$6.69  & 0.831    & 0.729    & 4553    & 3841   \\
 4x4grid       &          &     	47 &   3 & 0.1   & R$\geq$6.42  & 0.546    & 0.471    & 4433    & 3753   \\
 4x4grid       &          &     	49 &   3 & 0.2,1 & R$\geq$6.69  & 0.805    & 0.727    & 4433    & 3841   \\
 4x4grid       &          &     	49 &   3 & 1e-06 & R$\geq$5.04  & 0.559    & 0.465    & 4281    & 3561   \\
 4x4grid-avoid &          &     	45 &   3 & 0     & P$\leq$0.93  & 10.777   & 0.191    & 929     & 137    \\
\bottomrule
\end{longtable}
\subsubsection{Table for \(\varepsilon=0.0001\)}
\begin{longtable}{llrrllrrrr}

        \toprule
        Model & Const & $|S|$ & $|V|$ & $\delta$ & Prop & \multicolumn{2}{c}{Time (s)} & \multicolumn{2}{c}{Regions} \\
        \cmidrule(ll){7-8} \cmidrule(ll){9-10}
        & & & & & & nobig & big & nobig & big \\
        \midrule
        
 4x4grid       &          &     	47 &   3 & 0     & R$\geq$5.04  & 0.254    & 0.237   & 1257    & 1097    \\
 4x4grid       &          &     	49 &   3 & 0,0.8 & R$\geq$6.69  & 0.279    & 0.273   & 1385    & 1297    \\
 4x4grid       &          &     	49 &   3 & 0.1   & R$\geq$6.42  & 0.199    & 0.187   & 1369    & 1273    \\
 4x4grid       &          &     	49 &   3 & 0.2,1 & R$\geq$6.69  & 0.283    & 0.273   & 1385    & 1297    \\
 4x4grid       &          &     	47 &   3 & 1e-06 & R$\geq$5.04  & 0.185    & 0.167   & 1257    & 1097    \\
 4x4grid-avoid &          &     	47 &   3 & 0     & P$\leq$0.93  & 10.704   & 0.192   & 929     & 137     \\
\bottomrule
\end{longtable}
\subsubsection{Table for \(\varepsilon=0.01\)}
\begin{longtable}{llrrllrrrr}

        \toprule
        Model & Const & $|S|$ & $|V|$ & $\delta$ & Prop & \multicolumn{2}{c}{Time (s)} & \multicolumn{2}{c}{Regions} \\
        \cmidrule(ll){7-8} \cmidrule(ll){9-10}
        & & & & & & nobig & big & nobig & big \\
        \midrule
        
 4x4grid       &          &     	49 &   3 & 0     & R$\geq$4.99  & 0.062   & 0.06    & 153     & 121    \\
 4x4grid       &          &     	47 &   3 & 0,0.8 & R$\geq$6.62  & 0.058   & 0.057   & 129     & 105    \\
 4x4grid       &          &     	47 &   3 & 0.1   & R$\geq$6.36  & 0.054   & 0.052   & 137     & 121    \\
 4x4grid       &          &     	49 &   3 & 0.2,1 & R$\geq$6.62  & 0.058   & 0.058   & 121     & 105    \\
 4x4grid       &          &     	47 &   3 & 1e-06 & R$\geq$4.99  & 0.054   & 0.051   & 153     & 121    \\
 4x4grid-avoid &          &     	45 &   3 & 0     & P$\leq$0.94  & 0.171   & 0.097   & 137     & 73     \\
\bottomrule
\end{longtable}
\subsubsection{Table for \(\varepsilon=0.1\)}
\begin{longtable}{llrrllrrrr}

        \toprule
        Model & Const & $|S|$ & $|V|$ & $\delta$ & Prop & \multicolumn{2}{c}{Time (s)} & \multicolumn{2}{c}{Regions} \\
        \cmidrule(ll){7-8} \cmidrule(ll){9-10}
        & & & & & & nobig & big & nobig & big \\
        \midrule
        
 4x4grid       &          &     	47 &   3 & 0     & R$\geq$4.54  & 0.043   & 0.045   & 41      & 41   \\
 4x4grid       &          &     	49 &   3 & 0,0.8 & R$\geq$6.02  & 0.039   & 0.041   & 25      & 25   \\
 4x4grid       &          &     	47 &   3 & 0.1   & R$\geq$5.78  & 0.042   & 0.042   & 41      & 41   \\
 4x4grid       &          &     	47 &   3 & 0.2,1 & R$\geq$6.02  & 0.039   & 0.041   & 25      & 25   \\
 4x4grid       &          &     	47 &   3 & 1e-06 & R$\geq$4.54  & 0.04    & 0.041   & 41      & 41   \\
 4x4grid-avoid &          &     	45 &   3 & 0,0.8 & P$\leq$0.94  & 0.057   & 0.046   & 41      & 9    \\
\bottomrule
\end{longtable}

\end{scriptsize}

\newpage
\section{Experiments: Generalized PL versus standard PL in simple pMCs}
\label{appendix:exsimple}

We provide a table of the comparison between standard and generalized PL. In contrast to \cref{fig:genvsstandard}, we omit the benchmarks that are not supported by standard PL.

\begin{scriptsize}
\subsubsection{Table for \(\varepsilon=10^{-5}\)}
\begin{longtable}{llrrrlrrrr}

        \toprule
        Model & Const & $|S|$ & $|V|$ & $\delta$ & Prop & \multicolumn{2}{c}{Time (s)} & \multicolumn{2}{c}{Regions} \\
        \cmidrule(ll){7-8} \cmidrule(ll){9-10}
        & & & & & & st.\ PL & GPL & st.\ PL & GPL \\
        \midrule
        
 4x4grid       &           &     	47 &   3 & 0.1   & R$\geq$6.42  & 0.215   & 0.286   & 4361    & 4361    \\
 4x4grid       &           &     	47 &   3 & 1e-06 & R$\geq$5.04  & 0.219   & 0.289   & 4281    & 4281    \\
 4x4grid-avoid &           &     	45 &   3 & 0.1   & P$\leq$0.85  & 0.96    & 1.242   & 19585   & 19585   \\
 4x4grid-avoid &           &     	45 &   3 & 1e-06 & P$\leq$0.93  & 10.529  & 38.189  & 48889   & 49569   \\
 alarm         &           &     	31 &   2 & 0.1   & P$\geq$0.24  & 0.065   & 0.067   & 89      & 89      \\
 alarm         &           &     	31 &   2 & 1e-06 & P$\geq$0.04  & 0.064   & 0.067   & 69      & 69      \\
\bottomrule
\end{longtable}
\subsubsection{Table for \(\varepsilon=10^{-4}\)}
\begin{longtable}{llrrrlrrrr}

        \toprule
        Model & Const & $|S|$ & $|V|$ & $\delta$ & Prop & \multicolumn{2}{c}{Time (s)} & \multicolumn{2}{c}{Regions} \\
        \cmidrule(ll){7-8} \cmidrule(ll){9-10}
        & & & & & & st.\ PL & GPL & st.\ PL & GPL \\
        \midrule
        
 4x4grid       &           &     	47 &   3 & 0.1   & R$\geq$6.42  & 0.08    & 0.101    & 1369   & 1369    \\
 4x4grid       &           &     	47 &   3 & 1e-06 & R$\geq$5.04  & 0.075   & 0.095    & 1257   & 1257    \\
 4x4grid-avoid &           &     	45 &   3 & 0.1   & P$\leq$0.85  & 0.317   & 0.415    & 6305   & 6305    \\
 4x4grid-avoid &           &     	45 &   3 & 1e-06 & P$\leq$0.93  & 7.744   & 27.87    & 9961   & 9833    \\
 alarm         &           &     	31 &   2 & 0.1   & P$\geq$0.24  & 0.065   & 0.066    & 73     & 73      \\
 alarm         &           &     	31 &   2 & 1e-06 & P$\geq$0.04  & 0.064   & 0.067    & 57     & 57      \\
\bottomrule
\end{longtable}
\subsubsection{Table for \(\varepsilon=0.01\)}
\begin{longtable}{llrrrlrrrr}

        \toprule
        Model & Const & $|S|$ & $|V|$ & $\delta$ & Prop & \multicolumn{2}{c}{Time (s)} & \multicolumn{2}{c}{Regions} \\
        \cmidrule(ll){7-8} \cmidrule(ll){9-10}
        & & & & & & st.\ PL & GPL & st.\ PL & GPL \\
        \midrule
        
 4x4grid       &           &     	47 &   3 & 0.1   & R$\geq$6.36  & 0.023   & 0.027   & 137    & 137    \\
 4x4grid       &           &     	47 &   3 & 1e-06 & R$\geq$4.99  & 0.024   & 0.027   & 153    & 153    \\
 4x4grid-avoid &           &     	45 &   3 & 0.1   & P$\leq$0.85  & 0.045   & 0.062   & 489    & 489    \\
 4x4grid-avoid &           &     	45 &   3 & 1e-06 & P$\leq$0.94  & 0.08    & 0.262   & 265    & 265    \\
 alarm         &           &     	31 &   2 & 0.1   & P$\geq$0.23  & 0.064   & 0.066   & 49     & 49     \\
 alarm         &           &     	31 &   2 & 1e-06 & P$\geq$0.04  & 0.063   & 0.066   & 29     & 29     \\
\bottomrule
\end{longtable}
\subsubsection{Table for \(\varepsilon=0.1\)}
\begin{longtable}{llrrrlrrrr}

        \toprule
        Model & Const & $|S|$ & $|V|$ & $\delta$ & Prop & \multicolumn{2}{c}{Time (s)} & \multicolumn{2}{c}{Regions} \\
        \cmidrule(ll){7-8} \cmidrule(ll){9-10}
        & & & & & & st.\ PL & GPL & st.\ PL & GPL \\
        \midrule
        
 4x4grid       &           &     	47 &   3 & 0.1   & R$\geq$5.78  & 0.019  & 0.021  & 41      & 41      \\
 4x4grid       &           &     	47 &   3 & 1e-06 & R$\geq$4.54  & 0.019  & 0.02   & 41      & 41      \\
 4x4grid-avoid &           &     	45 &   3 & 0.1   & P$\leq$0.93  & 0.022  & 0.023  & 33      & 33      \\
 alarm         &           &     	31 &   2 & 0.1   & P$\geq$0.21  & 0.063  & 0.064  & 21      & 21      \\
 alarm         &           &     	31 &   2 & 1e-06 & P$\geq$0.03  & 0.063  & 0.065  & 17      & 17      \\
 brp           & (16, 2)   &    	324 &   4 & 0.1   & P$\leq$0.03  & 0.122  & 0.155  & 1185    & 1185    \\
\bottomrule
\end{longtable}

\end{scriptsize}

\newpage
\section{Experiment: Not-Well-Defined and Not-Graph-Preserving Regions}
\label{app:notwelldefined}

\begin{figure}[t]
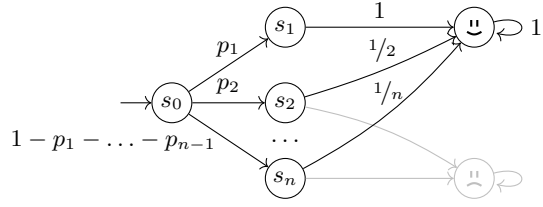

    \centering
    %
    \caption{parametric Markov chain}
    \label{fig:benchpmc}
\end{figure}

The pMC can be seen in \cref{fig:benchpmc}.
The corresponding iMC replaces each parametric transition with its corresponding interval in the region \(R_i\).
For \(R_1\), the corresponding MDP has \(2^{n-1}\) actions.
Each action in the MDP corresponds to taking some combination of the lower or the upper bounds for each interval of the corresponding iMC for the transitions to \(s_1, \ldots, s_{n-1}\), and one minus all of those probabilities for the transition to \(s_n\).

\begin{scriptsize}
\begin{center}
\begin{tabular}{rrrrr}
\toprule
 $n$ & Standard PL, $R_1$ & GPL, $R_1$ & GPL, $R_2$ & GPL, $R_3$ \\
 \midrule
 2   & 0.03       & 0.04       & 0.03       & 0.03       \\
 3   & 0.04       & 0.03       & 0.03       & 0.03       \\
 4   & 0.03       & 0.03       & 0.03       & 0.03       \\
 5   & 0.03       & 0.03       & 0.03       & 0.03       \\
 6   & 0.03       & 0.03       & 0.03       & 0.03       \\
 7   & 0.03       & 0.04       & 0.03       & 0.03       \\
 8   & 0.04       & 0.03       & 0.03       & 0.03       \\
 9   & 0.04       & 0.03       & 0.03       & 0.04       \\
 10  & 0.04       & 0.03       & 0.03       & 0.03       \\
 11  & 0.05       & 0.04       & 0.03       & 0.03       \\
 12  & 0.06       & 0.04       & 0.03       & 0.03       \\
 13  & 0.09       & 0.03       & 0.03       & 0.03       \\
 14  & 0.16       & 0.04       & 0.03       & 0.03       \\
 15  & 0.31       & 0.03       & 0.03       & 0.03       \\
 16  & 0.64       & 0.04       & 0.04       & 0.04       \\
 17  & 1.38       & 0.03       & 0.03       & 0.04       \\
 18  & 2.99       & 0.03       & 0.03       & 0.03       \\
 19  & 6.35       & 0.04       & 0.04       & 0.04       \\
 20  & 13.43      & 0.04       & 0.04       & 0.03       \\
 21  & 28.46      & 0.04       & 0.03       & 0.03       \\
 22  & 59.92      & 0.04       & 0.04       & 0.04       \\
 23  & 126.01     & 0.04       & 0.04       & 0.03       \\
 24  & MO         & 0.04       & 0.04       & 0.03       \\
 25  & MO         & 0.04       & 0.03       & 0.03       \\
 26  & MO         & 0.04       & 0.04       & 0.03       \\
 27  & MO         & 0.04       & 0.04       & 0.04       \\
 28  & MO         & 0.04       & 0.04       & 0.03       \\
 29  & MO         & 0.03       & 0.03       & 0.03       \\
 30  & MO         & 0.04       & 0.03       & 0.04       \\
 31  & MO         & 0.04       & 0.04       & 0.04       \\
 32  & MO         & 0.04       & 0.04       & 0.04       \\
\bottomrule
\end{tabular}
\end{center}
\end{scriptsize}

\end{toappendix}

\end{document}